\documentclass[aip,graphicx,reprint]{revtex4-1}

\usepackage{lipsum}
\usepackage{amsmath}
\usepackage{amsfonts}
\usepackage{amssymb}
\usepackage{graphicx}
\usepackage{xfrac}
\usepackage[colorlinks=true,citecolor=blue,urlcolor=magenta]{hyperref}
\usepackage{usebib}
\newbibfield{journal}
\newbibfield{volume}
\newbibfield{pages}
\newbibfield{year}
\newbibfield{url}
\bibinput{References}

\newcommand{\ket}[1]{\left| #1 \right>}
\renewcommand{\Re}{\operatorname{\mathbb{R}e}}
\renewcommand{\Im}{\operatorname{\mathbb{I}m}}
\newcommand{\RefCite}[1]{Ref. \onlinecite{#1}}
\newcommand{\IntdV}{\int\,d^3\mathbf{r}\,}
\DeclareMathOperator\Tr{Tr}
\newcommand{\avg}[1]{\left< #1 \right>}
\newcommand{\fullref}[1]{\citeauthor*{#1},  ``\usebibentry{#1}{title},'' \usebibentry{#1}{journal} \textbf{\usebibentry{#1}{volume}}, \usebibentry{#1}{pages} (\usebibentry{#1}{year})}

\newcommand{\ACSReprint}[2]{Reprinted with permission from \fullref{#1}. Copyright \usebibentry{#1}{year} American Chemical Society. Permission for further reuse should be directed to ACS.}
\newcommand{\ACSreprint}[2]{reprinted with permission from \fullref{#1}. Copyright \usebibentry{#1}{year} American Chemical Society. Permission for further reuse should be directed to ACS.}
\newcommand{\AIPReprint}[1]{Reprinted from \fullref{#1}, with the permission of AIP Publishing.}
\newcommand{\AIPreprint}[1]{reprinted from \fullref{#1}, with the permission of AIP Publishing.}
\newcommand{\APSReprint}[2]{Reprinted figure with permission from \fullref{#1}. Copyright \usebibentry{#1}{year} by the American Physical Society.}
\newcommand{\AAASReprint}[1]{From \fullref{#1}. Reprinted with permission from AAAS.}
\newcommand{\RCSreprint}[1]{republished with permission of the Royal Society of Chemistry from \fullref{#1}; permission conveyed through Copyright Clearance Center, Inc.}
\newcommand{\CCBYFourReprint}[1]{Reprinted from \fullref{#1} under the \href{https://creativecommons.org/licenses/by/4.0/}{CC BY 4.0} license.}
\newcommand{\CCBYFourreprint}[1]{reprinted from \fullref{#1} under the \href{https://creativecommons.org/licenses/by/4.0/}{CC BY 4.0} license.}
\newcommand{\CCBYThreeReprint}[1]{Reprinted from \fullref{#1} under the \href{https://creativecommons.org/licenses/by/3.0/}{CC BY 3.0} license.}
\newcommand{\CCBYNCNDFourReprint}[1]{Reprinted from \fullref{#1} under the \href{https://creativecommons.org/licenses/by-nc-nd/4.0/}{CC BY-NC-ND 4.0} license.}
\newcommand{\OSAReprint}[1]{Reprinted with permission from Ref. \onlinecite{#1} \textsuperscript{\copyright} The Optical Society.}
\newcommand{\SpringerNatureReprint}[2]{Reprinted with permission of Springer Nature Customer Service Center GmbH from \fullref{#1}, Copyright \usebibentry{#1}{year}.}

\draft 

\begin{document}


\title{Integrated Single Photon Emitters} 



\author{Junyi Lee}
\author{Victor Leong}
\author{Dmitry Kalashnikov}
\author{Jibo Dai}
\affiliation{Institute of Materials Research and Engineering, Agency for Science, Technology and Research (A*STAR), 2 Fusionopolis Way, \#08-03 Innovis, 138634 Singapore}
\author{Alagappan Gandhi}
\affiliation{Institute of High Performance Computing, Agency for Science, Technology and Research (A*STAR), 1 Fusionopolis Way, \#16-16 Connexis North, 138632 Singapore}
\author{Leonid Krivitsky}
\affiliation{Institute of Materials Research and Engineering, Agency for Science, Technology and Research (A*STAR), 2 Fusionopolis Way, \#08-03 Innovis, 138634 Singapore}


\date{\today}

\begin{abstract}
The realization of scalable systems for quantum information processing and networking is of utmost importance to the quantum information community. However, building such systems is difficult because of challenges in achieving all the necessary functionalities on a unified platform while maintaining stringent performance requirements of the individual elements. A promising approach which addresses this challenge is based on the consolidation of experimental and theoretical capabilities in quantum physics and integrated photonics. Integrated quantum photonics devices allow efficient control and read-out of quantum information while being scalable and cost effective. Here we review recent developments in solid-state single photon emitters coupled with various integrated photonic structures, which form a critical component of future scalable quantum devices. Our work contributes to the further development and realization of quantum networking protocols and quantum logic on a scalable and fabrication-friendly platform. 
\end{abstract}


\maketitle 
\tableofcontents


\section{Introduction}

The control and manipulation of physical objects at the quantum level has progressed considerably in the past decade. This quantum control promises fascinating advances to both technology and fundamental science. For example, use of quantum phenomena in data systems allows one to speed up computation and database search algorithms and to develop highly secure communication networks \cite{Kimble2008The, Sangouard2011Quantum}. A new class of devices are now in active development to fundamentally exploit the paradigm of quantum information and to make it accessible in practical applications.

A variety of physical systems have been identified as candidates for emerging quantum technologies, such as quantum dots, atomic defects in solids, atoms etc. Solid-state systems possess outstanding quantum optical properties that can be used to build practical quantum devices. Quantum information can be stored, for example, in the electron spin of a defect and the nuclear spin of nearby atoms with relatively long coherence times (a few ms) even at room temperature. During this time, it is feasible to record, manipulate, and read-out quantum information. Quantum logic can be implemented with incident microwave and RF fields, driving transitions between electron and nuclear sublevels. These blocks can interact with each other via photon-mediated interaction. Optical links connect the nodes as they enable reliable and fast transfer of quantum information. 

Building the quantum data system outlined above requires, among other things, efficient interfaces between solid-state single photon emitters (SPEs) and optical networks. A scalable and cost efficient approach towards implementing such interfaces relies on the use of integrated photonics technologies. The stationary nodes that encode the quantum information (for example, a spin state of an atomic-like defect) can be interconnected via optical waveguides made of low-loss materials. Optically resonant micro- and nanostructures can enhance the coupling of photons emitted from the stationary nodes into waveguides. Moreover, enhancement can also be obtained in waveguide QED due to slow light effects. Photons can then be routed to different nodes on the same chip or between different chips to create quantum entanglement between the nodes. Furthermore, one can use compact and highly sensitive photodetectors fabricated on the same chip for the read-out of quantum information. Generation, transport, manipulation and detection of quantum information can all be accomplished on this scalable, intrinsically stable and fabrication-friendly platform.

Besides applications in quantum information processing, the same physics and engineering concepts can be further applied in quantum metrology and sensing\cite{Maze2008Nanoscale, Balasubramanian2008}. Combining solid-state quantum systems with compact photonic devices will lead to the development of a new family of highly sensitive and compact  temperature, stress, inertia, electric and magnetic field sensors with high spatial resolution. These sensors will find applications in microelectronics, bio-chemical, and healthcare industries.

In this paper, we review recent experimental efforts in developing integrated solid-state single photon sources, which serve as a key enabling component for scalable quantum devices. Broader reviews of other necessary components in a quantum photonic chip are available elsewhere\cite{Elshaari2020Hybrid, kim2020hybrid}. Due to the multi-disciplinary nature of integrating solid state SPEs onto photonic chips, we have strove to make this review accessible to a broad audience with varying backgrounds by giving a theoretical overview of important SPE metrics and providing details related to the fabrication and integration of SPEs with resonant photonic structures. A shorter review covering solid state SPEs of slightly different systems can be found in \RefCite{Aharonovich2016Solid}.


We start off with a review of the nitrogen vacancy (NV) defect center in diamond as an illustrative example of a solid state SPE in  Section \ref{Sec:Theory}. After describing its basic photo-physical properties, we then provide a generic theoretical framework for the interaction of resonant photonic structures with quantum emitters. We then proceed with a review of experiments that have integrated NV centers in bulk diamonds with optical waveguides and resonant structures before discussing their prospects of larger scale integration with other photonic components and different material platforms (Section \ref{Sec:NVInBulk}). In Section \ref{Sec:ColorCenterND} we describe the integration of  color centers in nanodiamonds. The limitations and benefits of color centers in nanodiamonds versus those in bulk diamond are also discussed.

Section \ref{Sec:QuantumDots} is focused on integrated SPEs in Quantum Dots (QDs). Following a brief introduction into the photo-physics of QDs and their interaction with optical cavities, we describe methods for manipulating the QDs’ spin states. We then discuss experiments on interfacing multiple QDs coherently with the goal of generating entanglement between spatially separated QDs on the same chip.

In Section \ref{Sec:2DMaterials} we discuss SPEs in 2D materials, namely in  transitional metal dichalcogenides (TMDC) and hexagonal boron nitride (hBN). Following a brief overview of different types of 2D SPEs, we discuss their interfaces with resonant photonic structures. 

In Section \ref{sec:integ} we discuss various techniques for the integration of solid state SPEs with nanophotonics structures. This section describes dedicated nanofabrication and mechanical nano-manipulation methods.

Finally, we present two benchmarking tables in Section \ref{Sec:Conclusion} for comparing various experimentally integrated SPEs and resonators before concluding with a general outlook for the field. 



\section{Theoretical Background\label{Sec:Theory}}

Two-level quantum mechanical systems are natural candidates for true SPEs. At first sight, discrete energy levels within a solid-state system normally described by valence and conduction bands might seem at odds with intuition, but they can exist under special circumstances near a lattice defect or in quantum dots, where electrons are physically confined to such small spatial volumes that their eigen-energies are discrete. For many applications, however, it is not merely sufficient to have a two-level system since there are many other metrics to consider. In the following section, we use a nitrogen vacancy defect in diamond as an illustrative solid-state SPE to discuss these other considerations and to motivate the benefits of integrating solid-state SPEs with resonant photonic structures. Although we use the nitrogen vacancy defect in diamond as a concrete example, many of the challenges we point out are broadly applicable to quantum dots and 2-D materials, where these challenges can be similarly mitigated by their integration with photonic structures.

\subsection{The NV$^-$ center as an illustrative SPE\label{Sec:TheoryNV}}
\begin{figure}
\includegraphics[width=0.48\textwidth]{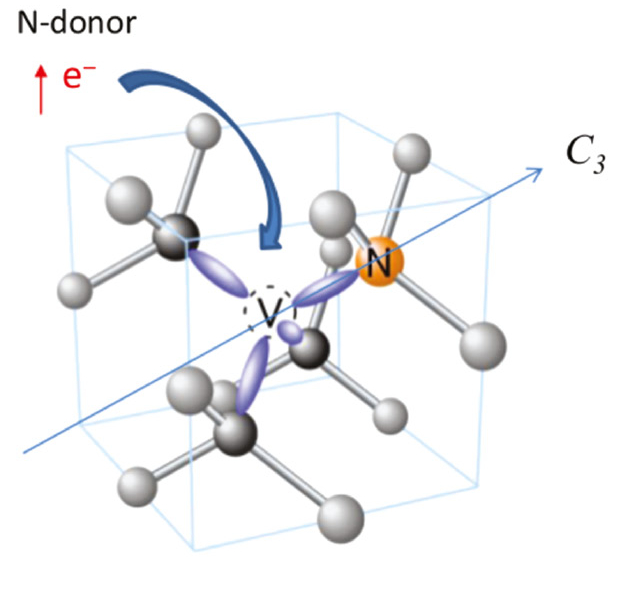}
\caption{Structure of the NV$^-$ center. Grey circles denote carbon atoms. The vacancy is denoted with a dashed vacant circle while the nitrogen substitutional defect is shown as an orange circle. $sp^3$ bonds are illustrated by grey rods while dangling bonds to the vacancy are drawn as purple ellipses. As shown in the figure, the NV center can have a $C_{3v}$ symmetry about any of the equivalent $<$111$>$ directions in a diamond that it is aligned with. A NV$^-$ state is negatively charged because it acquires and traps an additional electron from another donor. \CCBYFourReprint{Gali2019Ab}}
\label{Fig:NVStructure}
\end{figure}
\begin{figure}
\includegraphics[width=0.48\textwidth]{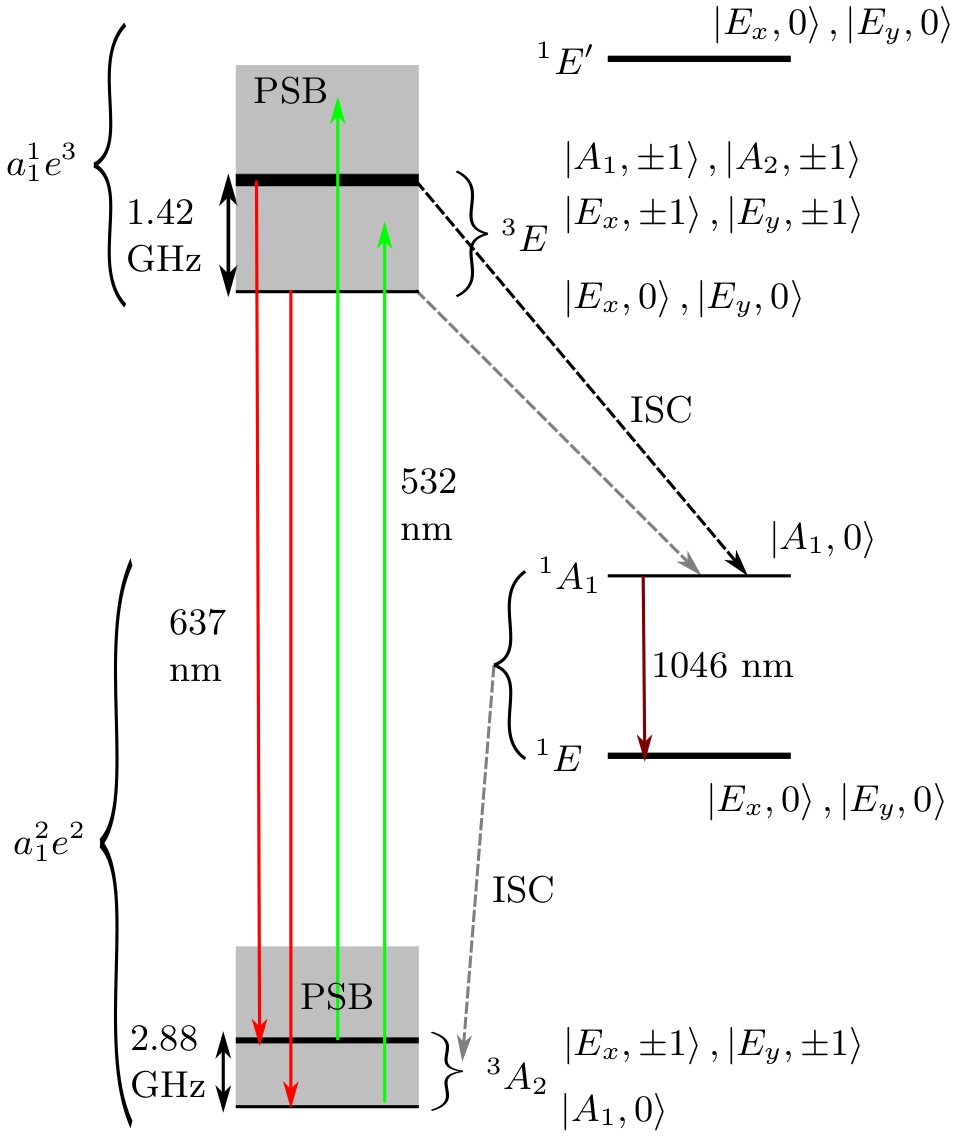}
\caption{Energy levels of the ground and first excited MO configuration of the NV$^-$. States labeled here with a $\ket{n,m_s}$ notation are spin-orbit states that transform according to a particular row of an irreducible representation of $C_{3v}$ (labeled by $n$; see \RefCite{Lenef1996Electronic, Doherty2011The}), which are the convenient basis states to use in the presence of spin-orbit/spin-spin interactions. Note that they are linear combinations of states with definite azimuthal spin quantum numbers $m_s$ and can hence have $m_s=\pm1$. PSB denotes the phonon side band. Solid lines with single arrow heads denote optical transitions, while solid lines with double arrow heads denote microwave transitions. Non-radiative inter-system crossings (ISC)s are denoted by dashed lines. A darker ISC line between the $m_s=\pm1$ states of $^3$E to $^1$A$_1$ is used to illustrate the faster ISC rate for that transition. Spacing of the energy levels are not drawn to scale.}
\label{Fig:NVEnergyLevels}
\end{figure}

A nitrogen vacancy (NV) center in diamond consists of a nitrogen atom (a substitutional defect) that is paired together with a neighboring vacant site (see Fig. \ref{Fig:NVStructure}). The substitution-vacancy pair can be aligned along any of the equivalent $<$111$>$ directions in the crystal and typically, all four possible orientations of the NV centers are found in equal proportions although relatively recent work have successfully created preferentially oriented NV centers \cite{Edmonds2012Production,Pham2012Enhanced,Fukui2014Perfect}. Although NV centers are known to exist in two distinct states (traditionally labeled as NV$^0$ and NV$^-$)\cite{Davies1992Vacancy}, it is the NV$^-$ state that has of late received the most attention due to its attractive optical and spin properties that have made it amenable to a variety of technological applications including quantum computation\cite{Jelezko2004Observation}, quantum information\cite{Hensen2015Loophole} and microscopic magnetic\cite{Taylor2008High}, electric\cite{Dolde2011Electric}, stress\cite{Kehayias2019Imaging}, inertia\cite{Ajoy2012Stable} and even thermal\cite{Neumann2013High} sensing. Figure \ref{Fig:NVEnergyLevels} shows the energy levels of the ground ($a_1^2 e^2$) and first excited ($a_1^1 e^3$) molecular orbital (MO) configurations\cite{Doherty2011The} of the NV$^-$ . Since the NV center has a $C_{3v}$ point symmetry with 2 one-dimensional irreducible representations ($A_1$ and $A_2$), and a two-dimensional irreducible representation ($E$)\cite{Lenef1996Electronic, Doherty2013The}, the nomenclature of the states and orbitals are typically given by their transformation properties under $C_{3v}$. The $^3$A$_2$ orbital singlet ground state has been amply confirmed by electron paramagnetic resonance (EPR) in the dark\cite{Redman1991Spin}, optical hole burning\cite{Reddy1987Two}, optically detected magnetic resonance\cite{Oort1988Optically,Brossel1952A} (ODMR), and Raman heterodyne measurements\cite{Manson1990Raman} to be a spin triplet that is split, at zero magnetic field, by $\approx$ 2.88 GHz\cite{Loubser1978Electron, Reddy1987Two} into a spin singlet A$_1$ (with $m_s=0$) and spin doublet E$_x$,E$_y$ (with $m_s=\pm1$) state due to spin-spin interactions\cite{Lenef1996Electronic, Martin1999Fine, Manson2006Nitrogen}.

Similarly, the orbital-doublet excited $^3$E state is also known to be a spin triplet via ODMR measurements\cite{Fuchs2008Excited,Neumann2009Excited}. The degeneracy of these states is likewise lifted by spin-spin and spin-orbit interactions\cite{Lenef1996Electronic} into E$_x$,E$_y$ states with $m_s=0$, and $A_1$,$A_2$,$E_x$ and $E_y$ states with $m_s=\pm1$ (see Fig. \ref{Fig:NVEnergyLevels}). Compared to the ground state however, the excited state is considerably more sensitive to shifts induced by lattice strains\cite{Tamarat2008Spin,Fuchs2008Excited,Neumann2009Excited} and the exact ordering of its states are more variable. Nevertheless, it is clear from EPR measurements that the zero-phonon line at 637 nm is associated with a spin-triplet excited state and that the triplet state consists of a spin-singlet (i.e. the $E_x$,$E_y$ states with $m_s=0$) and spin-doublet (consisting of $A_1$, $A_2$, $E_x$ and $E_y$ states with $m_s=\pm 1$) state that are separated by a zero-field splitting $D_{es}$ of $\approx$ 1.42 GHz \cite{Fuchs2008Excited, Neumann2009Excited}. Early uniaxial stress studies\cite{Davies1976Optical}, together with the measurements described above, indicate that the prominent zero-phonon line (ZPL) observed at 1.945 eV ($\approx$ 637 nm) is due to a $^3$A$_2\to\,^3$E transition (see Fig. \ref{Fig:NV_spectra}). Moreover, for a NV in a low strain environment, the ZPL emission is mostly linearly polarized with the plane of polarization depending on the NV's axial orientation, indicating that the transitions are mostly spin-conserving\cite{Davies1976Optical, Fu2009Observation}. 

\subsubsection{Photostability, saturated count rates and emission into ZPL\label{Sec:Photostability}}

Despite receiving less attention than the NV$^-$, the NV$^0$ state can affect the dynamics of NV$^-$ in important ways such as its photostability to which we now turn. An important characteristic of a SPE is its photostability. Although organic fluorescent dyes are an important class of single photon emitters, such sources have a significant disadvantage because of their susceptibility to photobleaching in which photochemical changes induced by the excitation light causes the emitter to degrade and permanently lose its fluorescence\cite{Demchenko2020Photobleaching}. This is, for obvious reasons, undesirable for many quantum information and communication applications. Fortunately, SPEs like defect centers in diamond, 2-D materials and quantum dots are considerably more robust. However, despite the excellent photostability of the NV$^-$ center, it is known that the fluorescence of NV$^-$ can nevertheless be significantly (but mostly reversibly) quenched\cite{Plakhotnik2011Nitrogen, Chapman2012Anomalous, Lai2013Quenching,Dumeige2004Photo} (sometimes called \textit{blinking}) when probed at high (typically pulsed) laser intensity and that this has been partially attributed to a spin-dependent ionization of NV$^-$ to NV$^0\,$\cite{Roberts2019Spin}. Moreover, oscillations between NV$^-$ and NV$^0$ have been observed as a function of the excitation wavelength\cite{Beha2012Optimum} and it is estimated that the NV center can be in the NV$^0$ state for $\sim$ 30\% of the time under usual operating conditions\cite{Waldherr2011Dark}. Nevertheless, these effects in NV may be mitigated by further annealing at 1200 $^\circ$C\cite{Naydenov2010Increasing, Orwa2011Engineering}, and more generally, the equilibrium concentration\cite{Wotherspoon2003Photo} and stability of the NV$^-$ state depends on the Fermi level, which may be altered, among other things, by doping (N is itself a deep donor), irradiation, heating, photo-excitation, surface-termination, and annealing conditions\cite{Davies1992Vacancy, Yoshimi1996Change, Collins2002The, Webber2012Ab, Fu2010Conversion, Hauf2011Chemical}. Similarly, although the exact mechanisms may differ, quantum dots in nanocrystals are also susceptible to blinking\cite{efros2016origin, Yuan2018Two}. 

\begin{figure}
\includegraphics[width=0.48\textwidth]{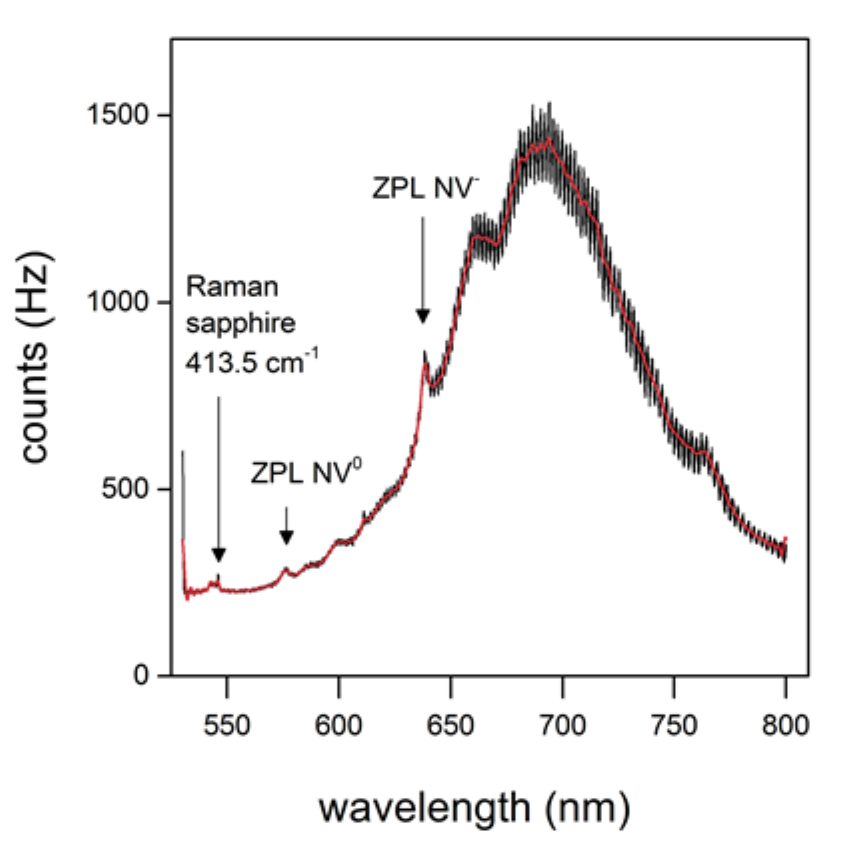}
\caption{Typical photoluminescence spectra of a NV center (on a sapphire substrate). Red-shifted emission to the ground state phonon sidebands constitute a large part of the spectrum and the NV ZPLs constitutes only a small fraction of the emission spectra. \SpringerNatureReprint{Brenneis2015Ultrafast}{2015}}
\label{Fig:NV_spectra}
\end{figure}

A related metric is the source's saturated count rate at which point further increasing the excitation power no longer induces significantly more fluorescence. Generally, most applications would benefit from a brighter source since there are always losses in any real world application and in particular, repeat-until-succeed quantum information schemes\cite{Lim2005Repeat, Lim2006Repeat, Bruschi2014Repeat} benefit from a higher rate of success with brighter sources. Table \ref{table:benchmarking} lists some experimentally measured count rates that give an idea of the brightness of various sources. We caution however that the numbers do not enable a fair comparison between different references since the measured count rates is highly dependent on experimental conditions such as the collection optics/position, detection efficiency, excitation intensity etc. that are highly variable from one experiment to another. 

\begin{figure}
\includegraphics[width=0.47\textwidth]{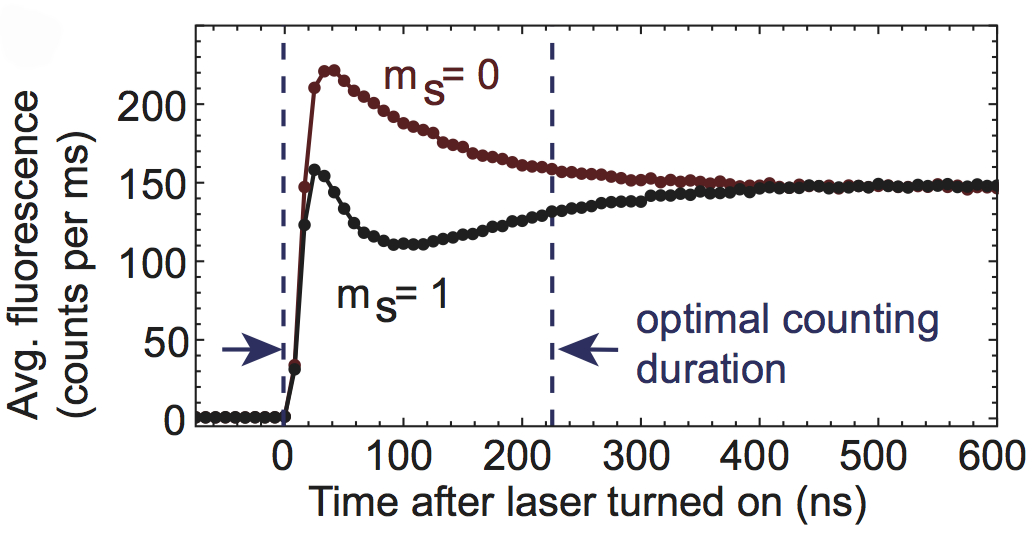}
\caption{Fluorescence from a NV$^-$ center after initialization to either a $m_s=0$ or $m_s=1$ state. The spin dependent fluorescence is due to a much faster non-radiative inter-system crossing (ISC) from the $m_s=\pm1$ excited states to the $^1$A singlet state (see Fig. \ref{Fig:NVEnergyLevels}), and consequently, fluorescence at the ZPL is due mostly to radiative decays from the $m_s=0$ excited state\cite{Choi2012Mechanism}. This spin-selective depopulation of $^3$E to $^1$A$_1$ also enables a $\sim$80\% polarization of the $m_s=0$ ground state by optical pumping\cite{Robledo2011Spin, Fuchs2010ExcitedState} that can then (if desired) be transferred to the $m_s=\pm1$ states by applying a microwave $\pi$ pulse. Data in this trace was obtained after averaging over $3\times10^7$ measurements. The optimal duration for photon counting can be obtained by using a maximum likelihood analysis to optimize the discrimination between $m_s=0$ versus $m_s=1$ states. \OSAReprint{Gupta2016Efficient}}
\label{Fig:NVFluorescenceTime}
\end{figure}

Physically, the count rate from a single emitter is inversely proportional to its excited state lifetime that can, as we further discuss in section \ref{Sec:theory_purcell}, be decreased by integrating the SPE with a resonator so that its rate of spontaneous emission into a resonant mode of the cavity is enhanced. This is useful for several reasons. Firstly, as Figure \ref{Fig:NV_spectra} illustrates, emission from the NV$^-$'s ZPL constitutes only a small fraction of its entire emission, with the majority of it coming from red-shifted transitions to the ground state phonon sidebands. The Debye-Waller factor, which quantifies this fraction, is particularly small for the NV$^-$ and is approximately 0.04\cite{Aharonovich2011Diamond, Zhao12Suppression}. This is undesirable for many quantum information applications since coherent information may be lost to the phonon reservoir when transitions to the phonon sideband occurs. A small Debye-Waller factor is also undesirable for many nano-sensing applications involving the NV$^-$ since many of them rely on the NV$^-$'s spin dependent ZPL emission to read out the NV$^-$'s spin state (see Fig. \ref{Fig:NVFluorescenceTime}). Fortunately, by integrating the SPE with a resonator that has been engineered to be resonant at the SPE's ZPL, high count rates into the desired ZPL transition can be achieved. 

\subsubsection{Indistinguishability\label{Sec:Indistinguishability}}

\begin{figure}
\includegraphics[width=0.48\textwidth]{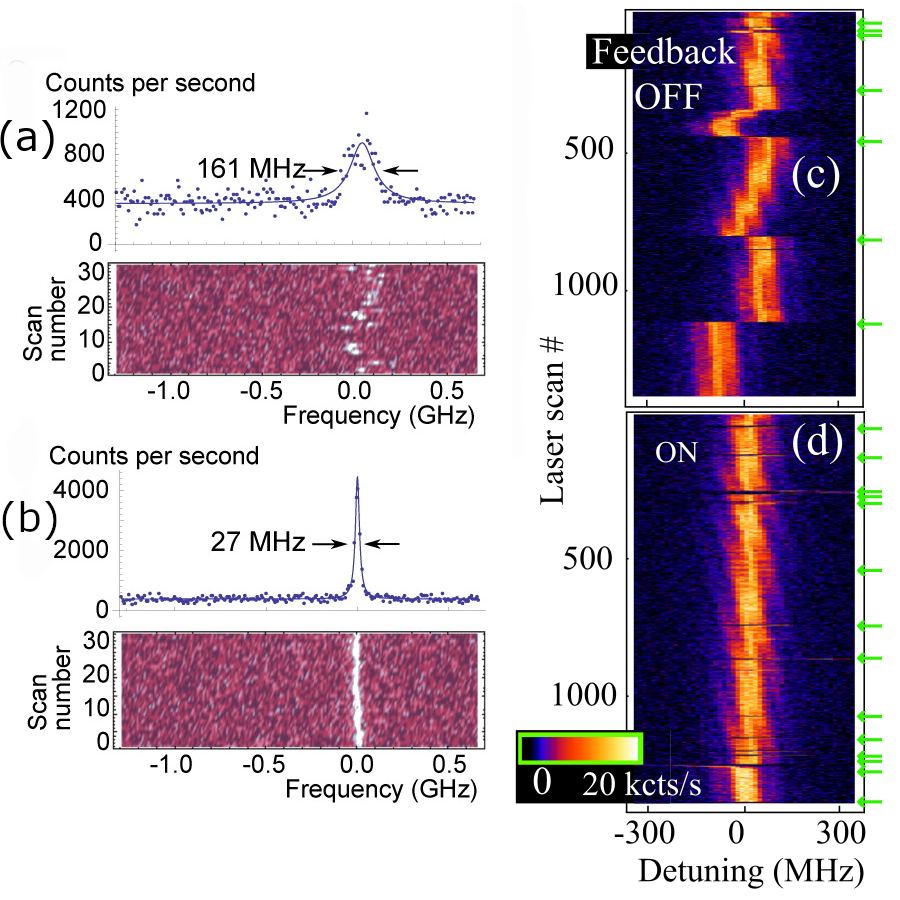}
\caption{(a), (b): Photoluminescence as a function of excitation frequency for a NV$^-$ in an appropriately processed diamond. The NV$^-$ is repumped at 532 nm for (a) and 575 nm for (b). Notice the significant decrease in spectral wandering for a resonant (of NV$^0$) repump. \ACSReprint{Chu2013Coherent}{2013}. (c), (d):  Photoluminescence as a function of excitation frequency for a NV$^-$ without (c) and with (d) active feedback of the ZPL transition. \CCBYThreeReprint{Acosta2012Dynamic}}
\label{Fig:NV_spectra_diffusion}
\end{figure}

Indistinguishability of photons is another important metric of SPEs designed for on-chip quantum information applications such as linear optical quantum computing\cite{Kok2007Linear}, quantum teleportation\cite{bennett1993teleporting, bouwmeester1997experimental, de2002experimental} and entanglement swapping\cite{zukowski1993event,pan1998experimental} that uses two-photon interference. In general, photons can be distinguished by their spectral/temporal shape as well as their polarization and time-of-arrival at a particular location. Although two identical but spatially separated two-level systems should in theory emit photons with the same spectral content, this is typically spoilt by the emitters' coupling to two different local environments. At short (compared to the radiative lifetime) time scales, interactions with the solid-state environment through, for example phonons, charge or spin noise\cite{kuhlmann2013charge, berthelot2006unconventional}, perturb the energies of the two-level system and induces dephasing of the optical transitions which homogeneously broaden the linewidth and decrease the indistinguishability of emitted photons. On the other hand, slower interactions (relative to the radiative lifetime) will induce spectral diffusion of the emission wavelength and inhomogeneously broaden the linewidth (see Figures \ref{Fig:NV_spectra_diffusion} and \ref{fig:qd_spec}a). Moreover, the excitation wavelength can also affect the distinguishability of emitted photons. 

In general, using a resonant excitation (637 nm for NV$^-$) is more advantageous to the creation of indistinguishable photons since it eliminates timing jitters (which decreases indistinguishability) associated with relaxation through phonons\cite{Kiraz2004Quantum}. Furthermore, higher frequency non-resonant excitation have a greater potential of ionizing defects around the SPE leading to larger charge fluctuations\cite{Bassett2011Electrical} that will in turn induce spectral diffusion. Consequently, resonant excitation is generally preferred for generating indistinguishable photons. However, we note that for the NV$^-$, resonant excitation cannot by itself generate ZPL photons continuously due to a permanent photoionization into the dark NV$^0$ state\cite{Siyushev2013Optically}. However, this can be alleviated by using a weak ($\sim$ 100 nW) repump laser that is resonant with the NV$^0$ ZPL (575 nm). Although a non-resonant 532 nm repump is also a popular choice, for reasons noted above, a weak resonant 575 nm repump is crucial to reducing longer term spectral diffusion by decreasing the probability of ionizing defects around the NV center\cite{Siyushev2013Optically} (see Fig. \ref{Fig:NV_spectra_diffusion}).

At short timescales, the photon's indistinguishability can be estimated by the metric\cite{Bylander2003Interference}
\begin{equation}
\xi\equiv\frac{T_2}{2 T_1}=\frac{1}{\Gamma\, T_1},
\label{Eq:P_def}
\end{equation}
where $T_1$ is the emitter's radiative lifetime and $T_2$ is the coherence time of the optical transition that is defined as\cite{Bylander2003Interference}
\begin{equation}
\frac{1}{T_2}=\frac{1}{2 T_1}+\frac{1}{T^*_2}=\frac{\Gamma}{2}.
\label{Eq:T2_def}
\end{equation}
$T^*_2$ here is the reciprocal of the dephasing rate $\Gamma^*=2/T^*_2$ that is causing $\textit{additional}$ (on top of the natural linewidth) homogeneous broadening while $\Gamma$ is the (angular) FWHM of the emission's homogeneously broadened spectrum. We note that in general $T_2\leq 2 T_1$ and therefore, $\xi\in[0,1]$. Experimentally, a common way to measure the indistinguishability of photons from a SPE is to measure the two-photon interference in a Hong-Ou-Mandel (HOM) type experiment\cite{Hong1987Measurement} in which two indistinguishable photons arriving at a 50/50 beam-splitter at the same time should always end up in the same output port. In such a setup, coincidence counts of photons at both output ports should decrease to zero for two indistinguishable photons. It can be shown that Eq. \eqref{Eq:P_def} gives the efficiency (or the normalized size) of a HOM dip with $\xi=1$ corresponding to perfect distinguishability of the photons\cite{Bylander2003Interference}. HOM interference between spatially separated defect centers have been experimentally demonstrated\cite{Sipahigil2012Quantum, Bernien2012Two, Sipahigil2014Indistinguishable} and as Table \ref{table:benchmarking} demonstrates, multiple integrated QDs have also demonstrated near ideal indistinguishability.

An intuitive way of understanding Eq. \eqref{Eq:P_def} is to see a transition with coherence time $T_2$ as emitting a photon wavepacket of temporal width $T_2/2$, which sets the width of a HOM interference dip since that is the maximum temporal overlap between two distinct photons. Moreover, there is a time jitter of order $T_1$ for the spontaneous emission to occur and therefore the probability of having two such distinct photon wavepacket interfere successfully is $\sim T_2/(2T_1)$. This suggests that one way of increasing the indistinguishability of photons from SPEs is to reduce their radiative lifetime $T_1$ by placing them into a resonant cavity. For example, this has been successfully done for quantum dots in micropillar cavities where a HOM dip was successfully measured\cite{santori2002indistinguishable, gazzano2013bright}. Moreover, such resonant cavities can be potentially used to implement other schemes for generating indistinguishable photons including cavity-assisted spin flip Raman transitions\cite{Kiraz2004Quantum, Sweeney2014Cavity}. Furthermore, cavities can be used to select a particular polarization, which is also important for indistinguishability, and a particular spatial mode, which can make outcoupling to an in-plane waveguide easier. Table \ref{table:benchmarking} summarizes some of the experimentally realized $T_1$ values of integrated SPEs. We have also provided $\Gamma^\dagger/(2\pi)$ values, which are the experimentally measured FWHM values that are not necessarily from homogeneously broadened lines. $\xi^\dagger$, which is calculated using Eq. \eqref{Eq:P_def} with $\Gamma\to\Gamma^\dagger$ is also tabulated as a measure of indistinguishability.  

Despite the utility of cavities described above, we acknowledge that they can only help with short term dephasing processes that homogeneously broaden the linewidth. For longer term fluctuations due, for example, to ionization of nearby defects that lead to local charge fluctuations\cite{Chu2013Coherent, Bernien2012Two} or drifting strains\cite{Acosta2012Dynamic} that shift the energies of the excited states, a different strategy is required. Passive solutions include carefully fabricating the material with tailored annealing and surface treatments\cite{Chu2013Coherent} while active solutions have also been investigated whereby the energies of the excited states are actively shifted via the Stark effect to stabilize the ZPL frequency\cite{Acosta2012Dynamic}. Using these strategies, near life-time limited linewidths ($\sim$ 13 MHz) have been obtained for NV$^-$ centers in bulk diamond at long time scales.

\subsubsection{Single photon purity}
\label{sec:purity}

Although a ``single photon'' can in principle be obtained by sufficiently attenuating a classical light source like a laser, there is a subtle but important difference between such attenuated sources and true SPEs: whereas a true SPE will never emit two photons at the same time, an attenuated source can occasionally deliver two photons in a single pulse. This is highly undesirable for some applications like quantum key distribution since security of the distributed key will be compromised in such cases and some of the key exchanged between the two parties will have to be discarded to ensure the security of the protocol\cite{Brassard2000Limitations, Lutkenhaus2000Security}. An important metric that is typically used to measure a source's single photon purity (in this sense) is the second order correlation function\cite{Glauber1963The}
\begin{equation}
g^{(2)}(x_1,x_2)=\frac{\Tr[\rho E^-(x_1) E^-(x_2) E^+(x_1) E^+(x_2)]}{\Tr[\rho E^-(x_1)E^+(x_1)]\Tr[\rho E^-(x_2)E^+(x_2)]},
\label{eq:g2_def}
\end{equation}
where $x_i$ are space-time coordinates, $\rho$ is the density matrix of the photons and $E^\pm(x_i)$ are positive/negative frequency components of the electric field operator. Typically, the field is assumed to be stationary and we are only interested in the time difference $\tau = x_2^0 - x_1^0$ so that Eq. \eqref{eq:g2_def} reduces to
\begin{equation}
g^{(2)}(\tau)=\frac{\avg{I(t) I(t+\tau)}}{\avg{I(t)}^2},
\label{eq:g2_reduced}
\end{equation}
where $I(t)$ denotes the intensity (or count rate) and the brackets $\avg{\dots}$ can be interpreted as a time average. Intuitively, Eq. \eqref{eq:g2_reduced} can be understood as the number of photons detected after a delay $\tau$ from the detection of a preceding photon at time $t$, normalized by the average count rate. Since the field is assumed to be stationary, $t$ drops out of the argument of $g^{(2)}$. Evidently, given that a true SPE can only emit a single photon at any particular instance of time, $g^{(2)}(0)$ should equal zero for a true SPE since the number of photons detected immediately after the detection of one photon should be exactly zero. In reality, additional background photons from other sources as well as finite time resolution and timing jitter in photodetectors and time-to-digital converters result in a non-zero $g^{(2)}(0)$. For SPEs like NV centers, quantum dots and defects in 2-D materials, it is typically not possible to optically resolve two closely separated emitters and a $g^{(2)}$ value of less than 0.5 is therefore typically used to discern whether emission is being collected from more than one emitter\cite{Kimble1977Photon}. Table \ref{table:benchmarking} lists some of the measured $g^{(2)}(0)$ values of various SPEs from experiment. In cases where corrected $g^{(2)}$ values are available, we give those that have been corrected for the timing response of the equipment used.

\subsection{Enhancement of ZPL emission using resonant cavities\label{Sec:theory_purcell}}

We motivated in sections \ref{Sec:Photostability} and \ref{Sec:Indistinguishability} how an enhanced spontaneous emission rate for SPEs is beneficial for numerous applications. In this section, we review how such an enhancement can be achieved by integrating SPEs with a resonant cavity. 

Classical electromagnetism shows that the time-averaged power radiated by a dipole emitter can be written as
\begin{equation}
P=\frac{\omega}{2}\Im[\mathbf{p}^*\cdot\mathbf{E}(\mathbf{r}_0)],
\label{Eq:DipolePower}
\end{equation}
where $\mathbf{p}$, $\omega$, and $\mathbf{E}(\mathbf{r}_0)$ are the dipole moment, angular frequency and electric field at position $\mathbf{r}_0$ respectively. The Purcell factor $F$ of a dipole emitter gives the enhanced emission rate of an emitter in an optical cavity, normalized with respect to its emission rate in free space (i.e. in the absence of the cavity). Using the well-known expression for $\mathbf{E}_0(\mathbf{r}_0)$, the electric field of a dipole in a homogeneous dielectric medium, the power radiated by a dipole in a homogeneous dielectric medium may be, from Eq. \eqref{Eq:DipolePower}, succinctly written as $P_0=\mu_0|\mathbf{p}|^2 n\omega^4/(12\pi c)$, where $n$ is the refractive index of the dielectric medium and $c$ is the speed of light\cite{Novotny2006Principles}. With these expressions for $P$ and $P_0$, the Purcell factor is given by $F=P/P_0$. One can also derive similar expressions for $P$ and $P_0$ using Fermi's golden rule. The quantum derivations will have an additional factor of 4, and this is related to fields from vacuum fluctuations\cite{Barnes2020Classical}. Nevertheless, the factor cancels out in the ratio of $F$ so that both classical and quantum derivations yield the same result.

When a dipole is placed in a structured dielectric medium like photonic crystal cavities, the corresponding electric field can be expressed as a sum of the dipole's own field $\mathbf{E}_0(\mathbf{r})$ and the scattered electric field $\mathbf{E}_s(\mathbf{r}_0)$. Using $F=P/P_0$ and the expression for $P_0$, it can be shown that
\begin{equation}
F=1+\frac{6\pi c}{\mu_0 n |\mathbf{p}|^2\omega^3}\Im[\mathbf{p}^*\cdot\mathbf{E}_s(\mathbf{r}_c)].
\label{eq:PurcellFactor}
\end{equation}
Equation \eqref{eq:PurcellFactor} may be evaluated by numerically simulating a dipole source in a finite-difference time-domain\cite{Taflove2000Computational} (FDTD) simulation of Maxwell's equations where in general the field created by the dipole consists of its own field and the scattered field, but the scattered field $\mathbf{E}_s(\mathbf{r})$ may be obtained by Fourier transforming the electric fields after the dipole excitation is switched off\cite{Ge2014Design}.

The fields $\mathbf{E}(\mathbf{r})$ and $\mathbf{H}(\mathbf{r})$ are the total fields in the presence of the dipole emitter with current density 
\begin{equation}
\mathbf{J}(\mathbf{r})=-i\omega\mathbf{p}\,\delta(\mathbf{r}-\mathbf{r}_0),
\label{Eq:Jmodel}
\end{equation}
where $\mathbf{p}$ is the dipole moment and $\delta(\mathbf{r})$ is the Dirac delta function. $\mathbf{E}(\mathbf{r})$ and $\mathbf{H}(\mathbf{r})$ obey Maxwell's equations for time harmonic fields
\begin{align}
\boldsymbol{\nabla}\times\mathbf{E}(\mathbf{r})&=i\omega\mu_0\mathbf{H}(\mathbf{r}) \label{Eq:CurlEHarmonic} \\
\boldsymbol{\nabla}\times\mathbf{H}(\mathbf{r})&=\mathbf{J}(\mathbf{r})-i\omega\epsilon(\mathbf{r})\mathbf{E}(\mathbf{r}). \label{Eq:CurlHHarmonic}
\end{align}
Let us assume the fields may be expanded using the quasi-normal modes of the optical cavity\cite{Kristensen2014Modes, Leung1994Completeness, Muljarov2016Exact}. Under single-mode conditions the expansions can be written as
\begin{align}
\mathbf{E}&=\sum_n\alpha_n\mathbf{E}_n=\alpha_0 \mathbf{E}_0 \label{Eq:EExpansion} \\
\mathbf{H}&=\sum_n\alpha_n\mathbf{H}_n=\alpha_0 \mathbf{H}_0, \label{Eq:HExpansion}
\end{align}
where $\alpha_0$ is the expansion coefficient for the single mode\cite{Sauvan2013Theory, Coccioli1998Smallest}. $\mathbf{E}_0$ and $\mathbf{H}_0$ are the electric and magnetic field of the quasi-normal single mode and they obey (in the absence of free currents)
\begin{align}
\boldsymbol{\nabla}\times\mathbf{E}_0(\mathbf{r})&=i\tilde{\omega}_0\mu_0\mathbf{H}_0(\mathbf{r}) \label{Eq:CurlEQuasiNormal} \\
\boldsymbol{\nabla}\times\mathbf{H}_0(\mathbf{r})&=-i\tilde{\omega}_0\epsilon(\mathbf{r})\mathbf{E}_0(\mathbf{r}), \label{Eq:CurlHQuasiNormal}
\end{align}
where $\tilde{\omega}_0$ is the complex frequency of the quasi-normal mode. Applying, Lorentz reciprocity theorem \cite{Sauvan2013Theory} to the set of fields ($\mathbf{E},\mathbf{H}$) and ($\mathbf{E}_0,\mathbf{H}_0$), we have $\int\,d^3\mathbf{r}\,\boldsymbol{\nabla}\cdot(\mathbf{E}\times\mathbf{H}_0-\mathbf{E}_0\times\mathbf{H})=0$. Subsequently, equations \eqref{Eq:CurlEHarmonic}, \eqref{Eq:CurlHHarmonic}, \eqref{Eq:CurlEQuasiNormal} and \eqref{Eq:CurlHQuasiNormal} can be combined as $i(\omega-\tilde{\omega}_0)\int\,d^3\mathbf{r}\,(\mathbf{E}\cdot\epsilon(\mathbf{r})\mathbf{E}_0-\mu_0\mathbf{H}\cdot\mathbf{H}_0)=\int\,d^3\mathbf{r}\,\mathbf{J}\cdot\mathbf{E}_0$. Using this result and equations \eqref{Eq:Jmodel}, \eqref{Eq:EExpansion} and \eqref{Eq:HExpansion}, it can be shown that the complex expansion coefficient $\alpha_0$ is
\begin{equation}
\alpha_0=-\frac{\omega\mathbf{p}\cdot\mathbf{E}_0(\mathbf{r}_0)}{(\omega-\tilde{\omega}_0)I},
\end{equation}
where $I$ is given by
\begin{equation}
I=\int\,d^3\mathbf{r}\, (\mathbf{E}_0\cdot\epsilon(\mathbf{r})\mathbf{E}_0-\mu_0\mathbf{H}_0\cdot\mathbf{H}_0).
\end{equation}
Using this integral, a mode volume $V$ for the single mode can be defined as
\begin{equation}
V\equiv\frac{I}{2\epsilon_0 n^2 [\mathbf{E}_0(\mathbf{r}_0)\cdot\mathbf{p}]^2},
\end{equation}
where $\epsilon_0$ is the vacuum permittivity and $n^2$ is the square of the refractive index. 

For cavities with high quality factor, the quasi-normal modes can be approximated to be normal modes where the integral $I$ and the mode volume $V$ are real-valued. If we assume that $\mathbf{E}_0\in\Re$,  we see from equations \eqref{Eq:CurlEQuasiNormal} and \eqref{Eq:CurlHQuasiNormal} that $\mathbf{H}_0\in\Im$, and therefore $I=\int\,d^3\mathbf{r}\, (\epsilon|\mathbf{E}_0|^2+\mu_0|\mathbf{H}_0|^2)=2\int\,d^3\mathbf{r}\,\epsilon|\mathbf{E}_0|^2$, where we have used the fact that $\IntdV \epsilon|\mathbf{E}_0|^2=\IntdV \mu_0|\mathbf{H}_0|^2$ which follows from the identity
\begin{equation}
\boldsymbol{\nabla}\cdot(\tilde{\mathbf{E}}\times\tilde{\mathbf{H}})=\tilde{\mathbf{H}}\cdot\boldsymbol{\nabla}\times\tilde{\mathbf{E}}-\tilde{\mathbf{E}}\cdot\boldsymbol{\nabla}\times\tilde{\mathbf{H}},
\label{Eq:VectorIdn1}
\end{equation}
and the requirement that $\IntdV \boldsymbol{\nabla}\cdot(\mathbf{E}_0\times\mathbf{H}_0^*)=0$ for normal modes in which there is no out flow of energy from the cavity. By assuming a real $I$ and using equations \eqref{Eq:EExpansion} and \eqref{Eq:DipolePower}, it may be shown that the maximum Purcell factor on resonance (i.e. $\omega=\omega_0=\Re[\tilde{\omega}_0]$) is given by
\begin{equation}
F_c=\frac{P}{P_0}=\frac{3}{4\pi^2}\left(\frac{\lambda_0}{n}\right)^3\frac{Q}{V},
\label{eq:FcHighQ}
\end{equation}
where $Q=\Im[\tilde{\omega}_0]/(2\Re[\tilde{\omega}_0])$ and $\lambda_0=c/(2\pi\omega_0)$. In the case of a slight deviation off resonance, it is straightforward to show that $F=F_c L_s(\omega)$, where $L_s(\omega)$ is the Lorentzian line shape function
\begin{equation}
L_s(\omega)=\frac{\omega_0^2}{\omega^2}\frac{\omega_0^2}{\omega_0^2+4Q^2(\omega-\omega_0)^2}.
\end{equation}
Sauvan \textit{et al.}\cite{Sauvan2013Theory} showed that for cavities with small quality factors, it is important to include the leakage part of the fields in the integral of $I$. Consequently, this leads to a complex volume, and in this case one can obtain a generalized $F$
\begin{equation}
F=F_c L_s(\omega)\left(1+2Q\frac{\omega-\omega_0}{\omega_0}\frac{\Im[V]}{2\Re[V]}\right),
\end{equation}
where $F_c$ is now
\begin{equation}
F_c=\frac{3}{4\pi^2}\left(\frac{\lambda_0}{n}\right)^3\frac{Q}{\Re[V]}.
\label{eq:FcLowQ}
\end{equation}

\begin{figure}
\includegraphics[width=0.47\textwidth]{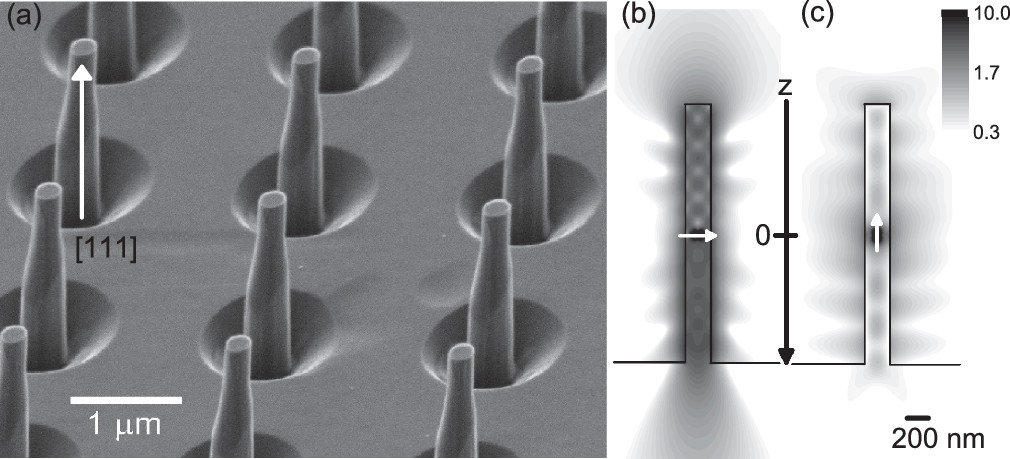}
\caption{(a) Scanning electron microscope (SEM) image of diamond nanopillars. (b), (c) FDTD simulations of the E-field's magnitude for orthogonal orientations of the optical dipole (shown with white arrows). \AIPReprint{Neu2014Photonic}}
\label{Fig:NanopillarDipoleOrient}
\end{figure}

Equations \eqref{eq:FcHighQ} and \eqref{eq:FcLowQ} indicate that to maximize the spontaneous emission at the ZPL, it is necessary to achieve a high $Q/V$ ratio at the ZPL wavelength. Typical cavity structures include microdisks\cite{srinivasan2007linear, Khanaliloo2015High}, micropillars\cite{song2019micropillar}, nanopockets\cite{Alagappan2018Diamond, Alagappan2020Purcell, Froch2020Photonic} and photonic crystals\cite{fan2010comparison}. The whispering gallery modes of microdisk cavities can have very high quality factors of $Q\sim10^5$, but relatively large mode volumes\cite{srinivasan2007linear}. For micropillar resonators with integrated Bragg mirrors, very high $Q>250,000$ and small $V<(\lambda/n)^3$ can be achieved, where $\lambda$ is the wavelength and $n$ is the refractive index\cite{song2019micropillar}. Nanowires\cite{Babinec2010A} and nanopillars\cite{Neu2014Photonic} (see Fig. \ref{Fig:NanopillarDipoleOrient} for an example), like micropillars, are also efficient vertical-emitting photon sources when coupled with a quantum emitter such as a quantum dot or NV$^-$ center. Yet although they are useful for applications requiring outcoupling of light from the plane of the device, they are not as well suited for planar routing of light. Photonic crystal cavities (PCCs) present an appealing compromise between high $Q$, low $V$, and efficient in-plane coupling. PCCs commonly take the form of a membrane of material with a periodic lattice of air holes, where selected holes have been displaced to form a defect in the photonic crystal bandgap. A prominent example is the L3 cavity with three missing holes in a line (see Fig. \ref{Fig:DiamondL3Cavity} for an example), which supports polarized single modes with a relatively wide spectral margin\cite{fan2010comparison}. For fabricated L3 cavities with embedded quantum dots (QDs) on a GaAs platform, $Q\sim10^4$ and $V\sim(\lambda/n)^3$ has been demonstrated\cite{lagoudakis2013deterministically, luo2019spin}, while an NV$^-$ center integrated with an all-diamond L3 cavity achieved a Purcell factor of 70\cite{Faraon2012Coupling}. A H1 PCC, which consists of a single central hole in a triangular 2D lattice, was also recently used to achieve a 43 fold spontaneous emission enhancement from an InGaAs quantum dot in a GaAs cavity\cite{Liu2018High}.

For optimal coupling to the cavity, it is important for the quantum emitters, which may be modeled as dipole emitters, to be correctly oriented and positioned to match the cavity's mode and polarization. For example, light emission from a strain-free NV$^-$ center has a single ZPL transition that may be modeled using a pair of orthogonal dipoles\cite{Jamali2014Microscopic} (with equal strength of dipole moment) perpendicular to the NV axis while SiV centers, which have four ZPL transitions at cryogenic temperatures\cite{Hepp2014Electronic, Neu2011Fluorescence}, may be modeled as single dipole emissions. A misalignment of the optical dipole's orientation with respect to the cavity can significantly affect the modes it excites as Fig. \ref{Fig:NanopillarDipoleOrient} shows. 

Cavities are not only useful for enhancing spontaneous emission but they are also useful in enhancing absorption, which can be beneficial under certain spin-to-charge\cite{Shields2015Efficient,Hopper2016Near}, spin-to-photocurrent\cite{Bourgeois2015Photoelectric, Brenneis2015Ultrafast} and magnetometry\cite{Jensen2014Cavity, Demeige2013Magnetometry} read-out schemes. Besides enabling enhanced emission/absorption and out-of-plane waveguiding of quantum emitters, integrated photonic structures can also of course provide for in-plane waveguiding through conventional waveguides or line-defect PCC waveguides. In the following sections, we review various examples of integrated SPEs and different integration techniques that have been employed.


\section{Integrated NV$^-$ centers in bulk diamond\label{Sec:NVInBulk}}

NV$^-$ centers have been successfully integrated with a variety of photonic structures in bulk diamond, which we here define as diamond substrates that are larger than nano-diamonds. Although nano-diamonds are in some ways easier to integrate with dissimilar photonic structures, they tend to suffer from poorer photostability and shorter coherence times due to their larger surface area to volume ratio that makes them particularly susceptible to surface effects. It is therefore desirable to integrate NV$^-$ centers in bulk diamond to other photonic structures. These structures can be fabricated from the same bulk diamond substrate or they could be made of a dissimilar material and coupled to NV$^-$ centers in a bulk diamond substrate evanescently. Moreover, with the advent of pick-and-place techniques, it is possible to envision NV$^-$ in diamond photonic structures that are in turn coupled to other dissimilar material systems that could offer additional functionalities. 

For fabricating all diamond photonic structures that contain only a single mode around the NV$^-$ center's ZPL wavelength (637 nm), it is necessary to use thin membranes of diamond ($n\approx2.4$ at 637 nm) that are $\sim$ 200 nm thick. Although such membranes may be obtained from nanocrystalline diamond films grown on a substrate\cite{Wang2007Observation, Wang2007Fabrication}, their optical quality is typically worse compared to bulk single-crystal diamonds due to increased absorption and scattering\cite{Wang2007Observation}. It is therefore preferable to obtain such thin diamond membranes from (typically oxygen plasma) reactive ion etching (RIE) of bulk single-crystal diamonds\cite{Luozhou2013Reactive}, which is a fabrication process that has been demonstrated to be compatible with moderately long NV$^-$ spin coherence lifetimes of $\gtrsim$ 100 $\mu$s\cite{Hodges2012Long} while also being consistent with low optical losses\cite{Babinec2010A,Faraon2012Coupling}. Moreover, RIE (with oxygen plasma) can be used to create surface-termination of the diamond membrane that encourages the conversion of NV$^0$ to NV$^-$ states\cite{Hauf2011Chemical}. To obtain good mode confinement, it is also typical to undercut the structures so as to achieve a large refractive index contrast between diamond and air. This may be achieved in several ways. For example, the diamond membrane can be first mounted on a sacrificial substrate, processed, and then made into a freestanding structure by a final isotropic etch step that removes the sacrificial substrate under the area of interest\cite{Faraon2012Coupling, Hausmann2013Coupling}. Alternatively, angular\cite{Burek2012Free, Burek2014High} and quasi-isotropic\cite{Khanaliloo2015Single, Khanaliloo2015High, Mouradian2017Rectangular} RIE etching can also be employed to create such freestanding structures. Instead of creating a freestanding structure, another typical variation is to further etch the substrate to create a pedestal, which would reduce the leaking of fields into the substrate\cite{Barclay2009Hybrid}.

Such all diamond photonic structures can then, in principle, be transferred to other dissimilar systems by using a pick-and-place technique to create a hybrid material platform. For example, this was demonstrated in \RefCite{Mouradian2015Scalable} where NV$^-$ containing diamond waveguides were transferred to a silicon platform containing SiN waveguides. GaP-diamond is another popular hybrid platform due to both the high refractive index of GaP ($\approx3.3$ at 637 nm) (compared to diamond $\approx2.4$ at 637 nm), and its relative ease of fabrication using standard semiconductor processing technology. In addition, unlike an all (bulk) diamond platform, which by inversion symmetry has a zero second-order non-linear susceptibility ($\chi^{(2)}$) (we note however, that diamond has a non-zero third-order non-linear susceptibility $\chi^{(3)}$ and that it has a relatively high non-linear refractive index that allows  it to be harnessed for non-linear four-wave mixing processes\cite{Hausmann2014Diamond}), GaP possesses a relatively large $\chi^{(2)}$ that allows it to be used in non-linear processes such as second harmonic generation\cite{Rivoire2009Second}. Moreover, unlike diamond which is forbidden by symmetry to have a bulk linear electro-optic coefficient\cite{Anastassakis1971Second}, GaP has a non-zero linear electro-optic coefficient ($r_{41}\approx-0.97$ pm/V at 633 nm\cite{Nelson1968Electro}) that enables it to be used for active electro-optic switching applications (as has been demonstrated in AlN material systems\cite{Xiong2012Aluminum, Xiong2012Low}), and as a III-VI semiconductor, GaP can potentially host on-chip integrated single-photon detectors as demonstrated on GaAs waveguides\cite{Sprengers2011Waveguide}. In the sub-sections below, we review some examples of integrated NV$^-$.

\subsection{Integration with waveguides}

\begin{figure}
\includegraphics[width=0.5\textwidth]{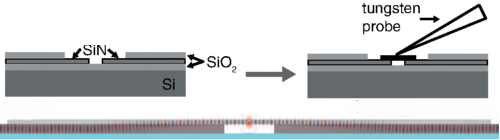}
\caption{Top: Schematic of single-mode diamond waveguides containing single NV$^-$ centers suspended over single-mode SiN waveguides assembled using a tungsten probe. Bottom: Electric field profiles from FDTD simulations showing the transfer of mode from the suspended diamond waveguide (grey) into the underlying SiN waveguide (brown). \CCBYThreeReprint{Mouradian2015Scalable} }
\label{Fig:DiamondToSiNWG}
\end{figure}

NV$^-$ centers have been integrated with waveguides in a variety of ways. One direct approach is to fabricate an all-diamond waveguide on a thin diamond membrane using a mask and RIE etch. Since the diamond membrane will have randomly dispersed native NV$^-$ centers, some of these diamond waveguides will, by chance, have NV$^-$ centers in the approximately correct location within the waveguides. These NV$^-$ integrated waveguides can then be post-selected and used to form more complex photonic circuits. This approach was taken in \RefCite{Mouradian2015Scalable} where tapered diamond micro-waveguides were fabricated from a 200 nm thick single crystal diamond membrane with a Si mask and RIE etch. In this case, the Si mask was separately fabricated using well developed silicon fabrication processes and then transferred onto a diamond substrate using a mask transfer technique. Due to the mature silicon technology, such masks can be fabricated with stringent tolerances and their patterns can then be transferred to the diamond substrate after a RIE etch. The resulting diamond waveguides are then characterized by photoluminescence measurements and those that are found to have a single NV$^-$ in the center of the waveguide, as verified by $g^{(2)}(0)$ measurements, are selected and placed on top of an air gap in between SiN waveguides by a probe (see Fig. \ref{Fig:DiamondToSiNWG}). Due to the air gap and taper of the diamond waveguides, up to 86\% of the NV$^-$'s ZPL emission can be coupled to the SiN waveguides\cite{Mouradian2015Scalable}. The background corrected saturated count rate from one end of the SiN waveguides was estimated to be $1.45\times10^6$ photons/s and a $g^{(2)}(0)$ as low as 0.07 was obtained. Photoluminescence excitation measurements of the NV$^-$ revealed a FWHM of 393 MHz and ODMR Hahn-echo measurements revealed a relatively long spin coherence time of $T_2\approx 120$ $\mu$s, which is, as in \RefCite{Li2015Coherent}, close to the spin coherence time of NV$^-$ centers in high quality bulk diamond crystals. This can likely be extended to the ms range if isotopically purified $^{12}$C carbon ($^{13}$C has a nuclear spin that decoheres the NV$^-$ spin) is used instead\cite{Balasubramanian2009Ultralong}.

\begin{figure}
\includegraphics[width=0.47\textwidth]{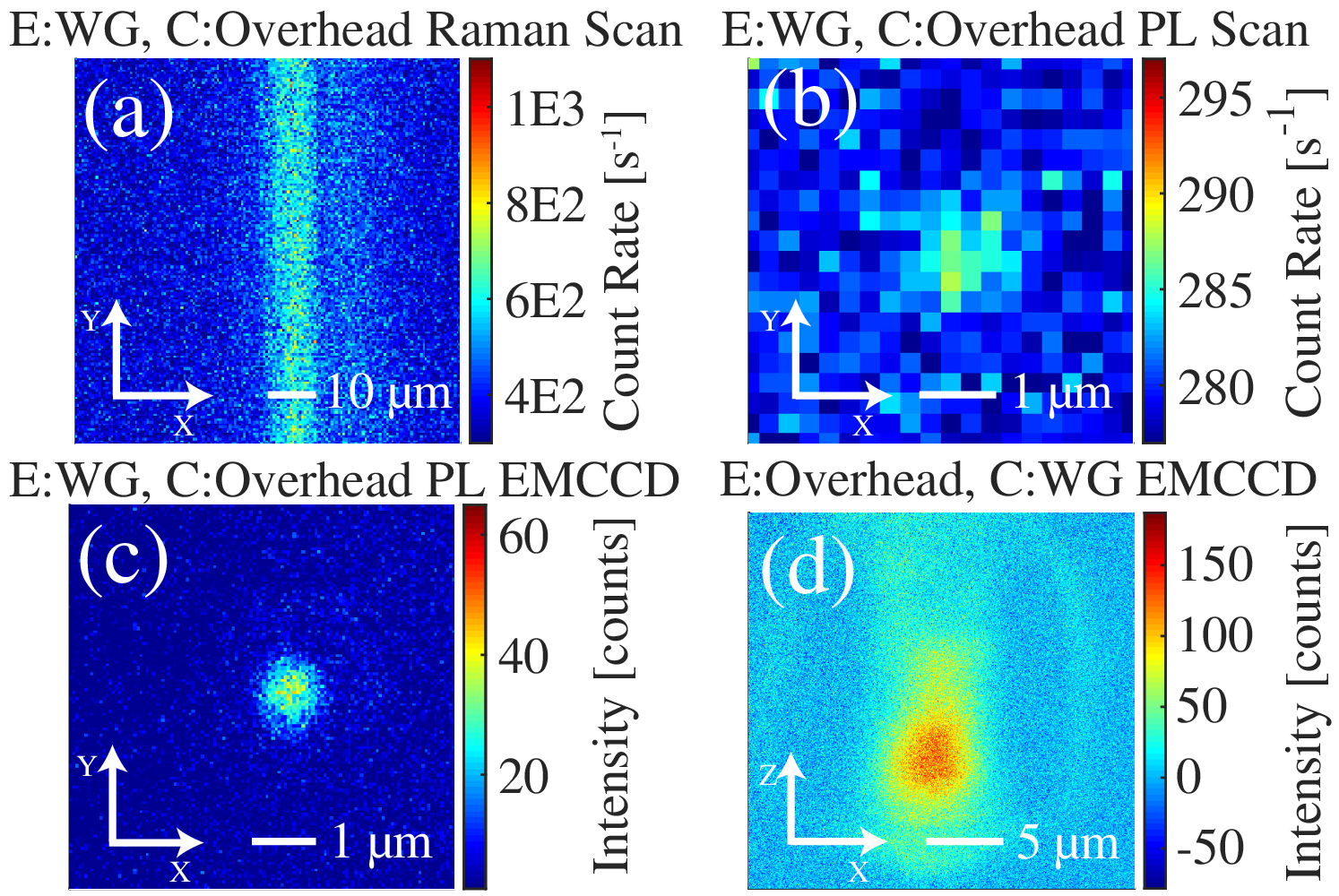}
\caption{Raman and NV fluorescence for various excitation (E)/collection (C) modes of an integrated laser written diamond waveguide and NV$^-$ center. (a) Raman scattered light collected above the waveguide after butt-coupled excitation of the waveguide, demonstrating waveguiding within the diamond. (b) Fluorescence of NV$^-$ above the waveguide after sending in excitation light from the end of the waveguide, demonstrating that the NV$^-$ is sufficiently close to the optical mode of the waveguide to interact with it. Fluorescence was measured using a home-built confocal microscope. (c) Same as in (b) but the fluorescence is measured using an electron multiplying charged coupled detector (EMCCD). (d) Fluorescence of the NV$^-$ measured at the end of the waveguide using an EMCCD after the NV$^-$ center was excited from above. \OSAReprint{Hadden2018Integrated}}
\label{Fig:LaserWrittenWaveguideNV}
\end{figure}

It is also possible to integrate NV$^-$ with diamond waveguides using fs-laser writing. As discussed in section \ref{Sec:LaserWriting}, fs laser pulses are capable of creating NV$^-$ centers in diamond. Moreover, as in the case of crystals like LiNbO$_3$\cite{Burghoff2006Efficient,Burghoff2007Origins} and sapphire\cite{Benayas2010Thermal}, it is possible to inscribe waveguides in diamond with fs laser writing. This may by accomplished by writing two parallel lines in diamond that results in graphitization of material within the focus leading to a decreased refractive index that in turn enables the confinement of an optical mode between the two laser written lines. In addition, the graphitized material, which has lower density, expands and causes stress-induced modification to the refractive index of the surrounding diamond that leads to vertical confinement of the optical mode\cite{Sotillo2018Polarized}. Importantly, the laser inscribed waveguides in diamond survive annealing at 1000 $^\circ$C$\,$\cite{Sotillo2017Visible}, which is commonly required for the formation of NV$^-$ centers, but is not necessarily guaranteed as in the case of laser inscribed waveguides in sapphire\cite{Benayas2010Thermal}. Since the same fs laser system can be used to both create NV$^-$ centers and write waveguides within bulk diamond, sub-micron relative positioning accuracy is possible between the NV$^-$ center and waveguide. Using this technique, single NV$^-$ centers, with a $31\pm9$ \% probability of creation per 28 nJ pulse, were successfully incorporated into the midst of laser written diamond waveguides and waveguiding of their spontaneous emission was confirmed\cite{Hadden2018Integrated} (see Fig. \ref{Fig:LaserWrittenWaveguideNV}). The measured $g^{(2)}(0)$ values of such NV$^-$ emission was as low as 0.07, confirming that single NV$^-$ centers were indeed deterministically created in the waveguides\cite{Hadden2018Integrated}.

NV$^-$ centers in bulk diamond have also been integrated with GaP waveguides, although there has not, to our knowledge, yet been a demonstration of deterministic \textit{single} NV$^-$ integration with GaP waveguides. However, NV$^-$ centers created by ion implantation in HPHT type Ib diamonds (at a depth of $\approx$ 100nm) have been evanescently coupled to a 120 nm thick GaP rib waveguide that was transferred onto diamond via epitaxial liftoff\cite{Yablonovitch1990Van} after removal from its underlying Al$_{0.8}$Ga$_{0.2}$P sacrificial layer atop a GaP substrate\cite{Fu2008Coupling}.  Evanescent coupling between the NV$^-$ and GaP waveguide was successfully observed when NV$^-$ emission was detected after sending in a 532 nm excitation beam through the GaP waveguide\cite{Fu2008Coupling}. The evanescent coupling requires that NV$^-$ centers are created close to the diamond's surface and for gaps between GaP and the diamond substrate to be minimized. Indeed, a significant disadvantage of a (bulk) hybrid platform compared to an all-diamond one is that the NV$^-$ centers cannot generally be placed in a maxima of the optical mode that would otherwise enable good optical coupling and an enhancement of spontaneous emission rates. To mitigate this, NV$^-$ centers may be coupled to optical resonators such as microdisks with sufficiently high quality factor\cite{Barclay2009Chip} and light within these resonators may then be outcoupled via coupling with another waveguide\cite{Thomas2014Waveguide, Gould2016Large} (see section \ref{Sec:diamond_large_scale_integration}). Despite the high refractive index of GaP, waveguiding in a GaP waveguide atop a diamond substrate can still be significantly lossy due to the reduced effective index of the guided mode and the moderately high refractive index of diamond. To enable waveguiding and to reduce losses due to mode leakage into the substrate, it is common to decrease the effective index of the substrate (from its bulk value) by etching it so as to create a diamond pedestal beneath the resonator \cite{Barclay2009Hybrid, Barclay2009Chip, Barclay2011Hybrid, Thomas2014Waveguide, Gould2016Large}. 

\subsection{Integration with resonators}

\subsubsection{Diamond ring resonators}

\begin{figure}
\includegraphics[width=0.47\textwidth]{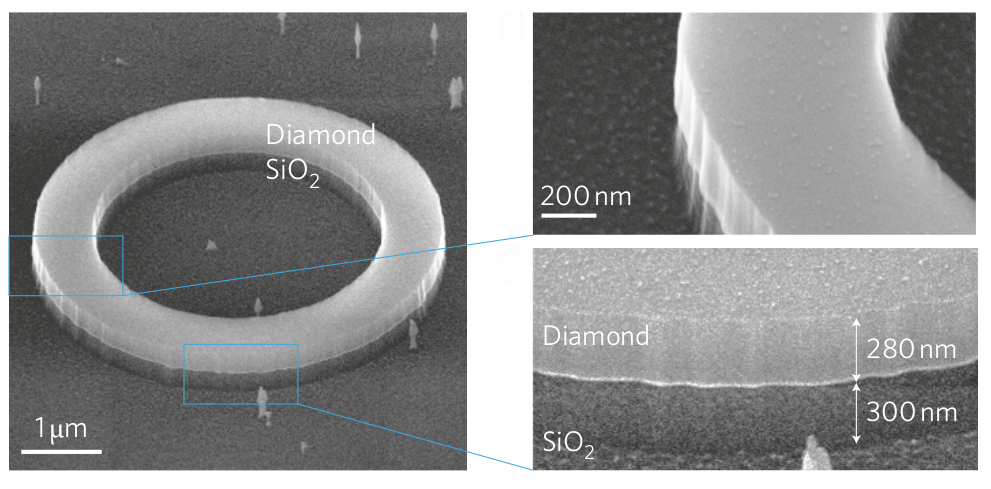}
\caption{A micro-ring resonator fabricated out of a single diamond crystal using EBL followed by RIE. Insets show the side wall roughness as well as how the SiO$_2$ substrate has been etched by 300 nm during the fabrication process. A NV$^-$ center coupled to the resonator was observed to have a spontaneous emission enhancement of $\sim$ 12. Although theoretical $Q$ values exceeded $10^6$, the achieved quality factors were only $\sim$ 5000 due to surface roughness of the ring. \SpringerNatureReprint{Faraon2011Resonant}{2011}}
\label{Fig:AllDiamondMicroRing}
\end{figure}

One of the earliest demonstration of NV$^-$ integration with a resonator came in 2011 with the successful coupling of a NV$^-$ with a 4.8 $\mu$m outer diameter and $700\times280$ (width$\times$height) nm diamond micro-ring resonator on top of a 300 nm high SiO$_2$ pedestal with mode volumes in the range of $\approx17 - 32$ $(\lambda/n)^3\,$ \cite{Faraon2011Resonant} (see Fig. \ref{Fig:AllDiamondMicroRing}). Photoluminescence measurements resolved roughly ten native NV$^-$ lines within the resonator. Characterization of the resonator was done at cryogenic temperatures ($<$ 10 K) and xenon was flowed through the cryostat, which allowed tuning of the cavity's resonance as the xenon condensed on the cavity and altered its resonance wavelength\cite{Mosor2005Scanning}. A spontaneous emission enhancement of $\approx$ 12 was obtained on resonance with a FWHM of $\approx$ 40 GHz and a radiative lifetime of 8.3 ns. The broad linewidth has been attributed to strain within the diamond.

\subsubsection{Diamond 2-D photonic crystal cavities}

\begin{figure}
\includegraphics[width=0.47\textwidth]{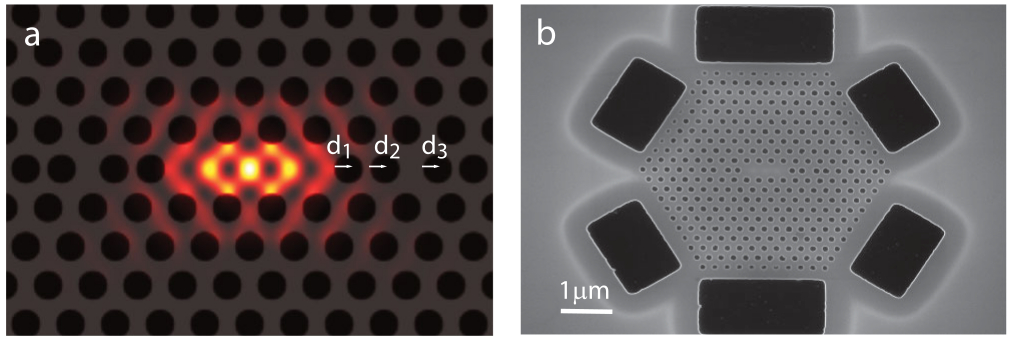}
\caption{(a) Simulated electric field energy density of a L3 cavity. (b) SEM image of a L3 cavity fabricated on a single crystal diamond membrane using EBL and RIE as described in the text. \APSReprint{Faraon2012Coupling}{2012} }
\label{Fig:DiamondL3Cavity}
\end{figure}

As discussed in section \ref{Sec:theory_purcell}, although ring resonators are capable of achieving high $Q$ factors, yet they tend to have large mode volumes and are therefore not as ideal in achieving Purcell factors. On the other hand, photonic crystals cavities provide a good compromise between having high quality factors and small mode volumes, which makes them particularly useful for enhancing a SPE's spontaneous emission rate. Accordingly, there has been various attempts at integrating NV$^-$ centers with PCCs. In \RefCite{Faraon2012Coupling}, an all-diamond suspended L3 cavity with theoretical mode volume of $\approx$ 0.88 $(\lambda_{mode}/n)^3$ and $Q=6000$ was designed and fabricated to have a resonance close to the NV$^-$'s ZPL (see Fig. \ref{Fig:DiamondL3Cavity}). Confocal characterization of a native NV$^-$ center that was successfully coupled to the cavity gave a measured $g^{(2)}(0)$ value of 0.38 and an on-resonant radiative lifetime of 4 ns. As in \RefCite{Faraon2011Resonant}, the cavity's resonance wavelength was tuned by flowing xenon in a cryogenic environment. High resolution photoluminescence excitation measurements revealed two distinct peaks from the coupled NV$^-$ center, with the FWHM of the main peak at $\approx$ 8 GHz. Given that $g^{(2)}(0)$ is 0.38, it's likely that the double peaks is due to strain-split branches of the same NV$^-$ and not to two spatially separated NV$^-$.

\subsubsection{Diamond 1-D photonic crystals cavities}

\begin{figure}
\includegraphics[width=0.47\textwidth]{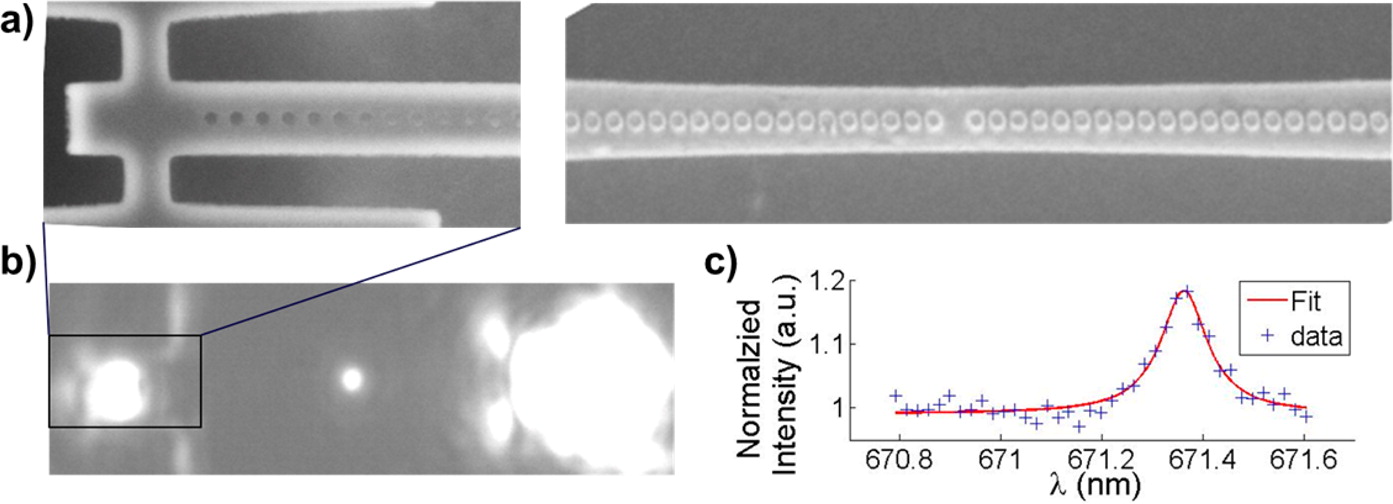}
\caption{(a) Left: SEM top view of the outcoupling region of a suspended diamond 1D nanobeam. Right: SEM top view of a suspended diamdond 1D nanobeam. Note the tapering of the width in the middle. (b) Transmission measurement with a supercontinuum source coupled in from the right. Notice the localization of the mode in the center of the nanobeam. (c) Transmission spectrum indicating $Q\sim6000$. \ACSReprint{Hausmann2013Coupling}{2013}}
\label{Fig:NV_1D_nanobeam}
\end{figure}

Besides 2-D PCCs, NV$^-$ centers have also been successfully integrated with 1-D PCCs. A relatively common 1-D PCC is a nanobeam that consists of a suspended diamond waveguide that contains periodic holes in it with some (optical) defects introduced near its center. Typically, the defect consists of either missing holes or holes with slightly different periodicity near the center. However, it is also possible to introduce a defect by increasing or decreasing the width of the waveguide in the middle\cite{Quan2011Deterministic}. In \RefCite{Hausmann2013Coupling} 1-D PCCs were created out of suspended 500 nm wide diamond waveguides that had 130 nm diameter air holes in it with a periodicity of $165-175$ nm and a 400 nm tapered width in the middle (see Fig. \ref{Fig:NV_1D_nanobeam}). Photoluminescence excitation and white light transmission spectra give a spectrometer limited $Q$ of above 6000 (simulated $Q$ was $\approx5\times10^5$) and the mode volume is estimated to be 1.8 $(\lambda/n)^3$. In this case, a two pronged strategy was employed to tune the cavity's resonance to the ZPL of native NV$^-$ centers within the nanobeams. First, a coarse tuning by means of controlled oxygen plasma etching was employed to blue shift the resonance, and then the device is later placed in a cryogenic environment (4 K) and xenon gas was introduced, as above, to red shift the resonance more precisely. A spontaneous emission enhancement of 7 was observed on resonance and $g^{(2)}$ values of 0.2 can be obtained. However, we note that the FWHM of the plotted photoluminescence is rather large at $\sim$ 490 GHz.

In \RefCite{Li2015Coherent}, 1-D nanobeams were also fabricated from a diamond membrane, albeit with a new Si mask transfer and RIE etch technique. NV$^-$ centers were then created by $^{15}$N implantation and annealing. The cavities had a theoretical mode volume of 1.05 $(\lambda/n)^3$ and quality factors ranging in the 1000s. For a nanobeam with $Q=1700\pm300$, a Purcell factor of 8 and 15 was achieved, for the $E_x$ and $E_y$ branch of a single NV$^-$'s ZPL respectively. The measured $g^{(2)}$ value was 0.28. As before, the enhancement was lower than expected from the $Q/V$ ratio, but this is here attributed to a poor alignment of the NV axis and non-ideal spatial position in the cavity. In another nanobeam with $Q$ factor of $3300\pm50$, a larger Purcell factor of 62 was achieved but in this case, there were multiple NV$^-$ centers present (as judged by multiple spectrally distinct ZPL transitions). Interestingly however, the single NV$^-$ center was observed to retain a long spin coherence time of $\sim$ 230 $\mu$s, which is similar to the spin coherence time of NV$^-$ centers in the parent unprocessed diamond. This indicates that the fabrication process using a Si mask did not adversely degrade the properties of the NV$^-$ center and is promising for future applications requiring long spin coherence times in cavity coupled SPEs. 

\subsection{Larger scale integration\label{Sec:diamond_large_scale_integration}}

\begin{figure*}
\includegraphics[width=0.75\textwidth]{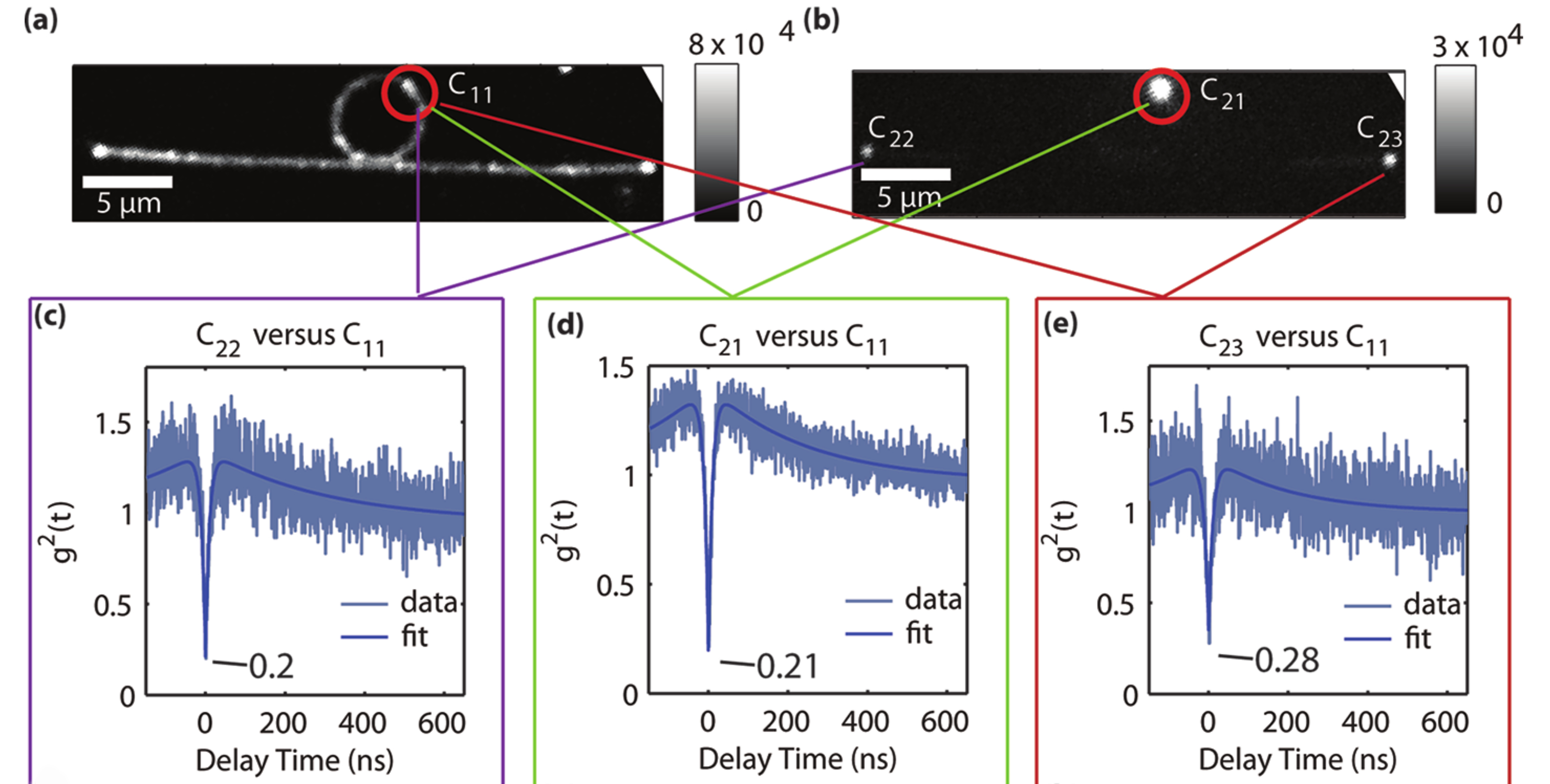}
\caption{(a), (b) Confocal images of the micro-ring resonator, waveguide and grating coupler. (c), (d), (e) $g^{(2)}$ measurements made using correlation from light collected at $C_{11}$, $C_{21}$, $C_{22}$ and $C_{23}$ in (a), (b). \ACSReprint{Hausmann2012Integrated}{2012} }
\label{Fig:integrated_NV_gratingcoupler}
\end{figure*}

Larger scale integration has also been achieved on an all-diamond platform consisting of micro-ring resonators coupled with waveguides and grating couplers\cite{Hausmann2012Integrated, Faraon2013Quantum}. The first demonstration\cite{Hausmann2012Integrated} was characterized at room temperature with a confocal microscope that had two independent collection arms, with one of the collection arm also being used to excite the cavity. This allowed the structure to be excited at one location while emitted photons were collected at a different location. The micro-ring resonator had a outer diameter of 40 $\mu$m and a $1000\times410$ nm cross-section. Fluorescence collected from the output of both gratings gave a saturated count rate of $(15\pm0.1)\times10^3$ Hz with a saturated pump power of $(100\pm4)$ $\mu$W. FDTD modeling suggests that the total collection efficiency is $\sim$ 15\%, and therefore it appears that there remains room for significant improvement. In addition, coincidence counts of photons collected from the ring and each grating under simultaneous excitation of the ring gave a $g^{(2)}$ value of $\sim$ 0.24, indicating that a single native NV$^-$ center in the ring had successfully outcoupled to the gratings (see Fig. \ref{Fig:integrated_NV_gratingcoupler}). Moreover, the fluorescence spectrum suggest a loaded $Q$ of $(3.2\pm0.4)\times10^3$ at 665.9 nm. However, no attempt was made to determine if the ring had successfully enhanced the spontaneous emission rate of the coupled NV$^-$ center. 

This was accomplished in a slightly later demonstration of a very similar system consisting of a 4.5 $\mu$m outer diameter micro-ring resonator, a ridge waveguide about $\approx$ 100 nm away, and grating couplers\cite{Faraon2013Quantum}. In this experiment, a spontaneous emission enhancement of $\approx$ 12 was reported after applying the same technique as in \RefCite{Faraon2011Resonant} to tune the cavity's resonance to native NV$^-$ centers in the resonator. Transmission measurements of the mode used to enhance the NV$^-$ center(s) ZPL line gave a coupled $Q$ factor of 5500 and the mode volume is estimated to be $\approx$ 15 $(\lambda/n)^3$. Excitation of native NV$^-$ center(s) in the bulk material showed that when the cavity was on resonance, approximately 25 times more photons was collected from the grating than from the bulk. However, given that no $g^{(2)}$ measurements were performed, it is not clear if only one NV$^-$ center was excited in each case, and it is therefore difficult to make an unambiguous comparison.

\begin{figure*}
\centering
\includegraphics[width=\textwidth]{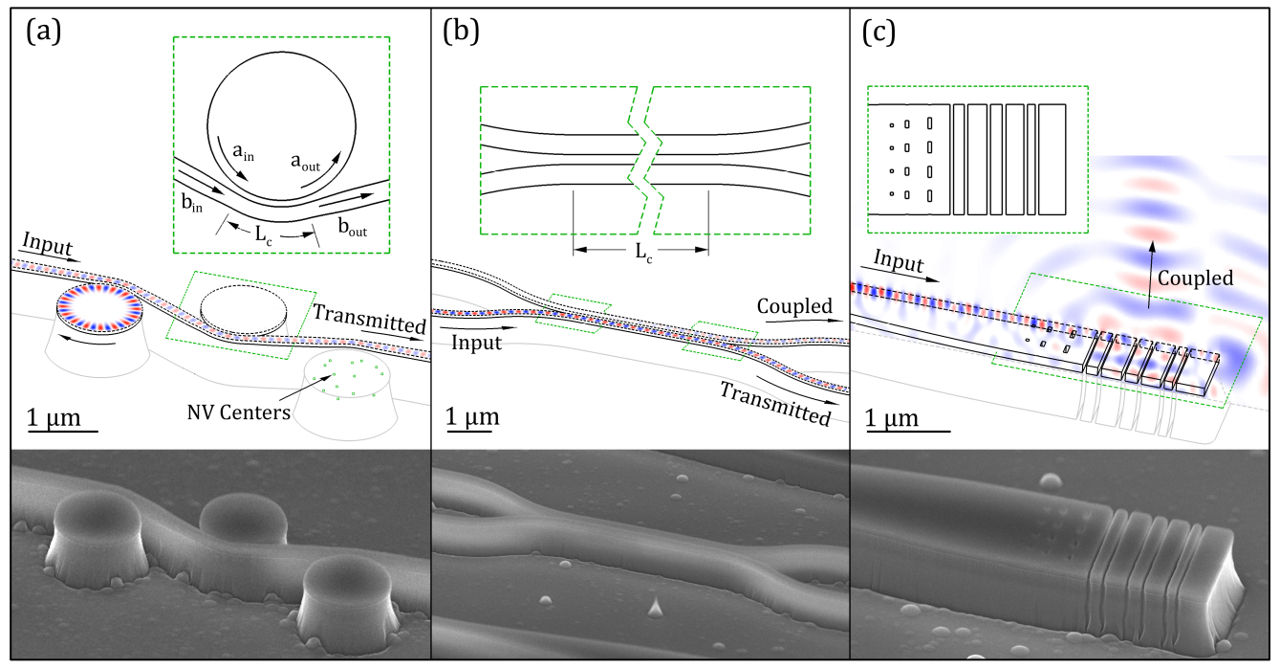}
\caption{Examples of integrated hybrid GaP-diamond systems where a thin 125 nm film of GaP was epitaxially grown and transferred onto a diamond substrate before being patterned with EBL and RIE. Above: schematic view overlaid with FDTD simulations, bottom: SEM images of integrated devices. (a) waveguide-coupled disk resonators, (b) directional coupler, (c) grating coupler. Reprinted with permission from \RefCite{Gould2016Large} \textsuperscript{\copyright} The Optical Society.}
\label{Fig:GaPIntegrated}
\end{figure*}

Larger scale integration has also been achieved on a GaP-on-diamond hybrid architecture\cite{Thomas2014Waveguide, Gould2016Large}. In \RefCite{Gould2016Large}, a 125 nm layer thick of GaP was transferred onto a diamond substrate, that had previously been implanted and annealed to produced NV$^-$ centers approximately 15 nm below the surface. The GaP was then patterned using electron-beam lithography and etched through using a Cl$_2$/N$_2$/Ar RIE. To obtain better mode confinement, the diamond substrate was further etched using O$_2$ RIE to get a $\approx$ 600 nm high diamond pedestal. GaP disk resonators, waveguides, directional couplers and grating couplers on a diamond pedestal were fabricated using this approach\cite{Gould2016Large} (see Fig. \ref{Fig:GaPIntegrated}). The coupled disk resonators were measured via transmission measurements to have loaded $Q$ factors in the range of $2500-10000$ and a full range of coupling ratios were obtained by varying the directional couplers' coupling region's length, which consisted of two 160 nm ridge waveguide spaced 80 nm apart (see Fig. \ref{Fig:GaPIntegrated}b). Emission from NV$^-$ below the ridge waveguide was successfully outcoupled by the grating couplers but unfortunately, no NV$^-$ ZPL line was observed at the grating couplers  when the coupled disk resonators were excited by 532 nm light due to a mismatch of the cavity's resonance (no attempt was made to tune the resonance here).

\subsection{Deterministic integration}

Most of the examples we have cited thus far relied on native NV$^-$ centers that were randomly dispersed in the diamond membrane. Given that optimal placement of the NV$^-$ center within the cavity is crucial to obtaining ideal spontaneous emission enhancement, deterministic integration of NV$^-$ centers is an important technological milestone. A 1-D deterministic integration of NV$^-$ centers in the vertical dimension was first attempted by delta-doping\cite{Ohno2012Engineering} a high purity chemical vapor deposition (CVD) grown diamond membrane with a thin ($\sim$ 6 nm) layer of nitrogen impurities\cite{Lee2014Deterministic}. This produces a layer of NV$^-$ centers that is well localized in the vertical dimension. Following this, 1-D suspended nanobeams are fabricated on the membrane with theoretical $Q$ values of $\sim$ 270,000 and mode volumes of $\sim$ 0.47 $(\lambda/n)^3$. However, measured photoluminescence spectra indicated that the highest $Q$ obtained experimentally was $\sim$ 24,000, which is generally attributed to imperfections in fabrication. A spontaneous emission enhancement of 27 was obtained by cooling the diamond to cryogenic temperatures (4.5 K) and flowing nitrogen into the cryostat to tune a cavity with mode of $Q$ $\sim$ 7000 to the NV$^-$'s ZPL. An on-resonance life-time of $10.43\pm0.5$ ns was also measured, which when combined with the off resonance life-time of $22.34\pm1.1$ ns, and a Debye-Waller factor of 0.03, gives a similar Purcell factor of 22. It is worth noting that the enhancement is somewhat smaller than expected based on the structure's $Q/V$ ratio. Compared to \RefCite{Hausmann2013Coupling}, the $Q/V$ ratio is larger by a factor of 30 but the enhancement is only $\sim$ 4 times larger. Similarly, the $Q/V$ ratio of this nanobeam is $\sim$ 5 times larger than the L-3 cavity in \RefCite{Faraon2012Coupling} but the enhancement is actually $\sim$ 3 times \textit{smaller}. The surprisingly low enhancement in this case is attributed to the fact that there is likely to be several NV$^-$ centers in the nanobeam and their linewidths has been estimated to be $\sim$ $150-260$ GHz, which is, for comparison, $\sim$ 4 times larger than in \RefCite{Faraon2012Coupling}. A consequence of this large linewidth is that the NVs' ZPL is poorly coupled to the cavity's resonance since the cavity mode is considerably narrower. The low enhancement could also plausibly be due to poor alignment of the NV axis (which affects the polarization of its emission) to the cavity.

Deterministic placement of NV$^-$ in 2-D photonic crystals have also been attempted using ion implantation through a hole in an atomic force microscopy (AFM) tip\cite{Riedrich2015Nanoimplantation, Jung2019Spin}. In \RefCite{Riedrich2015Nanoimplantation}, NV$^-$ centers were created at the center of 2D PCCs with measured $Q$ of $150-1200$ and estimated $V_{mode}\approx 1(\lambda/n)^3$. The created NV$^-$s had relatively large spectral diffusion limited FWHMs of $\sim$ 250 GHz at 10 K and were expected to have lateral spatial resolution of $<$ 15 nm and vertical spatial resoultion of $\approx$ 3 nm. Unfortunately, the spatial resolution of the NV$^-$ centers were only measured to be less than 1 $\mu$m and it is not obvious if the expected resolution was actually obtained. Moreover, the NV$^-$ creation yield was quite low at $0.8\pm0.2$ \%.

Higher deterministic single NV$^-$ creation yield can be obtained by using hard Si masks for \textit{both} implantation and diamond patterning where a single NV-cavity system yield of $26\pm1$ \% was obtained\cite{Schukraft2016Precision}. By having a high Si mask aspect ratio for the ``implantation'' holes, etching rate for the underlying diamond substrate is negligible but N ions during implantation are still able to implant into the diamond due to different conditions for etching and implantation\cite{Schukraft2016Precision}. This allows for the use of a single mask for both diamond patterning and NV$^-$ positioning, thereby eliminating any loss of accuracy due to re-alignment of separate masks. Unfortunately, there was no definite measurement of the overall spatial resolution obtained although a 1-D nanobeam with $Q=577$ and coupled single NV$^-$ center was reported. Nevertheless, considering that spatial positioning of NV$^-$ to about $\sim$ 10 nm have been obtained using masks\cite{Ohno2014Three, Scarabelli2016Nanoscale}, this scalable approach to deterministically position NV$^-$ centers within photonic structures is promising.

\subsection{Outlook}

Moving forward, we believe that there is still much room for improvement in deterministically integrating high quality single NV$^-$ centers to complex photonic circuits using diverse strategies that have been developed over the years. Although most of the work discussed above which demonstrated coupling of NV$^-$ centers to all-diamond photonic structures relied on randomly positioned native NV$^-$ centers, we note that spatial positioning of NV$^-$ centers to about $\sim$ 10 nm in all three dimensions\cite{Ohno2014Three, Scarabelli2016Nanoscale} has already been separately demonstrated. In these demonstrations, delta doping of CVD grown diamond (see section \ref{Sec:irradiation_and_annealing}) is typically used together with a mask for accurately creating vacancies via irradiation that then lead to NV$^-$ formation in the thin nitrogen doped layer after annealing. Previously, such mask consisted of spin-coated resist\cite{Ohno2014Three} and other additional layers\cite{Scarabelli2016Nanoscale} that can be difficult to coat with even thickness over a large area. However, the recent development of mask transfer techniques (see section \ref{Sec:pick_and_place_microprobe}) open up the possibility of using high quality Si masks that can be positioned with sub-micron or even nm scale accuracy on the diamond substrate and then later removed mechanically. It is therefore possible to imagine using a single mask to both create NV$^-$s and photonic structures at deterministic positions\cite{Schukraft2016Precision}. Successful NV$^-$ integrated photonic structures, as characterized by optical measurements, can then be picked-and-placed (as in \RefCite{Mouradian2015Scalable}) by a microprobe to integrate with other photonic structures of a potentially dissimilar material that will further unlock other functionalities. This allows NV$^-$ centers to be created in high quality single crystal diamond (potentially with isotopically enriched $^{12}$C) where they can have long optical and spin coherence times, while still being able to be efficiently routed and processed by other photonic elements on a chip. 


\section{Color centers in nanodiamonds\label{Sec:ColorCenterND}}

As discussed in the preceding section, NV centers are not the only defects in diamond that exhibit discrete energy levels with optical transitions although they are arguably the most studied defect. Recently, defect centers consisting of group-IV elements such as silicon vacancy (SiV)\cite{sipahigil+7:16, kumar+3:17}, germanium-vacancy (GeV)\cite{bhaskar+12:17} and tin-vacancy (SiV)\cite{tchernij+13:17, iwasaki+6:17} centers in diamond have been of particular interest due to symmetries in their configuration that leads to a higher Debye-Waller factor and narrower spectral lines that increases the indistinguishability of their emitted photons. Moreover, although there has, as the preceding section shows, been a great deal of work in bulk diamond, there has also been considerable work in integrating color centers in nanodiamonds to hybrid photonic structures. We note however, that NV centers in nanodiamonds are less photo-stable and tend to have significantly larger inhomogeneously broadened ZPL linewidths as compared to their bulk counterparts. Although this makes them less suitable for many quantum computing/processing applications, nanodiamonds are more suited for bio-sensing/labeling applications\cite{hegyi+1:13, schirhagl+3:14, childress+2:14}, and interestingly enough, the spontaneous emission rates of NV centers in nanodiamonds can also be enhanced by encasing them in phenol-ionic complexes\cite{kerem+4:15}. Moreover, nanodiamonds with a high concentration of SiV centers can also be used as temperature sensors\cite{sumin:19}. In this section, we give examples of other color centers in nanodiamonds that have been coupled to photonic structures.

\subsection{\label{sec:nanodiamond_fab}Fabrication}

Nanodiamonds are synthesized by various techniques such as detonation, laser assisted synthesis, HPHT high energy ball milling of microcrystalline diamond, hydrothermal synthesis, CVD growth, ion bombardment on graphite, chlorination of carbides and ultrasonic cavitation\cite{zishan}. In the laboratory, the detonation method and HPHT growth are commonly employed to synthesize NV containing nanodiamonds on a large scale\cite{marina:19}. CVD growth is another promising technique that has successfully synthesized single NV centers in nanodiamonds\cite{rabeau+6:07}. More recently, a new metal-catalyst free method to synthesize nanodiamonds with varying contents of NV and SiV centers produced high-quality color centers with almost lifetime-limited linewidths\cite{jantzen+9:16, toan+8:17}. 


In \RefCite{bradac:10}, the authors reported the first direct observation of NV centers in discrete 5 nm nanodiamonds at room temperature. Although the luminescence of those NV centers was intermittent (i.e. they undergo blinking), the authors were able to modify the surface of the nanodiamonds to mitigate the undesirable blinking. In another work, the authors showed the size reduction of nanodiamonds by air oxidation and its effect on the nitrogen-vacancy centers that they host\cite{Gaebel+4:11}. The smallest nanodiamond in their samples that still hosted a NV center was about 8 nm in size.

SiV centers in nanodiamonds have subsequently been investigated \cite{neu+6:11, muller+9:14, haussler:19}. \RefCite{neu+6:11} describes the first ultrabright single photon emission from SiV centers grown in nanodiamonds on iridium. The SiV centers were grown using microwave-plasma-assisted CVD and those single SiV$^-$ defects achieved a photon count rate of about 4.8 Mcounts/s (at saturation). Bright luminescence in the 730-750 nm spectral range were observed using confocal microscopy. No blinking was observed but photobleaching occurred at high laser power. Enhanced stability might be gained by controlling the surface termination of the nanodiamonds, as was shown for the case of NV centers \cite{Fu2010Conversion}.

Residual silicon in CVD chambers often results in the formation of $\text{SiV}^-$ centers in most CVD-grown nanodiamonds\cite{neu+6:11, toan+8:17}. Likewise, due to silicon-containing precursors, many HPHT-synthesized nanodiamonds also include $\text{SiV}^-$ centers\cite{crane+7:19}. In \RefCite{haussler:19}, the authors demonstrated optical coupling of single $\text{SiV}^-$ centers in nanodiamonds and were able to manipulate the nanodiamonds both translationally and rotationally with an AFM cantilever.

Fabrication of other color centers such as GeV centers in nanodiamond were also recently demonstrated. For example, single GeV centers in nanodiamonds were successfully fabricated by the authors in \RefCite{siampour+3:18} after they introduced Ge during HPHT growth of the nanodiamonds. More generally, in \RefCite{bradac+4:19}, the authors studied a larger variety of group IV color centers in diamond, including SiV, GeV, SnV and PbV centers.

We note that it is possible to control the size and purity of the HPHT nanodiamonds down to 1 nm \cite{stehlik:15}. In other works, the size of nanodiamonds are typically tens of nanometers \cite{schietinger+2:08, santori+5:10, zheng+10:18}, which make nano manipulation of them feasible. For example, emission from single NV centers hosted in uniformly-sized single-crystal nanodiamonds with size $30.0\pm 5.4$ nm have been reported \cite{zheng+10:18}.


Although high count rates are in general achievable for NV and SiV color centers in nanodiamonds\cite{aharonovich+6:09}, these high count rates were sometimes reported to be correlated to blinking\cite{Schroder2011Ultrabright}. Compared to SiV centers in the bulk, SiV centers in nanodiamonds have significantly less reproducible spectral features and can feature a broad range of ZPL emission wavelengths and linewidths\cite{lindner+9:18}. More generally, the linewidths of SiV centers in nanodiamonds have been shown to depend on the strain of the diamond lattice\cite{lindner+9:18}. Nevertheless it is sometimes possible to obtain nearly lifetime-broadened optical emission in SiV centers in nanodiamonds at cryogenic temperatures\cite{li+4:16, jantzen+9:16}, and indeed nearly lifetime limited zero-phonon linewidths have been obtained in both NV and SiV centers in nanodiamonds. For example, despite spectral diffusion and spin-nonconserving transitions, zero-phonon linewidths as small as 16 MHz has been reported for NV centers in type Ib nanodiamond at low temperature \cite{shen+2:08}.

For GeV centers in HPHT nanodiamond, the stability of its ZPL emission wavelength and linewidth has been attributed to the symmetry of its molecular configuration, although a large variation of lifetimes was also reported\cite{Minh+7:19}. The authors there estimate a quantum efficiency of about 20\% for GeV centers in HPHT nanodiamonds.

\subsection{\label{sec:nanodiamond_integration}Integration with photonic structures}

\begin{figure}
\includegraphics[width=0.48\textwidth]{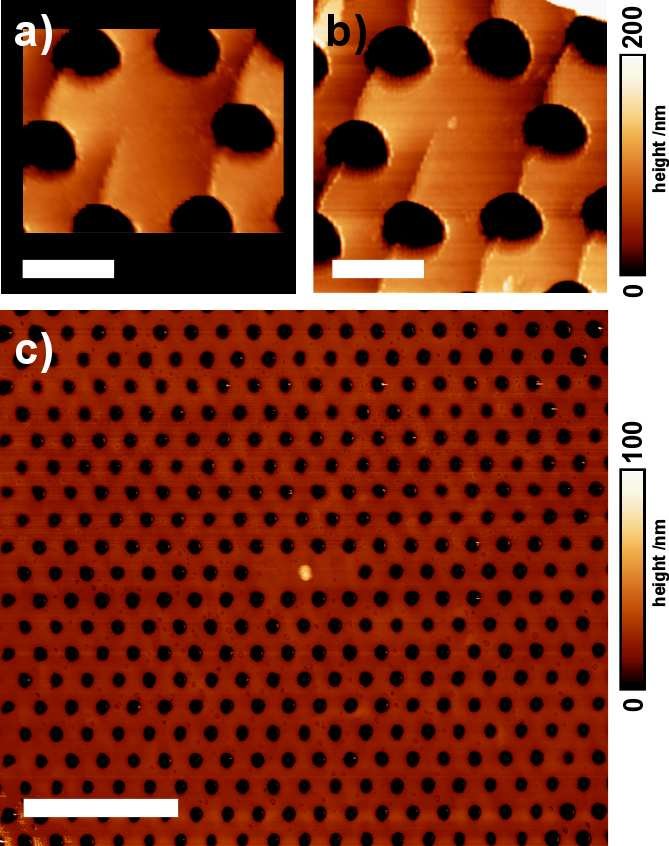}
\caption{(a) AFM image of the core of a photonic crystal fiber. (b) A pre-selected nanodiamond is then transferred to center of the photonic crystal fiber using a pick-and-place technique. (c) A nanodiamond in the center of a gallium phosphide photonic crystal membrane cavity. \AIPReprint{schell:11}}
\label{Fig:PickAndPlaceNanodiamond}
\end{figure}

As mentioned in section \ref{Sec:NVInBulk} above, a hybrid GaP-diamond platform is attractive for multiple reasons and there has been work involving not just bulk GaP-diamond systems but also hybrid GaP-nanodiamond systems. For an extensive review, see \RefCite{marina:19}. Purcell enhancement of the ZPL emission by a factor of 12.1 has been reported in a hybrid nanodiamond-GaP platform where the ZPL of an NV center is coupled to a single mode of a PCC \cite{wolters:10}. In that work, both the nanodiamond and cavity are first pre-selected and the resonance of the cavity is then tuned to the ZPL of the NV center by locally oxidizing the GaP with a focused blue laser\cite{wolters:10}. Finally, the pre-selected nanodiamond is then transferred to the GaP cavity using a pick-and-place technique \cite{schell:11, oliver:11} (see Fig. \ref{Fig:PickAndPlaceNanodiamond}). Alternatively, a GaP PCC may be transferred using a micro-PDMS adhesive on a tungsten probe (briefly discussed and illustrated in section \ref{Sec:pick_and_place_microprobe} and Fig. \ref{Fig:PDMSTransferSchematic}) to a pre-selected nanodiamond containing a NV$^-$ center of desirable properties\cite{englund+6:10}. 

Nanodiamonds were also integrated with silica micro-resonators to achieve cavity QED (cQED) effects. In one early attempt, diamond nanocrystals were attached to silica micro-resonators by dipping silica micro-disks with diameters of 20 $\mu$m into an isopropanol solution containing suspended nanodiamonds with a mean diameter of 70 nm\cite{santori+5:10}.  Initially, the micro-disks had a quality factor of $Q=40000$ at room temperature, but after deposition of the nanodiamonds, low temperature measurement showed that the quality factor decreased significantly to around $2000-3000$. By condensing nitrogen gas to tune the cavity modes, the authors observed that a single NV center could couple to two cavity modes simultaneously. However, there was no significant change in the spontaneous emission rate, which was probably due to, in addition to the emitters' large linewidth, the resonator's large mode volume and limited quality factor. In \RefCite{barclay+5:09}, a tapered fiber is used to both pick up and position NV containing nanodiamonds onto a high-Q SiO$_2$ micro-disk cavity. The same tapered fiber could then also be used to characterize light transmission through the system. Coupling in the strong cQED regime has also been achieved between NV centers in nanodiamonds and silica micro-spheres resonators\cite{park+2:06}.

Besides silica resonators, there has also been work on polystyrene micro-sphere resonators. Nanodiamonds with a mean diameter of 25 nm can be attached to polystyrene micro-spheres with diameters of $\sim$ 5 $\mu$m by first dispersing both on a cover slip and then using near-field scanning optical microscopy (NSOM) tips to bring the micro-spheres close to a pre-selected nanodiamond containing a single NV$^-$ center\cite{schietinger+2:08}. Touching a nanodiamond with a micro-sphere then attaches the former to the latter. Using this technique, the authors demonstrated coupling of two single NV centers found in two different nanodiamonds to the same micro-sphere resonator\cite{schietinger+2:08}.

Silicon carbide is another material that can be integrated with diamond due to its similarity with diamond. For example, the authors in \RefCite{marina:19} developed a scalable hybrid photonics platform which integrates nanodiamonds with 3C-SiC micro-disk resonators fabricated on a silicon wafer. By condensing argon gas on the structure, the authors were able to continuously red shift the resonator's resonance and tune it to the color center's emission to observe an enhancement of the center's spontaneous emission.

\begin{figure}
\includegraphics[width=0.47\textwidth]{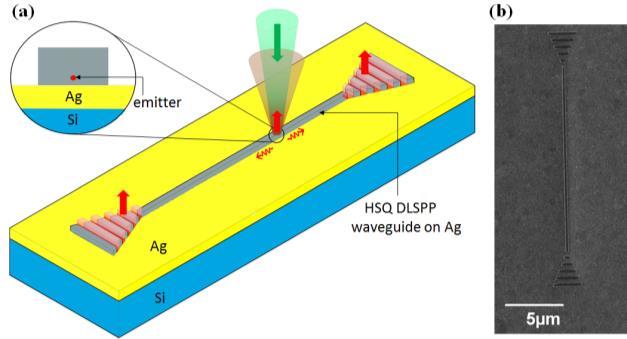}
\caption{(a) Dielectric-loaded surface plasmon polariton waveguide (DLSPPW) circuits are built around nanodiamonds containing single photon emitter with enhanced emission rate. (b) Scanning electron micrograph of a hydrogen silsesquioxane (HSQ) waveguide fabricated on silver-coated silicon substrate. \ACSReprint{siampour+2:17}{2017}}
\label{Fig:DLSPP_nanodiamond}
\end{figure}

It is also possible to enhance the spontaneous emission rate of a quantum emitter coupled to waveguiding structures like dielectric-loaded surface plasmon polariton waveguides (DLSPPWs) where the significantly confined mode volume of the surface plasmon polariton can enable Purcell factors above unity (see Fig. \ref{Fig:DLSPP_nanodiamond}). Experiments involving embedded nanodiamonds with NV centers in a DLSPPW consisting of a hydrogen silsesquioxane (HSQ) waveguide on top of a silver film demonstrated a spontaneous emission enhancement of up to 42 times\cite{siampour+2:17, siampour+2:17:2}. In a similar vein, a GeV center embedded within a similar DLSPPW was successfully excited by 532 nm light propagating within the waveguide and achieved a three-fold enhancement in its spontaneous emission rate due to the small mode volume within the waveguide\cite{siampour+3:18}. Although likely to be less useful than coupling to DLSPPWs due to higher losses, coupling of single NV centers in nanodiamonds to silver nanowires can enable interesting studies of surface plasmon polaritons (SPP) as in \RefCite{kolesov+7:09} where a wave-particle duality was demonstrated for SPPs excited by single photons from a nanodiamond.

The spontaneous emission rate of a quantum emitter can be significantly enhanced when coupled to a plasmonic nano-antenna. For example, enhancement factors of up to 90 times was observed for a NV center within a nanodiamond that was coupled to a nanopatch antenna\cite{simeon:18}. Even higher enhancement of up to 300 times has been theoretically proposed by coupling SiV centers in a nanodiamond to a specific geometry of gold dimers \cite{xiang+3:17}. 
 
\begin{figure}
\includegraphics[width=0.47\textwidth]{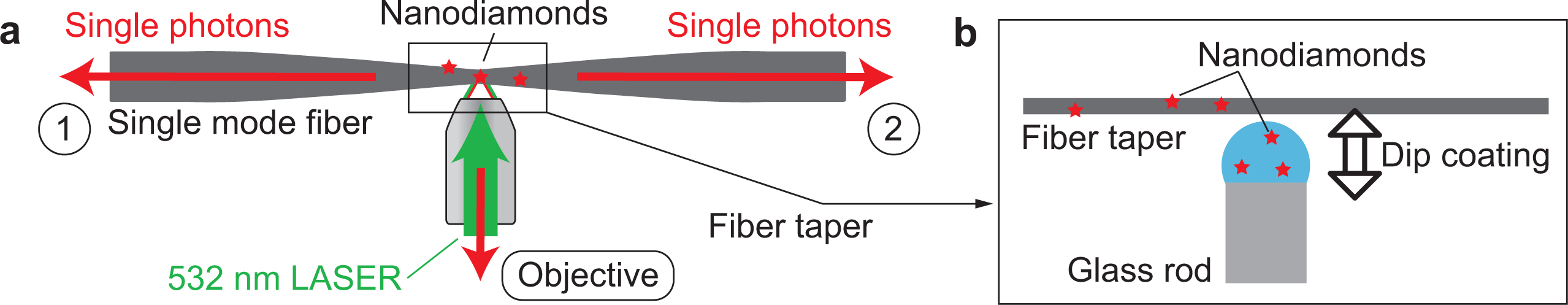}
\caption{(a) Nanodiamonds were attached to the fiber taper, and single photons emitted by the color centers were collected either through the objective or the fiber ends. (b) The tapered fiber was dipped into a small droplet of a nanodiamond solution on the facet of a thin glass rod. The nanodiamonds were attached to the fiber taper as it is moved through the droplet. \OSAReprint{schroder+5:12}}
\label{Fig:Nanodiamond_tapered_fibre_system}
\end{figure}

Besides static resonators such as micro-disks and micro-spheres, a fiber-based micro-cavity technique, where a tunable cavity is typically formed by the combination of a fiber and macroscopic mirror, can also be applied to NV and SiV centers in diamond to enhance their efficiency, brightness and single photon purity\cite{kaupp+5:13, albrecht+4:13, kaupp+9:16, benedikter+9:17}. Lastly, it is also possible to directly couple the color centers in a nanodiamond to an optical fiber. For example, in \RefCite{schroder+4:11} pre-selected NV containing nanodiamonds were placed directly on a fiber facet to create an alignment free single photon source. High coupling efficiency was also reported in a nanodiamond-tapered fiber system\cite{schroder+5:12, liebermeister:14, vorobyov:16}  (see Fig. \ref{Fig:Nanodiamond_tapered_fibre_system}), and in \RefCite{Henderson+10:11}, NV containing nanodiamonds were successfully embedded in tellurite soft glass. 

\section{Quantum Dots\label{Sec:QuantumDots}}

\subsection{\label{sec:qdot_intro}Introduction to Quantum Dots}

A quantum dot (QD) is a small nanometer-sized three-dimensional inclusion of a narrower bandgap material within a wider bandgap matrix. The 3D confinement potential of the QD leads to a discretization of energy levels and gives it localized, atom-like properties. 
By controlling and manipulating these properties, QDs can be utilized in many aspects of quantum technologies, such as SPEs or as qubit systems. QDs have also been studied extensively and developed for numerous optoelectronic applications, including light-emitting diodes (LEDs)\cite{wood2010colloidal}, photovoltaic devices\cite{sargent2012colloidal}, and flexible displays\cite{choi2018flexible}. 

Compared to other atomic systems (e.g. trapped ions) used in early experimental realizations of quantum logic \cite{monroe1995demonstration}, QDs are embedded within a solid-state medium and thus do not require bulky and complicated vacuum systems and optical trapping setups. Moreover, QD-based devices can take advantage of well-established growth techniques, e.g. molecular beam epitaxy (MBE)\cite{snyder1991effect,xin1996formation,fafard1996inas} or metalorganic vapor phase epitaxy (MOVPE)\cite{oshinowo1994highly,heinrichsdorff1997room,kim2002self}, which allow for monolithic growth with monolayer precision. Coupled with the ability to electrically control these devices\cite{yuan2002electrically,yu2019electrically,pedersen2020near}, QDs have attracted extensive research efforts in developing and realizing QD photonic devices.

QDs used in integrated photonics applications are typically based on III-V materials, 
especially In(Ga)As in (Al, Ga)As \mbox{matrices}.
The most common QD growth approach uses the Stranski-Krastanov mechanism\cite{mirin19951}: as QD material is successively deposited and reaches a critical thickness, strain energy from mismatched lattice constants drive the formation of 3D nano-islands through a self-assembly process which allows for more efficient strain relaxation. 
The downside of self-assembly is that the QDs are randomly positioned,
but site-controlled growth techniques have been developed to gain deterministic control over the QD positioning\cite{surrente2017dense,strauss2017resonance} and their coupling to nanophotonic structures\cite{rigal2018single}.

QD devices have numerous applications in quantum integrated photonics. They can serve as tunable, high-quality single-photon sources 
that can be integrated into nanophotonic structures
such as waveguides\cite{mnaymneh2020chip} and beamsplitters\cite{kim2017hybrid}.
To complement this, photonic device components for photon manipulation, such as  modulators\cite{bhasker2018intensity}, frequency sorters\cite{elshaari2017chip}, and frequency converters\cite{singh2019quantum}, have been developed.
High-speed near-infrared (NIR) detectors\cite{umezawa2014investigation, wan2017monolithically} based on QDs have also been demonstrated in recent years. 
By controlling the QD spin, spin-photon interfaces can also be realised, allowing the QD to be used as a quantum memory, as well as a range of additional applications such as single-photon switching\cite{lodahl2017quantum}. 

In this work, we will focus more on examples and applications of QDs integrated on photonic platforms; a broad-spectrum overview can be found in another recent review\cite{arakawa2020progress}.

\subsection{As a Single-Photon Emitter}

\begin{figure}
	\includegraphics[width=0.8\columnwidth]{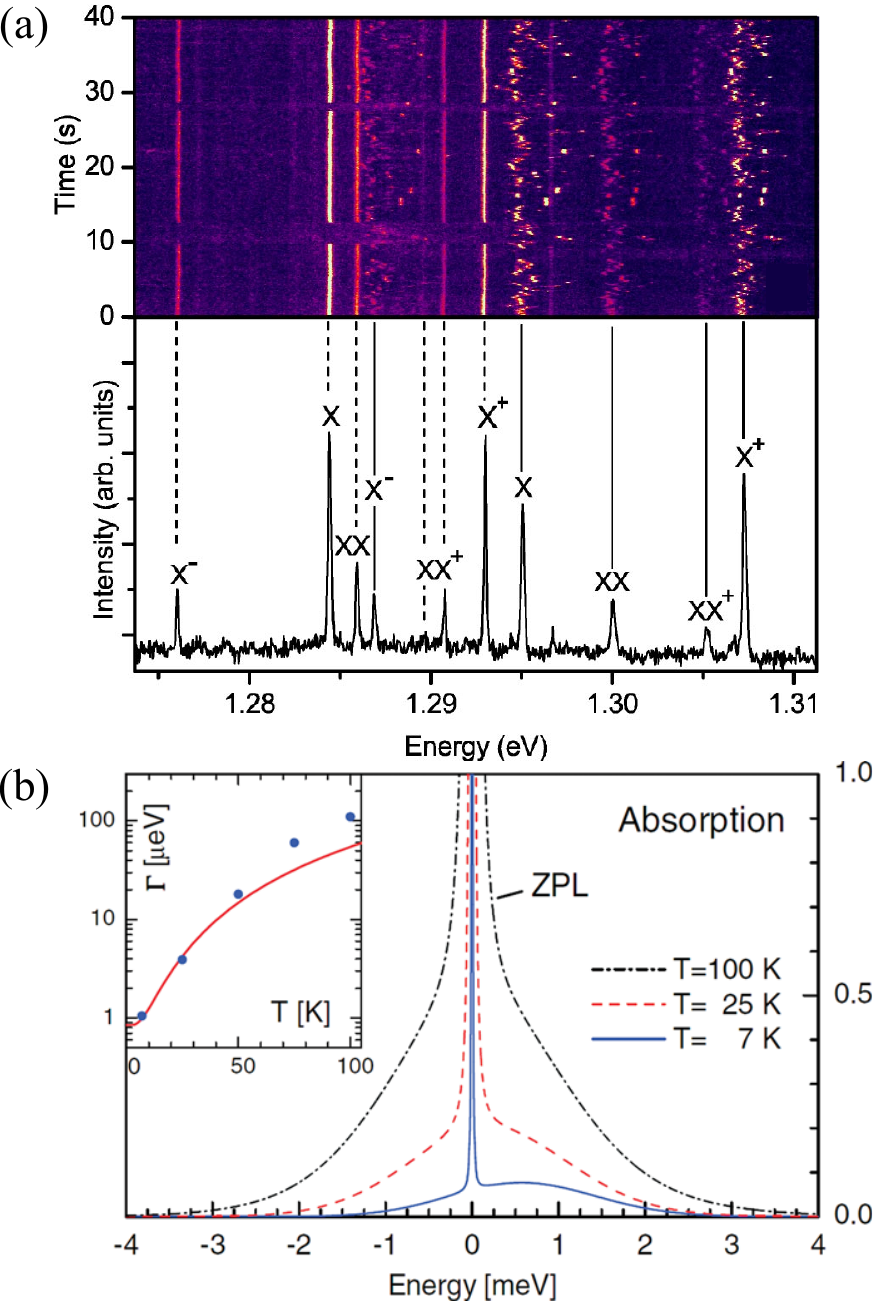}
	\caption{\label{fig:qd_spec}
		(a) The bottom part shows a typical spectrum of multiple InAs/GaAs QDs.
		The top part shows the temporal evolution of the spectrum showing spectral wandering of the QD transitions. \APSReprint{rodt2005correlation}{2005}
		(b) Calculated absorption spectra of an InAs QD.
		At lower temperatures, the broadband phonon interactions are suppressed, while the zero-phonon line (ZPL) is asymmetrically broadened. The inset compares the calculated broadening with experimental data (circles). \APSReprint{muljarov2004dephasing}{2004}
	}
\end{figure}

SPEs can be realized from QDs by utilizing the radiative recombination from an excitonic state of a single QD\cite{michler2000quantum,dietrich2016gaas}.
The first demonstrations of single-photon emission from QDs were performed under optical pumping\cite{michler2000quantum}, and then by electrical pumping\cite{yuan2002electrically}.
Beyond single-photon generation, multiphoton generation via demultiplexing of high-brightness integrated QDs has also achieved four-photon coincidence rates of $>$1\,Hz\cite{hummel2019efficient}.
QD-based SPEs have been extensively studied, and in-depth discussions can be found in other review articles\cite{shields2010semiconductor,senellart2017high}. 

The photon statistics of QD single-photon sources can be degraded by imperfections such as multi-photon emission from multi-exciton states, or if light is also collected from nearby QDs. 
As discussed in Section~\ref{sec:purity}, we will use the $g^{(2)}(0)$ value as a measure of the source's single-photon purity.

For non-resonant excitation, multi-photon emission can result from the QD capturing additional carriers after the first photon emission, which can subsequently recombine.
Therefore, to obtain a low $g^{(2)}(0)$, relaxation into the QD and the radiative cascade causing recombination should occur on a longer timescale than the decay of the initial carriers\cite{flagg2012dynamics, giesz2013influence}. 
With resonant, pulsed excitation, $g^{(2)}(0)$ values close to zero have been demonstrated\cite{he2013demand,ding2016demand,somaschi2016near},
while the lowest reported $g^{(2)}(0)$ values of below $10^{-4}$ have been achieved with two-photon excitation\cite{schweickert2018demand,hanschke2018quantum}. 
However, we note that these lowest values were not obtained from QDs integrated with on-chip planar waveguides; 
demonstrations with integrated QDs have reported more modest $g^{(2)}(0)$ values due to factors such as increased background emission from cavity modes\cite{davanco2017heterogeneous} (see also Table~\ref{table:benchmarking}).

The radiative cascade of high-energy carriers also results in a temporal uncertainty (i.e. jitter) of photon emission\cite{flagg2012dynamics}, which leads to decreased indistinguishability for higher excitation powers\cite{gazzano2013bright}. 
However, this can be overcome with strictly resonant pumping schemes\cite{he2013demand,wei2014deterministic}. 
Moreover, resonant pumping and adding a weak auxiliary continuous wave reference beam to the excitation beam of the QD can help to suppress charge fluctuations\cite{konthasinghe2012coherent} that would otherwise lead to spectral diffusion. 

To suppress the effects of phonon interactions, one can operate at cryogenic temperatures although we acknowledge that for InGaAs QDs at 4\,K, PSB emissions can still represent $\sim$10\% of emission (see Fig. \ref{fig:qd_spec}a). Also, spectral filtering of the QD ZPL can yield high indistinguishability close to unity\cite{thoma2016exploring}, albeit at the expense of photon rates. 

Since the first Hong-Ou-Mandel (HOM) two-photon interference experiment with QDs reported indistinguishabilities of $\sim$70\%\cite{santori2002indistinguishable}, near-unity values have been consistently achieved\cite{wei2014deterministic,wang2016near,loredo2016scalable,reindl2019highly} (e.g. 0.995$\pm$0.007, \RefCite{wei2014deterministic}).
Recent reports have also reproduced high HOM visibility values QDs integrated with nanophotonic waveguides\cite{kalliakos2016enhanced,kirvsanske2017indistinguishable,dusanowski2019near,schnauber2019indistinguishable}.

\subsection{Spin-Photon Interfaces}

By accessing and manipulating the their spin, QDs can provide not only photonic qubits, but spin qubits as well. Various level structures can be exploited for qubit encoding, and rapid spin initialization, manipulation, and read-out can be achieved with short optical pulses (in the nanosecond range)\cite{warburton2013single}. Such spin-photon interfaces can enable many quantum information processing tasks, 
such as deterministic spin-photon entanglement and mediating strong photon-photon interactions.

\begin{figure}
	\includegraphics[width=0.85\columnwidth]{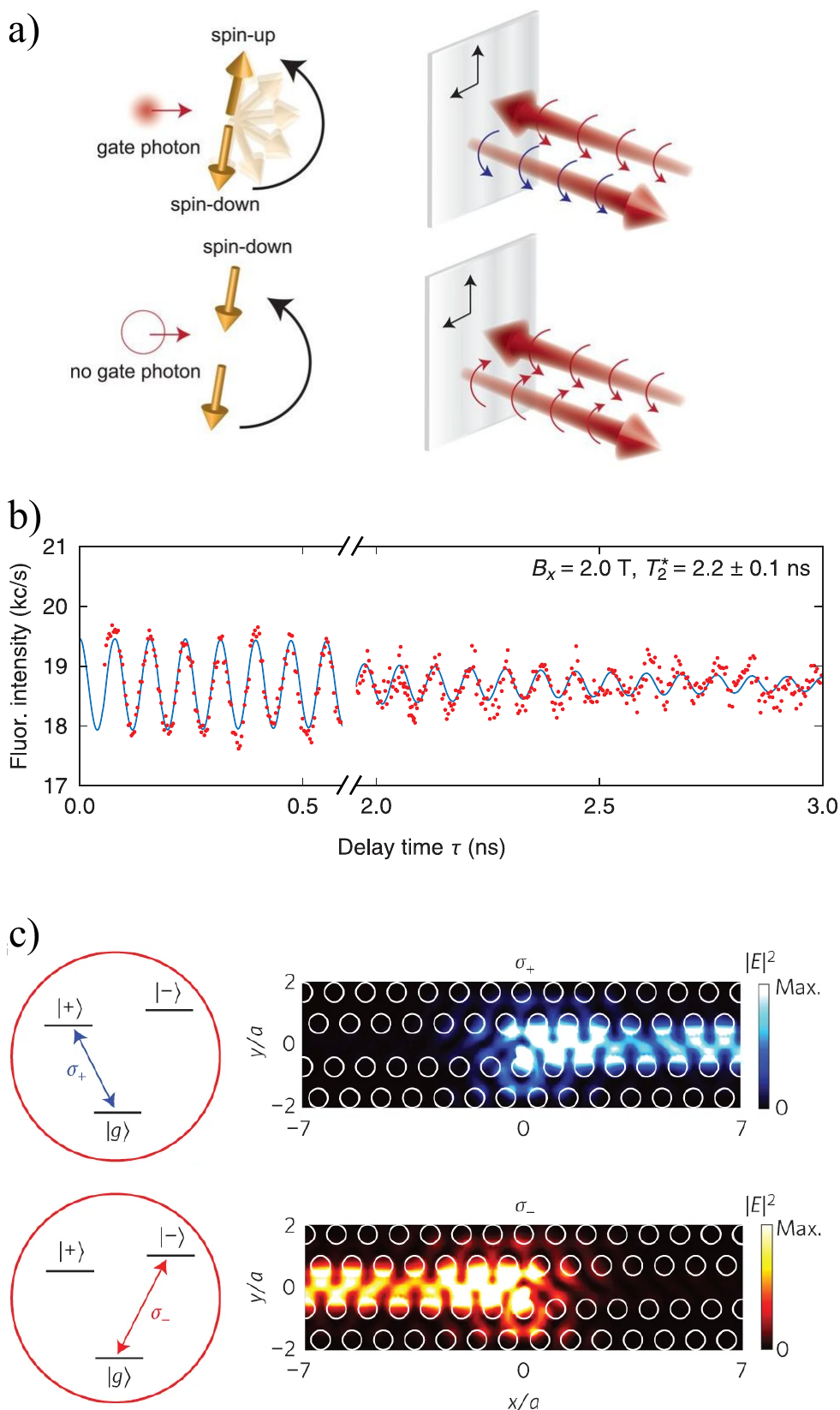}
	\caption{\label{fig:qd_spinphoton}
	(a) Schematic of the single-photon switch and transistor. A gate photon controls the state of the spin, and then the spin determines the polarization of the signal field. \AAASReprint{sun2018single}
	(b) Ramsey interference of a QD spin embedded in a nanobeam waveguide, with a dephasing time $T_2^*$\,=\,2.2\,$\pm$\,0.1\,ns, and a contrast of 0.04 for the first period. \APSReprint{ding2019coherent}{2019}
	(c) [Left] QD level scheme featuring two circularly polarized exciton transitions. [Right] Calculated directional emission patterns of $\sigma_+$ and $\sigma_-$ transitions. \SpringerNatureReprint{sollner2015deterministic}{2015}
	}
\end{figure}

The strong nonlinearity at a single-photon level has led to demonstrations of photon blockade\cite{faraon2008coherent} and tunable photon statistics via the Fano effect\cite{foster2019tunable}.
Single-photon switches and transistors have been realized via a QD spin\cite{sun2018single,javadi2018spin,luo2019spin} (see Fig.~\ref{fig:qd_spinphoton}a).
Coherent control of the QD spin has also been achieved, with \RefCite{ding2019coherent} demonstrating Ramsey interference
with a dephasing time $T_2^*$\,=\,2.2\,$\pm$\,0.1\,ns (see Fig.~\ref{fig:qd_spinphoton}b).

The strong light confinement in nanophotonic waveguides also opens up the possibilities of chiral, or propagation-direction-dependent, quantum optics\cite{lodahl2017chiral}. This can be used to deterministically induce unidirectional photon emission from quantum dot spin states, i.e. $\sigma_\pm$ transitions emit in different directions\cite{sollner2015deterministic,coles2016chirality} (see Fig~\ref{fig:qd_spinphoton}c).
This can help to realize complex on-chip non-reciprocal devices such as single-photon optical circulators\cite{tang2019chip}.
Although chirality has only been demonstrated to date using waveguides,
recent theory papers have shown that chirality with significant Purcell enhancement should be possible using a ring resonator geometry\cite{martin2019chiral,tang2019chip}.

\subsection{Interfacing Multiple QDs}

\begin{figure*}
	\includegraphics[width=0.95\textwidth]{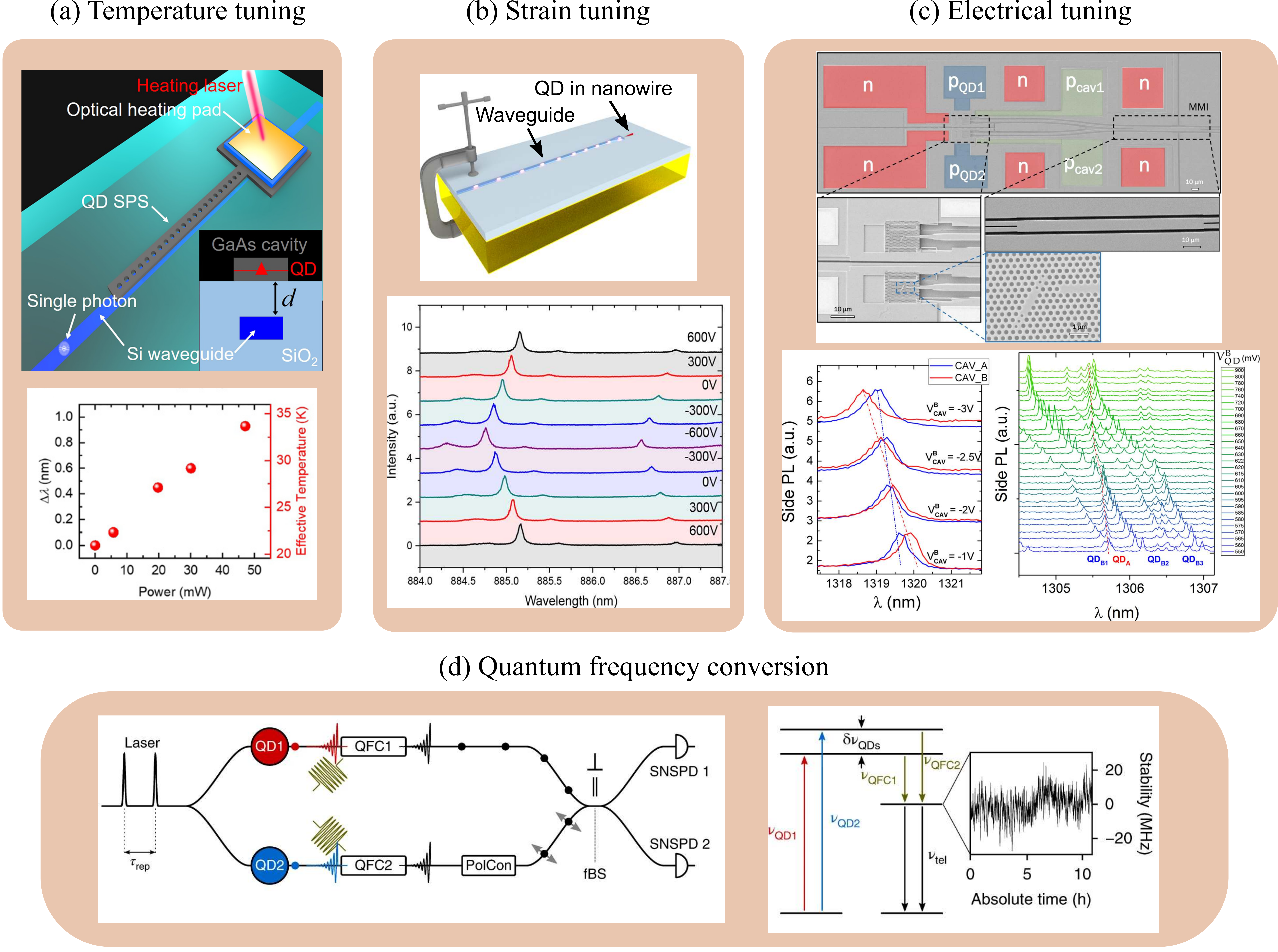}
	\caption{\label{fig:qd_tuning}
	(a) [Top] Schematic of a QD device temperature-tuned by laser irradiation on a heating pad.
	[Bottom] Dependence of the QD detuning on the heating laser power. \AIPReprint{katsumi2020situ}
	(b) [Top] Artistic representation of a waveguide-coupled QD photon source fabricated on a strain-tunable substrate.
	[Bottom] Emission spectra of the nanowire QD as a function of the applied voltage to the piezoelectric substrate. \ACSReprint{elshaari2018strain}{2018}
	(c) [Top] Annotated SEM picture of a device with two integrated QD devices. The QDs and cavities can be electrically tuned individually.
	[Bottom] Spectra showing the electromechanical tuning of the cavities (left) and the Stark tuning of the QDs (right). \CCBYFourReprint{petruzzella2018quantum}
	(d) [Left] Schematic of a device where the emission of two QDs are individually tuned via quantum frequency conversion (QFC).
	[Right] Energy diagram depicting the QFC scheme. \SpringerNatureReprint{weber2019two}{2019}
	}
\end{figure*}

Hybrid quantum photonics platforms aim to integrate multiple quantum sources, including dissimilar quantum systems, onto the same device. Two-photon interference has been demonstrated between QDs and other quantum emitters, including atomic vapors\cite{vural2018two}, a Poissonian laser\cite{bennett2009interference,prtljaga2016chip}, parametric down-conversion source\cite{polyakov2011coalescence, huber2017interfacing}, and frequency combs\cite{konthasinghe2014resonant}. The rest of this section will focus on the interfacing of multiple QDs on the same photonics circuit.

To obtain high intereference visibility, the emitted photons have to be identical, but it is experimentally challenging to find two QDs with almost identical emission energies, linewidths, and polarization. While much effort has been invested in fabricating highly reproducible QDs\cite{ollivier2019reproducibility}, it is often necessary to employ tuning mechanisms, both for the QDs and the cavity structures they are embedded in, to match the emission properties and ensure a high indistinguishability of the photons. 

The photonics community has been actively developing multiple techniques for tuning on-chip resonators\cite{du2016mechanically,radulaski2018thermally,li2019photon}.
However, certain methods such as wet-chemical etching\cite{hennessy2005tuning} and gas condensation\cite{Mosor2005Scanning} (unless used in conjunction with local heating) are not suited for tuning individual devices on a chip or array.
We emphasize here that a scalable solution for fully integrated quantum photonics would require that the tuning can be applied locally and independently to individual emitter devices.

\paragraph{Temperature}
Temperature tuning can affect the bandgap structure, which strongly tunes the QD energy; it can also alter the refractive index and cause physical expansion, which would shift the resonances of a cavity device coupled to the QD. The simplest way to achieve this is to change the sample temperature in the cryostat, but this does not allow for the tuning of individual devices\cite{yoshie2004vacuum, gevaux2006enhancement}. Instead, local temperature changes can be applied via electrical heaters or laser irradiation\cite{faraon2007local, faraon2009local,katsumi2020situ} (see Fig. \ref{fig:qd_tuning}a). A recent report has also employed temperature tuning of two QDs in a nanophotonic waveguide to achieve superradiant emission\cite{kim2018super}.

\paragraph{Strain}
QD energy is sensitive to strain tuning, and strain sensors have been demonstrated by detecting energy shifts at the $\mu$eV level for InGaAs QDs embedded in a photonic crystal membrane\cite{carter2017sensing}. However, to achieve larger tuning ranges, strain can be induced via piezoelectric crystals, and a tuning rate of $\sim$1\,pm/V~\,\cite{elshaari2018strain} has been achieved (see Fig. \ref{fig:qd_tuning}b). A difficulty with this approach is the relatively large fabrication overhead of integrating piezoelectric materials. However, another recent work has shown that strain tuning can be achieved via laser-induced local phase transitions of the crystal structure, which circumvents this issue\cite{grim2019scalable}.

\paragraph{Electric field}
The application of electric fields across the QD can be used to control the energy of the QD excitonic lines via the quantum-confined Stark effect\cite{thon2011independent,pagliano2014dynamically,hallett2018electrical}.
The application of a forward bias voltage leads to a blue-shift of the QD emission wavelength of several nm. This can be complemented by independently tuning the cavity mode of the photonic crystal structure, e.g. via electromechanical actuation\cite{petruzzella2015fully,petruzzella2018quantum} (see Fig. \ref{fig:qd_tuning}c).
Beyond wavelength tuning, recent work has also shown that electrical control of QDs is crucial to obtain the best optical properties for integrated QDs\cite{pedersen2020near}.

\paragraph{Frequency conversion}
An alternative strategy is to tune the emitted photons via on-chip quantum frequency conversion (QFC)\cite{singh2019quantum}. \RefCite{weber2019two} performs QFC separately on the output of two QDs to convert them from 904\,nm to the telecom C-band, achieving a two-photon interference visibility of 29$\pm$3\% (see Fig. \ref{fig:qd_tuning}d).


\section{2-D materials\label{Sec:2DMaterials}}

\subsection{Introduction into 2D Materials}

Single photon sources in 2D materials have unique advantages compared to other quantum emitters in 3-D bulk material. Confined in atomically thin material,  they can potentially have high photon extraction efficiencies and their emission properties can be controlled by a variety of effects including strain, temperature, pressure and applied electric and magnetic field. Indeed, single photon sources in monolayered 2D material can have almost unity out-coupling efficiency as none of the emitters are surrounded by high refractive index material and their emitted light are consequently not affected by Fresnel or total internal reflection  \cite{journal1,journal2}. In addition, 2D materials can be easily transferred and integrated with photonic structures or other 2D materials to form synergistic  heterostructures that combine the advantages of various materials together in one unified structure\cite{journal3}. In this review of single photon sources in 2D materials we restrict ourselves to transitional metal dichalcogenides (TMDC) and hexagonal boron nitride (hBN) although we acknowledge that there are other important examples of 2D materials including graphene, anisotropic black phosphorus and borophene \cite{journal4,journal5,journal6}. An important feature of single photon emitters in these 2D materials is that similar to NV$^-$ centers in diamonds and QDs, they can be used for efficient spin-photon quantum interfaces by tailoring the light-matter interactions due to the broken inversion symmetry \cite{journal7,journal8}. The zero field splitting in TMDC materials can be up to $\sim$ 0.7 meV, which is about 50 times higher than InAs/GaAs self-assembled quantum dots, and it has a surprisingly large anomalous g-factor of $\sim$ 8-10 that can potentially allow for extremely fast coherent spin coupling\cite{journal9}. On the other hand, hBN has a considerably smaller zero-field splitting of 0.00145 meV and a more modest g-factor of 2\cite{journal10}.

\subsection{Single photon emitters in Transitional Metal Dichalcogenides (TMDC)}

\begin{figure}[t]
\includegraphics[width=0.95\columnwidth]{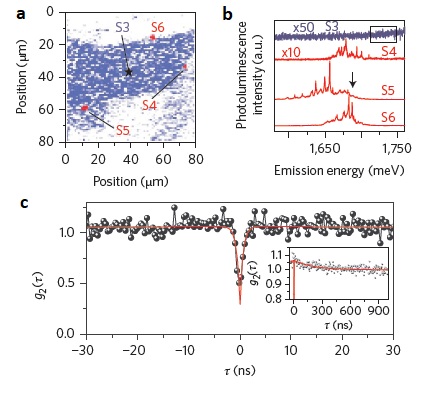}
\caption{\label{fig:Figure1} Single photon emitters in TMDC materials: a) contour plot of photoluminescence scanning experiment for WSe\textsubscript{2} flake with emission spots at the edges of the flake; b) these spots possess narrow line emission spectra; c) measurement of g\textsuperscript{2}(0) for one of the narrow emission lines, the value below 0.5 clearly indicates the single photon properties of the emitter. \SpringerNatureReprint{journal43}{2015}}
\end{figure}

A monolayer of TMDC can be described as a MX\textsubscript{2} sandwich structure with M being a transition metal atom (e.g., Mo, W) enclosed between two lattices of chalcogen atom X(e.g., S, Se, Te) \cite{journal11,journal12}. Depending on the choice of elements and the number of layers present, TMDC materials can have widely varying electrical, optical, chemical, thermal and mechanical properties \cite{journal14,journal15,journal16,journal18,journal20,journal21}. Although TMDCs have strong in-plane covalent bonds, they are only weakly bonded in between the layers by Van der Waals forces, which allows them to be easily exfoliated to form monolayer flakes. Alternatively, single layer TMDCs can be fabricated using CVD or MBE\cite{journal23,journal24}. Despite the fact that multilayer TMDCs have indirect bandgaps, monolayer TMDCs are actually direct bandgap semiconductors, which enables them to have enhanced interactions with light\cite{journal14,journal25,journal26,journal27}. There are two distinctive properties that are associated with monolayer TMDCs: strong excitonic effects and valley/spin-dependent properties. The latter can be attributed to the fact that there is no inversion center for a monolayer structure, which opens up a new degree of freedom for charge carriers, i.e. the $k$-valley index, that brings new valley-dependent optical and electrical properties into play \cite{journal20,journal28,journal29,journal30}. In contrast, TMDC's strong excitonic effects is due to strong Coulomb interactions between charged particles (electrons and holes) and reduced dielectric screening, which result in the formation of excitons with large binding energies (0.2 to 0.8 eV), charged excitons (trions), and biexcitons \cite{journal32, journal34,journal35,journal36,journal40}. These manifest themselves by broadband photoluminescence (tens of nanometers) in the visible and near-IR ranges at room and cryogenic temperatures. At the same time, TMDCs can possess quantum-dot like defects, which exhibit themselves in the photoluminescence spectrum as a series of sharp peaks with peak intensities up to several hundred times larger than the excitonic photoluminescence with linewidths around 100 $\mu$eV and excited state lifetime from 1 to several ns\cite{journal9, journal43,journal44,journal45} (Fig.~\ref{fig:Figure1}). For exfoliated samples these defects are usually associated with local strain and typically appear at cracks or edges of the flake while for grown samples they are mostly due to impurities and can appear everywhere. A number of works have shown that these defects emit in the single photon regime and can be controlled by induced strain, applied temperature, electric and magnetic fields \cite{journal9,journal45,journal46,journal47,journal48}. However, a significant drawback is that the single photon emission quenches at temperatures above 20 K although some recent research have shown that special treatment of TMDC flakes can lead to a redistribution of the energy levels and enable emission at room temperature \cite{journal49,journal50}.

\subsection{Single photon emitters in Hexagonal Boron Nitride (hBN)}

\begin{figure}[t]
\includegraphics[width=0.95\columnwidth]{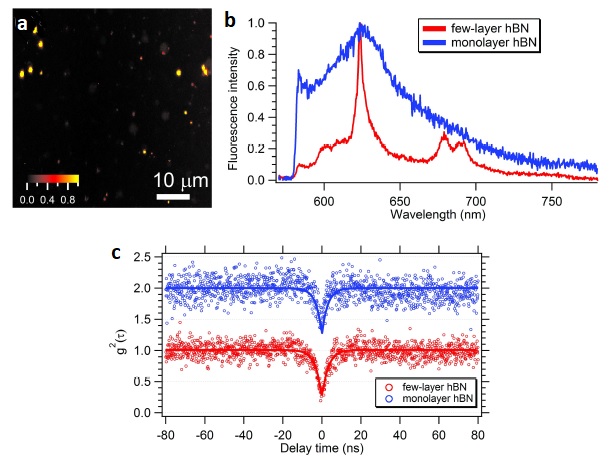}
\caption{\label{fig:Figure2} Single photon emitters in hBN nanoflakes: a) scanning confocal map of hBN nanoflake with bright luminescent spots corresponding to single defects; b) room temperature emission spectrum for single defects in mono- and few-layer hBN nanoflakes; c) measurements of g\textsuperscript{2}(0) for single defects in mono- and few-layer hBN nanoflakes. The curves are shifted for clarity demonstrating g\textsuperscript{2}(0)$<$0.5. \SpringerNatureReprint{journal51}{2016} }
\end{figure}

hBN monolayers are structurally similar to TMDC but whereas single photon emitters (SPEs) in TMDCs are associated with localized excitons, SPEs in insulating hBN are, similar to color centers in diamond, attributed to atomic-like defects of the crystal structure\cite{journal51,journal52,journal53}. These defects in hBN are some of the brightest single photon sources in the visible spectrum and have large Debye-Waller factors with good polarization contrasts. Like NV$^-$ centers, their electronic levels are within the band gap ($\sim$ 6eV), resulting in stable and extremely robust emitters at room temperature over a wide spectral bandwidth ranging from green to near infrared with most emitters emitting around the yellow-red region\cite{journal54} (Fig.~\ref{fig:Figure2}). These SPEs in hBN are generally characterized by short excited state lifetime (several ns), absolute photon stability, and high quantum efficiency \cite{journal51, journal52,journal54}. Close to Fourier transform limited linewidths below 100 MHz have been recorded with resonant excitation at cryo and room temperatures\cite{PhysRevB.98.081414, PhysRevB.101.081401}.Recent research indicates that various types of defects are responsible for the multiplicity of observed ZPL emissions, including nitrogen vacancy defects (NV), carbon substitutional (of a nitrogen atom) defects, and oxygen related defects\cite{journal53,journal55}. Interestingly, the asymmetric linewidths of some of these ZPLs have been attributed to the existence of two independent electronic transitions\cite{journal56}.

\subsection{Deterministic Creation and Control of Single Photon Emitters in 2D Materials\label{Sec:2DCreation}}

\begin{figure}[t]
\includegraphics[width=0.95\columnwidth]{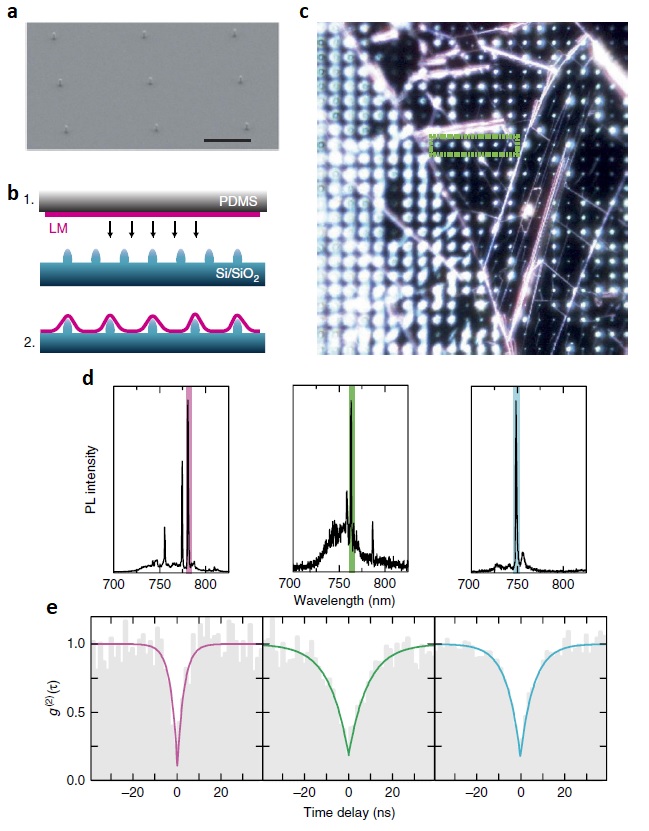}
\caption{\label{fig:Figure3} Deterministic creation of single photon emitters by strain: a) SEM image of nanopillar substrate, fabricated using electron beam lithography, black scale bar corresponds to 2 $\mu$m; b) illustration of the fabrication method; c) dark-field optical image of mono-layer WSe\textsubscript{2} flake on top of nanopillar array. Green rectangle indicates the 6 adjacent nanopillars where single photon emitters were measured; d) examples of measured spectras at different pillars demonstrating narrow emission lines; e) measurements of g\textsuperscript{2}(0) for the emission lines in (d). \CCBYFourReprint{journal48} }
\end{figure}

The random distribution of SPEs in 2D materials is a significant obstacle that prevents their integration with photonic structures. Deterministically creating SPEs in 2D materials is therefore an important technological goal. One way to do so is to induce SPEs by introducing strain to the material. This concept was successfully realized by several groups that transferred 2D materials onto the tops of metallic or dielectric nanopillar arrays\cite{journal47,journal48,journal57} (Fig.~\ref{fig:Figure3}). The SPE (as confirmed by measured g\textsuperscript{2}(0)$<$0.1) yield of this technique exceeds 90~$\%$ and the quality of these engineered SPEs was reported to be even higher than naturally occurring defects with 10 times less spectral diffusion  ($\sim$ 0.1 meV)\cite{journal48}. An even simpler (but less scalable) way to create SPEs with strain was suggested by Rosenberger \textit{et al.}: place a deformable polymer film below the 2D material of interest and apply mechanical force to the film using an atomic force microscopy (AFM) tip\cite{journal58}. Although this approach is less scalable, it can enable one to tune the SPE's optical properties through careful strain engineering with the AFM tip.

Remarkably, emission from SPE in 2D material defects can be controlled through the application of a voltage\cite{journal45,journal46}. For example, in \RefCite{journal46}, photo and electro-luminescence were observed from point defects in a 2D material (WSe$_2$) sandwiched between hBN layers that served as tunnelling barriers between WSe$_2$ and its graphene electrodes. Moreover, Schwarz \textit{et al.} showed that it was possible to tune the emission wavelength of the SPE ($\sim$ 0.4 meV/V) by changing the applied bias voltage, which makes the platform amenable to a host of technological applications.

\subsection{Enhancement of Emission from 2D Materials by Coupling into Resonant Modes}

\begin{figure*}
\includegraphics[width=0.95\textwidth]{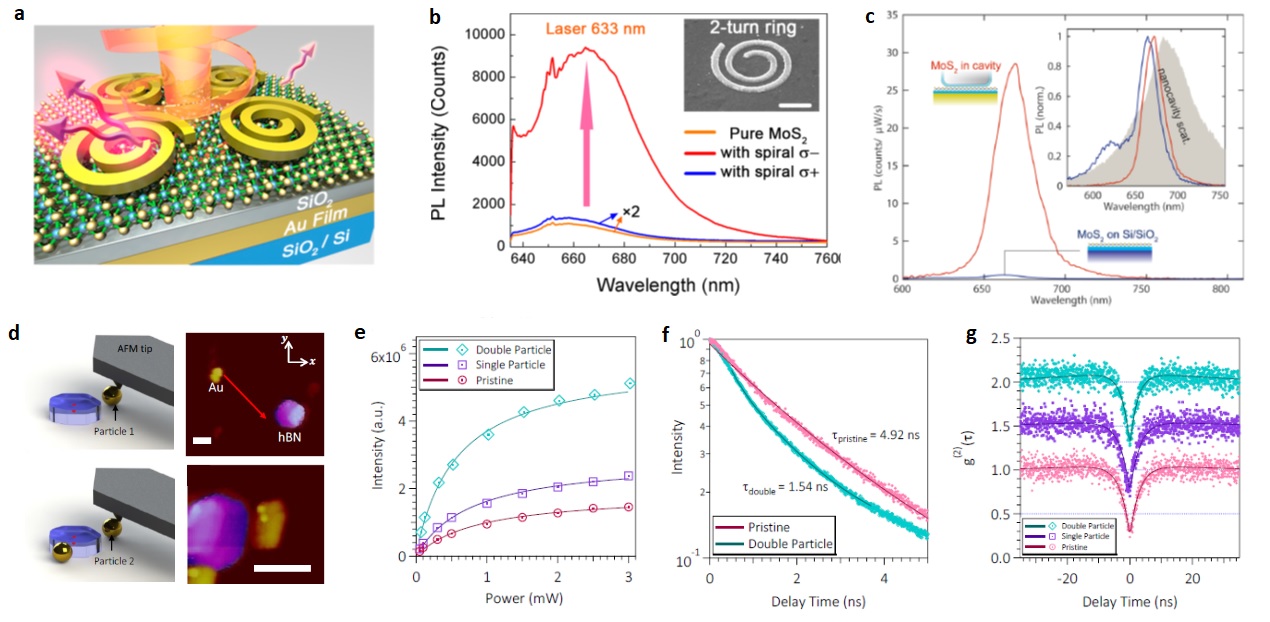}
\caption{\label{fig:Figure4} Enhancement of spontaneous emission with nanocavities: (a) Schematic of spiral ring structure with circularly polarized light excitation. The photoluminescence intensity is enhanced when the MoS\textsubscript{2} monolayer couples with plasmonic spiral structures; (b) Spectra of MoS\textsubscript{2} monolayer with and without spiral structures, under the excitation of different circularly polarized light, inset shows the cross section view of a 2-turn spiral ring. The scale bar is 400 nm. Figures (a) and (b) are \ACSreprint{journal63}{2017} (c) Spectra of MoS\textsubscript{2} monolayer on a SiO\textsubscript{2}/Si substrate (blue) and in the nanocavity (red) obtained using a diffraction-limited excitation spot. The inset presents normalized spectra for the two cases along with the scattering spectrum for a typical nanocavity (gray). \ACSReprint{journal59}{2015} (d) Deterministic positioning of gold nanospheres with hBN nanoflake by AFM tip, white scale bar is 250 nm. (e) A comparison between time-resolved measurements of pristine and double particle arrangement. (f) A comparison of fluorescence saturation curves between the pristine, single particle, and double particle arrangements. (g) Measurements of g\textsuperscript{2}(0) for pristine, single particle, and double particle arrangements. The curves are shifted for clarity. Figures (d) - (g) are \RCSreprint{journal64} }
\end{figure*}

Although the tunability and large oscillator strengths of SPEs in 2D materials make them attractive in photonic applications, their sub-nanometer thickness results in a small light-matter interaction length that limits their efficiency. However, as noted in section \ref{Sec:theory_purcell}, this disadvantage may be mitigated by coupling them with resonant photonic structures where both their absorption and emission can potentially be enhanced\cite{journal59}. Fortunately, the atomic thickness of 2D materials makes them especially amenable to integration with photonic structures such as planar photonic crystal cavities, ring resonators, and optical microcavities. 

The first demonstration of this came from the coupling of photoluminescence from SPEs in TMDC materials to dielectric and plasmonic nanocavities\cite{journal60, journal61,journal62,journal63} (Fig.~\ref{fig:Figure4}, a - c). In \RefCite{journal60}, a coupling efficiency of over 80\% was demonstrated and a spontaneous emission enhancement of over 70 times was reported. Intriguingly, the photoluminescence enhancement can be controlled via an optical spin orbit coupling, which depends on both the resonant nanoparticles' geometry as well as the incident laser's polarization and power\cite{journal63} (Fig.~\ref{fig:Figure4}, a, b). Subsequently, single photon emission from a single emitter in a hBN nanoflake was successfully coupled to both one and two resonant gold nanospheres\cite{journal64} (Fig.~\ref{fig:Figure4}, d - g). These nanospheres were brought into close proximity with a pre-characterized SPE (verified by measuring $g^{(2)}(0)<0.5$) by means of an AFM tip and the emitter was observed to have a photon flux of about 6 MHz that corresponded with a 3-fold Purcell enhancement\cite{journal64}.

A natural structure for SPEs in 2D materials to couple to are noble metal nanopillars since they can kill two birds with one stone by first facilitating the deterministic creation of SPEs (as discussed in section \ref{Sec:2DCreation} above), and then enhancing the created SPEs' spontaneous emission through the SPEs' coupling with surface plasmon resonances of the metallic nanopillars. This has been successfully implemented for both TMDCs and hBN at cryogenic and room temperatures where increased brightness, shorter lifetimes and enhanced spontaneous emission of the SPEs were all reported\cite{journal65,journal66,journal67,journal68}. Coupling of the emitter to plasmonic modes results in linearly polarized emission that depends on the geometry of the nanopillars and the orientation of the optical dipole\cite{journal66}. A record high Purcell enhancement of 551 times was achieved with metallic nanocubes\cite{journal67}.

\subsection{Coupling and Transfer of Emission from 2D Materials into Photonic Structures}

\begin{figure*}
\includegraphics[width=0.95\textwidth]{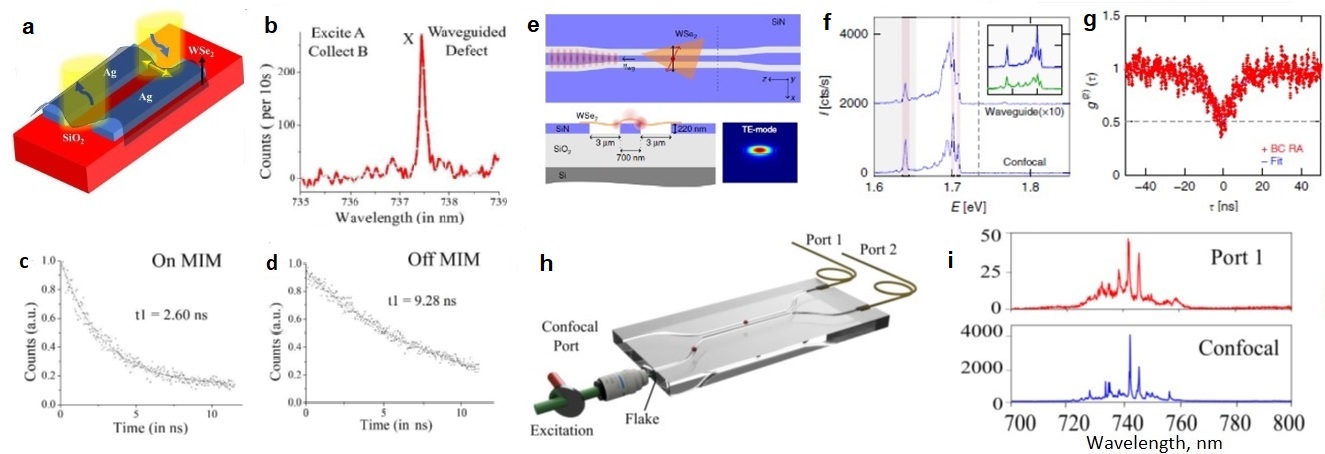}
\caption{\label{fig:Figure5} Coupling and transport of single photon emission from 2D materials in photonic structures: (a) Scheme of an MIM waveguide covered by a WSe\textsubscript{2} monolayer. The yellow dipole is a quantum emitter, the blue arrows denote the excitation and collection points; (b) The spectra of the defect with the excitation spot fixed at the location of the defect at the waveguide and collection spot moved to the far end of the waveguide; (c) and (d) Lifetime measurements for emitters located on the MIM waveguide and out of it. Figures (a) -- (d) were \AIPreprint{journal70} (e) Illustration of the Si\textsubscript{3}N\textsubscript{4}  photonic device with a WSe\textsubscript{2} flake integrated on top of a 220 nm thick single mode Si\textsubscript{3}N\textsubscript{4}  waveguide, separated by 2 air trenches from the bulk Si\textsubscript{3}N\textsubscript{4} ; (f) Confocal and waveguide-coupled spectrum of the emitter excited by 702 nm pump. The waveguide spectrum is multiplied by 10 and offset by 2000 cts/sec for improved visualization. The inset shows confocal spectra obtained by either green excitation at 532 nm (green curve) or excitation at 702 nm (blue curve); (g) Background-corrected measurements of g\textsuperscript{2}(0) for the coupled emitter made in confocal geometry. Figures (e) -- (g) were \CCBYFourreprint{journal72} (h) Scheme of Ti in diffused lithium niobate directional coupler with a WSe\textsubscript{2} flake at the input facet. Emitters are excited in confocal geometry. Their emission is detected by a confocal microscope and through the two output ports (Port 1 and Port 2); (i) The spectra measured through the waveguide output port and confocally when the emitter is excited at the facet. Figure (h) and (i) were \CCBYFourreprint{journal71}}
\end{figure*}

To fully integrate SPEs in 2D materials onto an on-chip photonic platform, it is also necessary to couple emission from SPEs into waveguiding photonic structures. To this end a few groups have successfully coupled emission from SPEs in 2D materials into the surface plasmon polariton modes of silver based waveguides. For example, localized SPEs formed from the intrinsic strain gradient formed along a WSe$_2$ monolayer when it was deposited on top of a silver nanowire were efficiently coupled to the guided surface plasmon modes of the nanowire\cite{journal69}. A coupling efficiency of 39\% was measured for a single SPE by comparing the intensity of the laser excited SPE and the emission intensities at both ends of the silver nanowire\cite{journal69}. Separately, S. Dutta \textit{et al.} demonstrated the coupling of single emitters in WSe{$_2$} to propagating surface plasmon polaritons in silver-air-silver, Metal-Insulator-Metal (MIM) waveguides\cite{journal70} (Fig.~\ref{fig:Figure5}, a - d). The waveguides were fabricated using EBL, followed by metal deposition of Cr and Ag, and then a liftoff in acetone with subsequent protection by a 4 nm buffer layer of oxide. As before, strain gradients on the monolayer due to the waveguide generated sharp localized SPEs that were intrinsically close to the plasmonic mode. Due to the sub-wavelength confinement of the surface plasmon polariton modes, a 1.89 times enhancement of the SPE's radiative lifetime was observed under illumination at 532 nm at cryogenic (3.2 K) temperatures and bright narrow lines associated with the SPE's emission were measured.

Although surface plasmon polariton modes on silver waveguides may help to enhance the spontaneous emission of a SPE, the metal interface results in significantly lossy propagation that is undesirable. Such high propagation losses can be circumvented by using dielectric waveguides instead. In \RefCite{journal72}, integration of a monolayer WSe$_2$ flake with a 700 nm wide Si$_3$N$_4$ waveguide that was patterned using standard EBL techniques was achieved by carefully picking up and releasing a bulk exfoliated flake using a GelPak stamp (Fig.~\ref{fig:Figure5}, e - g). Confocal scans of the WSe$_2$ monolayer with a 532 nm excitation laser at cryogenic temperatures indicated that several SPEs were sufficiently close to the Si$_3$N$_4$ waveguide to enable coupling to it. Although some luminiscence was measured at the end of the waveguide, which provided proof of a non-zero coupling, the SPE's coupling to the waveguide is strongly dependent on the orientation of its optical dipole and consequently, the photoluminiscence spectra for a confocal scan can be significantly different from that obtained at the end of the waveguide. Brighter emission and saturation counts of up to 100 kHz can be obtained by exciting the SPE at close to the free exciton wavelength ($\approx$ 702 nm) with a tunable Ti:Sapphire laser. Besides enabling brighter emission, excitation at 702 nm also produces less background fluorescence. Measurements of the background subtracted $g^{(2)}$ correlation function in confocal geometry revealed a anti-bunching dip with $g^{(2)}(0)=0.47$, that suggest the existence of a SPE. 

Finally, we note that there has been successful integration of SPEs in 2D materials to an on-chip beamsplitter in the form of a lithium niobate directional coupler\cite{journal71} (Fig.~\ref{fig:Figure5}, h - i). Indeed, emission from an excited SPE in a strain engineered WSe$_2$ monolayer was coupled into one input port of the directional coupler and its photoluminiscence, which consisted of strong emission lines corresponding to emission from the WSe$_2$ flake, was successfully measured at the other output port of the directional coupler (Fig. \ref{fig:Figure5}e), demonstrating the desired operation of the beam splitter and showing that an on-chip Hanbury Brown and Twiss measurement is possible.

\section{Integration Approaches}
\label{sec:integ}

Numerous techniques have been developed to directly integrate solid-state quantum emitters with on-chip nanophotonic structures.
Doing so would allow for dense integration of these emitters on a large scale,
and also provide potential benefits in coupling efficiencies, device stability, and ease of control. 

In this section, we will provide an overview of hybrid integration approaches, and discuss their applicability to the solid-state emitters presented in this review. 
A detailed reference on integration methods for hybrid quantum photonics can be found in \RefCite{kim2020hybrid}.

\subsection{Random Dispersion}

A simple integration method is to forego deterministic positioning, and rely on the random placement of the quantum emitters.
For QDs\cite{mirin19951,wan2017monolithically} and 2D materials\cite{journal9, journal43,journal44,journal45}, the emitters may already be randomly distributed in their as-grown state.

In the case of NV$^-$ centers, type Ib diamonds, which by definition have significant singly dispersed nitrogen impurities\cite{Dyer1965Optical, Chrenko1971Dispersed, Mainwood1979Substitutional}, naturally host an ensemble of randomly positioned NV$^-$ centers. Such diamonds can be found naturally or manufactured using a High-Pressure High-Temperature process\cite{Dyer1965Optical}. Similarly, type IIa diamonds, which by definition have much lower concentrations of nitrogen impurities compared to type Ib diamonds, also host a sparse ensemble of randomly positioned NV$^-$ centers. Although rare in nature, type IIa diamonds can be grown using chemical vapor deposition\cite{Werner1998Growth}. Photonic structures fabricated on diamond membranes will therefore have randomly positioned NV centers, and indeed, although the yield for such structures is poor with sub-optimal coupling between the photonic structure and NV center, many early experiments relied on such a random dispersal technique. Other color centers in diamond do not form naturally and have to be integrated using more deterministic techniques.

Colloidal QDs and nanodiamonds containing color centers can also be randomly dispersed onto photonic structures via drop casting or spin coating\cite{yang2017room}. Nanodiamonds have also been dip coated directly onto single mode optical fibers \cite{schroder+5:12}. In this technique, the tapered fiber is dipped into a droplet of nanodiamond-containing solution on the tip of a glass rod. The tapered fiber is then pulled by a linear stage along the axis of the fiber. However, SPEs in the form of nanoparticles often have a large surface area, which may lead to optical instabilities such as blinking or bleaching. 
This is due to the stronger influence of surface states and enhanced Auger recombination\cite{efros2016origin}. 
Moreover, random dispersion is not suitably scalable for quantum photonic applications where efficient, deterministic coupling between emitters and the photonic circuit is crucial.

To improve the positioning precision, and thus the coupling efficiency, a lithography-based masking method can be used to selectively deposit dispersed emitters on top of the photonic structures\cite{chen2018deterministic} (Fig. \ref{fig:integ}a). Despite its limitations, randomly positioning of emitters can still be a useful method to rapidly prototype hybrid quantum photonics platforms.

\subsection{Targeted Creation}

\subsubsection{Irradiation and annealing in diamond\label{Sec:irradiation_and_annealing}}

Deterministic positioning of NV centers can be achieved by deliberately creating vacancies in type Ib diamonds with irradiation of either a focused ion\cite{Martin1999Generation, Waldermann2007Creating, McCloskey2014Helium}, proton\cite{Wee2007Two} or electron\cite{Martin1999Generation,Kiflawi2007Electron, Becker2018Nitrogen} beam. Since vacancies in diamond can, with an activation energy of $\approx$ 2.3 eV migrate during annealing\cite{Davies1992Vacancy, Mainwood1994Nitrogen, Collins2009The} at $\gtrsim$ 600 $^\circ$C, the diamond is then typically annealed after irradiation to allow the vacancies to diffuse and be ``captured'' by an existing nitrogen impurity\cite{Davies1976Optical,Davies1977Charge,Davies1992Vacancy}. The spatial resolution of this technique is therefore not just dependent on the resolution of the focused ion or electron beam, but also on the concentration of nitrogen impurities in the diamond, which determines how far a vacancy has to diffuse before being captured by a nitrogen impurity. In fact, despite having a significantly larger concentration of nitrogen compared to type II diamonds, the spatial resolution of this technique in type Ib diamonds can still be limited by the concentration of nitrogen impurities\cite{Martin1999Generation}. The diffusion length $l_D$ of the vacancies may be estimated using $l_D\sim\sqrt{D\Delta t}$, where $D$ here is the diffusion coefficient and $\Delta t$ is the anneal time. $D$ may be obtained via the Arrhenius equation
\begin{equation}
D=D_0\exp\left(-\frac{E_a}{k_B T}\right),
\end{equation}
where $E_a$ is the activation energy, $k_B$ is the Boltzmann constant, $T$ is the temperature, and $D_0$ is the maximal diffusion constant calculated to be around $3.7\times10^{-6}$ cm$^2$/s for vacancies near the diamond surface\cite{Hu2002The}. For typical conditions, the diffusion length is likely to be $\sim$ 100 nm, and indeed, vacancies have been observed to diffuse by a few hundred nm in the vertical (along the irradiation axis) direction\cite{Santori2009Vertical} and a vacancy diffusion limited transverse spot size of less than 180 nm was obtained using a focused Ga$^+$ ion beam\cite{Martin1999Generation}. Although there has been reports of vacancy transverse diffusion lengths that are in the tens of $\mu$m range\cite{Gippius1999Formation,Steeds2009Long} that seem to defy the simple $l_d\sim\sqrt{D\Delta t}$ estimate, we note that this could potentially be explained by the scattering of ions/electrons on masks\cite{Toyli2010Chip, Spinicelli2011Engineered, Pezzagna2011Creation, Sangtawesin2014Highly} if they were used on the diamond's surface\cite{Orwa2012An}. Nevertheless, we note that if the mask is carefully designed, transverse spatial resolution in the tens of nanometers can be achieved using an implantation and annealing approach\cite{Scarabelli2016Nanoscale} (see Fig. \ref{Fig:NVImplantWithMask} and section \ref{Sec:NVImplantAnneal}). Furthermore, we note that the choice of radiation used can significantly affect the vertical distribution of vacancies. In general, the heavier ions deposit most of their energy within a narrow band and creates vacancies at a more well defined depth whereas the lighter electrons tend to create a more uniform depth profile of vacancies\cite{Acosta2009Diamonds}.

\begin{figure}
\includegraphics[width=0.47\textwidth]{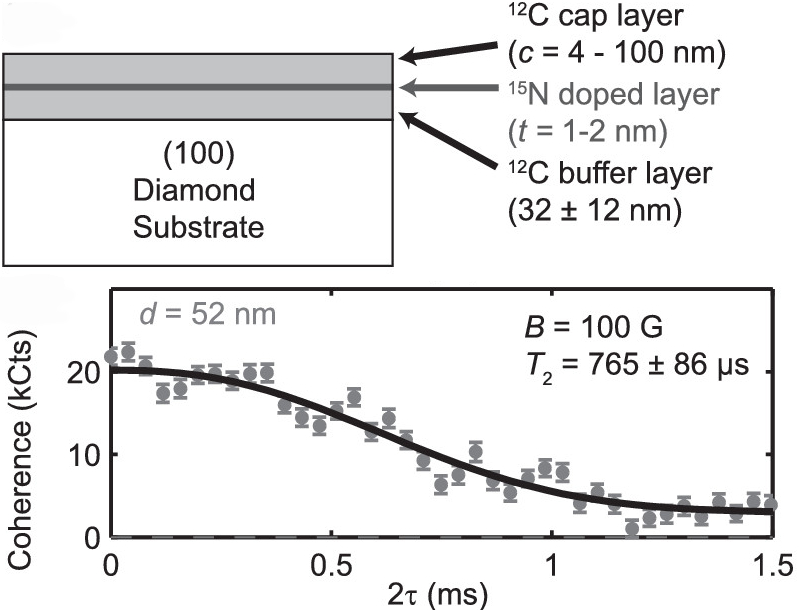}
\caption{Top: NV$^-$ centers created by CVD growth and nitrogen delta doping of a bulk diamond substrate. Vacancies were created by electron irradiation at 2 MeV. Bottom: NV$^-$ centers created in this way at a depth of 52 nm exhibited a Hahn-echo $T_2$ coherence time of 765 $\mu$s at room temperature. Using a similar procedure, \RefCite{Ishikawa2012Optical} achieved a $T_2$ of $\approx$ 1.7 ms in isotopically enriched $^{12}$C diamond. \AIPReprint{Ohno2012Engineering}}
\label{Fig:NdopedCVDDiamond}
\end{figure}

An alternate pathway to creating NV$^-$ centers is to grow an isotopically pure $^{12}$C (which has no nuclear spin) diamond layer on an existing substrate using plasma assisted CVD, and then introducing nitrogen gas during the last stages of the growth\cite{Ohno2012Engineering, Ishikawa2012Optical, Ohashi2013Negatively} (see Fig. \ref{Fig:NdopedCVDDiamond}). It is possible using this procedure to make single NV$^-$ centers with a long spin coherence time of $T_2\approx1.7$ ms\cite{Ishikawa2012Optical}. To achieve 3-D localization of NV$^-$ centers using such an approach, the depth of the nitrogen doped layer, which determines the depth of the NV centers, can first be carefully controlled by slowing the growth rate down to $\sim$ 0.1 nm/min\cite{Ohno2012Engineering}, which allows a depth precision of a few nm (delta doping). Transverse localization ($\lesssim$ 450 nm) of long coherence NV$^-$ centers with $T_2\approx 1$ ms can then be achieved by using a Transmission Electron Microscope (TEM) to create vacancies within the nitrogen doped layer followed by annealing\cite{McLellan2016Patterned}. The long coherence time of those NV centers are not limited by either $^{13}$C nuclear spins or lattice damage induced by the electrons (which is thought to be small compared to ion implantation) but rather the presence of other nitrogen impurities that were not converted into NV centers \cite{McLellan2016Patterned}. A variation to using electron irradiation via a TEM is to use irradiation of $^{12}$C\cite{Ohno2014Three} or He$^+\,$\cite{Kleinsasser2016High} ions to create vacancies. Compared to electron irradiation, using ions allows for a more localized layer of vacancies, which will reduce the unwanted creation of NV centers in the substrate of the CVD grown diamond. However, we note that using such an approach can possibly decrease the $T_2$ of the NV centers due to increased lattice damage caused by the ions.

\subsubsection{Implantation and annealing in diamond\label{Sec:NVImplantAnneal}}

A slightly different approach is to start with a type IIa diamond which does not contain significant amounts of nitrogen impurities and to introduce both the vacancy and nitrogen impurity at the same time by implanting N$^+$ (or N$_2^+$) ions with a focused ion beam and then annealing at $\gtrsim$ 600 $^\circ$C. Using this approach, it is possible to fabricate single NV$^-$ centers with transverse spatial resolution of tens of nm and a yield of $\sim$ 50\% using 2~MeV N$^+$ ions with a beam diameter of 300 nm\cite{Meijer2005Generation}. The NV$^-$ yield, which is defined as the ratio of active NV$^-$ centers to the number of implanted N$^+$ ions, is proportional to the ion beam's energy with a particularly strong slope in the keV region\cite{Pezzagna2010Creation}. This is most likely due to the fact that the number of vacancies an ion generates is also proportional to its energy and \textsc{SRIM}\cite{Ziegler2004SRIM} calculations show that the NV$^-$ yield show a very similar energy dependence\cite{Pezzagna2010Creation}. A less energetic beam should therefore be used to decrease the NV$^-$ yield (for single NV$^-$ creation) but a less energetic beam also results in shallower NV$^-$ centers within the diamond, which can be undesirable for some applications. 

\begin{figure}
\includegraphics[width=0.47\textwidth]{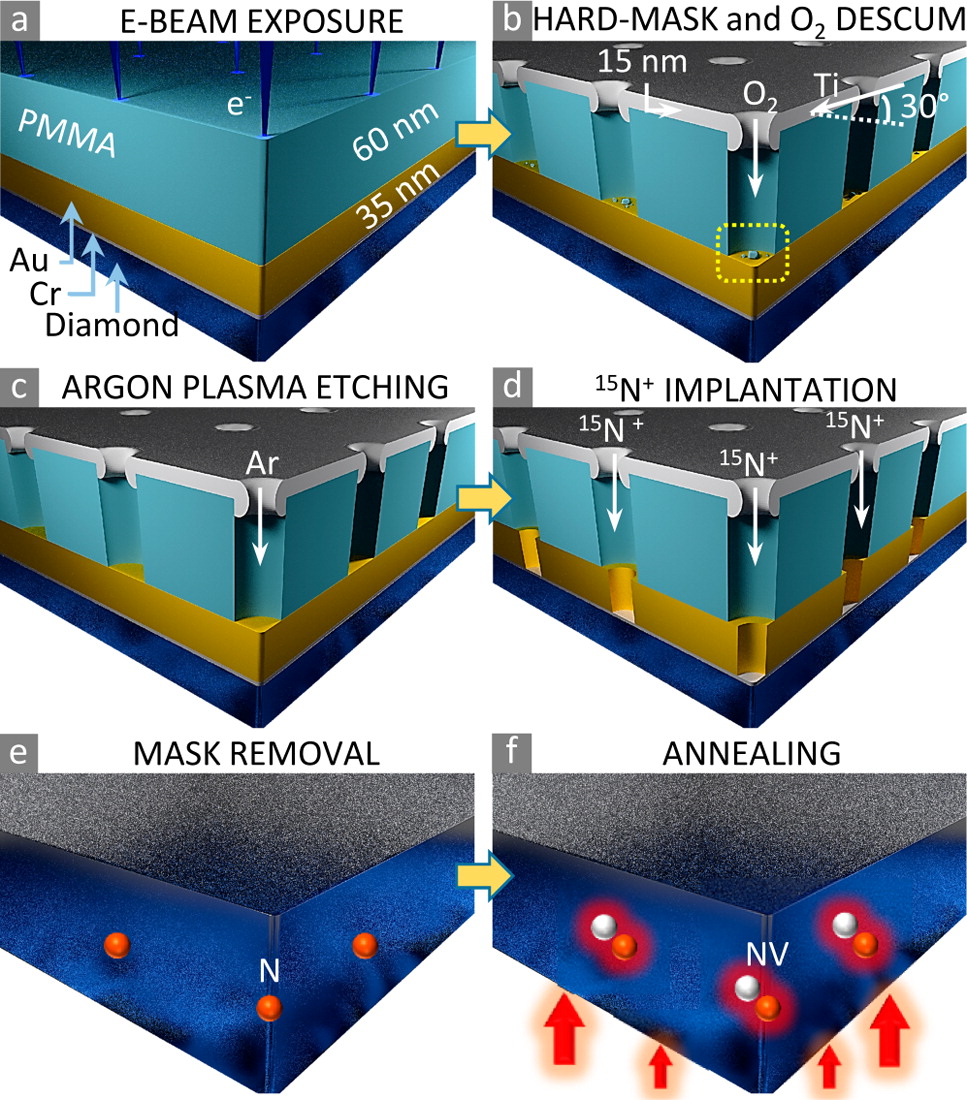}
\caption{Fabrication of NV$^-$ center by implantation of $^{15}$N ions on a masked diamond substrate. (a) Electron beam lithography (EBL) is used to pattern a polymethyl methacrylate (PMMA) mask spin coated on top of an Au layer that sits on a 8 nm thick Cr adhesion layer. (b) After development of the PMMA resist, a hard Ti mask is deposited at 30$^\circ$ to reduce the aperture size and to protect the resist's top surface from an O$_2$ plasma reactive ion etch (RIE) that removes any resist residue from the bottom of the aperture. (c) The pattern is then transferred to the Au mask by Ar plasma etching. Note that the Cr layer is not etched. (d) 10 keV $^{15}$N ions were implanted to introduce nitrogen atoms at $\approx$ 7.5 nm below the surface. (e) Mask is removed by wet chemical treatment. (f) Sample was annealed at 1000 $^\circ$C to to form NV$^-$ center with spatial resolutions of $\sim$ 13 nm. \ACSReprint{Scarabelli2016Nanoscale}{2016}}
\label{Fig:NVImplantWithMask}
\end{figure}

If a focused ion beam is not available, high spatial resolution can also be achieved by the appropriate use of a mask and (unfocused) N$^+$ ion beam implantation. In \RefCite{Scarabelli2016Nanoscale}, NV$^-$ spatial resolution of $\sim$ 10 nm in all three directions were accomplished using N$^+$ ion implantation with the mask shown in Figure \ref{Fig:NVImplantWithMask}.

\subsubsection{Laser writing and annealing in diamond\label{Sec:LaserWriting}}

While traditional irradiation or implantation of particles typically create a trail of vacancies following the implanted particle's path (with increased straggling for lighter particles), irradiation of the diamond by focused femtosecond (fs) laser pulses can create vacancies at a more localized depth. Moreover, their transverse spatial resolution can be better than the diffraction limit due to the non-linear processes involved. In addition, fs laser pulses from the same optical system can also be used to fabricate other photonic structures on the same diamond, opening a convenient avenue of integrating photonic structures with NV centers on the same diamond. 

Implementing fs laser machining in diamond is complicated by the fact that there is a significant mismatch between the refractive indices of diamond ($\approx2.4$ at 790 nm) and air or (more commonly) immersion oil ($\approx$ 1.5). This can result in considerable aberration of the focal point within diamond and cause a significant elongation of the focal volume (which reduces the peak field intensity and localization of vacancies) in the beam's direction of propagation. Indeed, an early experiment attempting to create NV centers using fs lasers focused the beam \textit{above} the diamond's surface instead of within it and relied on the ionization of O$_2$ and N$_2$ molecules in air to generate free electrons and ions that are subsequently accelerated by the light's electric field into the diamond\cite{Liu2013Fabrication}. 

\begin{figure}
\includegraphics[width=0.47\textwidth]{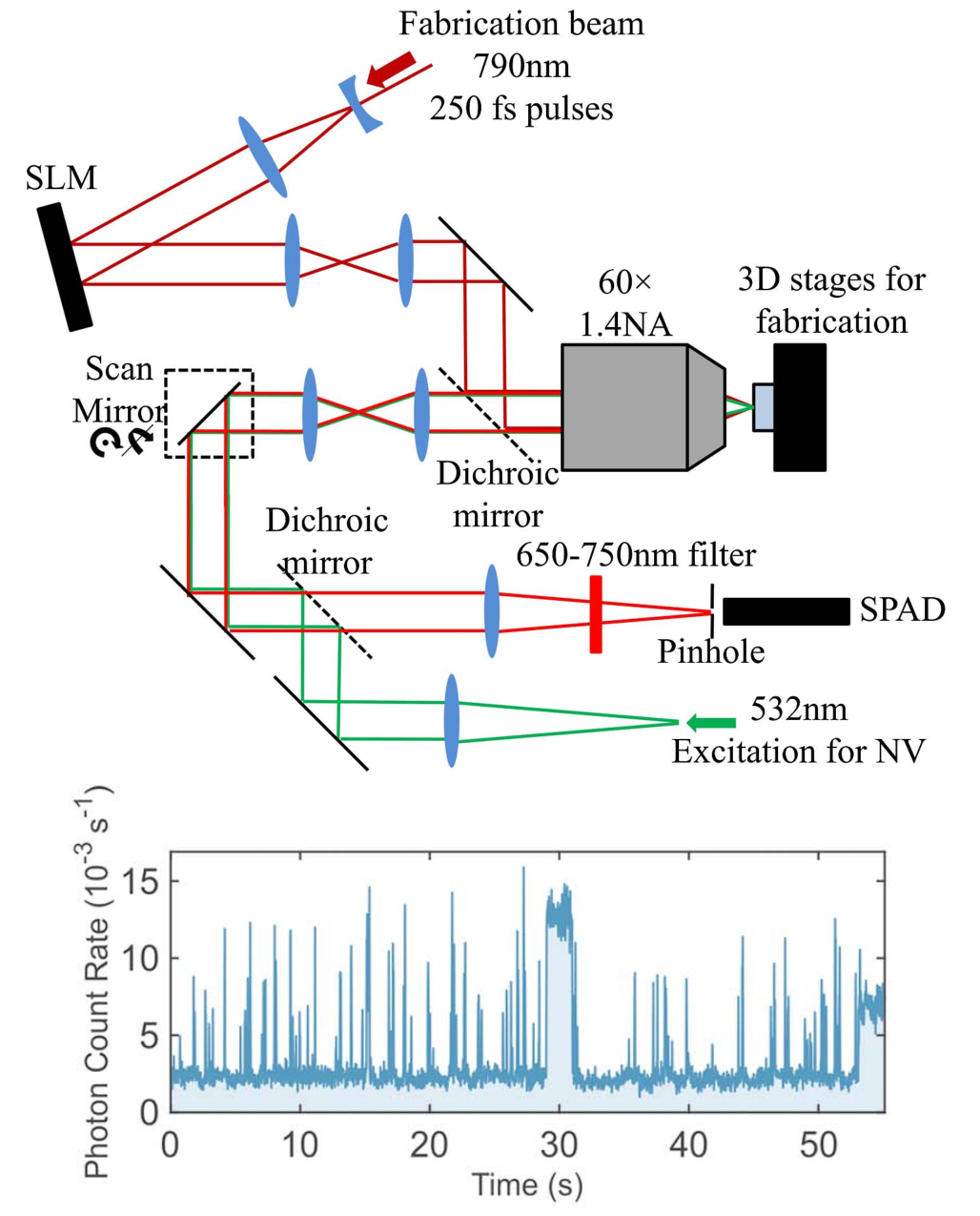}
\caption{Top: Optical setup enabling both vacancy creation and annealing using the same fs laser with a SLM to correct for aberrations. Vacancy creation uses a single 27 nJ pulse while annealing occurs with a 1 kHz pulse train consisting of 19 nJ pulses. Real time monitoring of photo luminescence is also possible by incorporating a separate optical arm consisting of both a 532 nm excitation laser and a single photon avalanche detector with a long-pass filter. Bottom: Fluorescence counts measured during the annealing pulses showing first the creation and destruction of a NV$^-$ center ($\approx$ 28 to 30 ns) followed by the creation of a second NV$^-$ center ($\approx$ 52 ns). \CCBYFourReprint{Chen2019High}}
\label{Fig:NVLaserWrite}
\end{figure}

\begin{figure*}
	\includegraphics[width=0.95\textwidth]{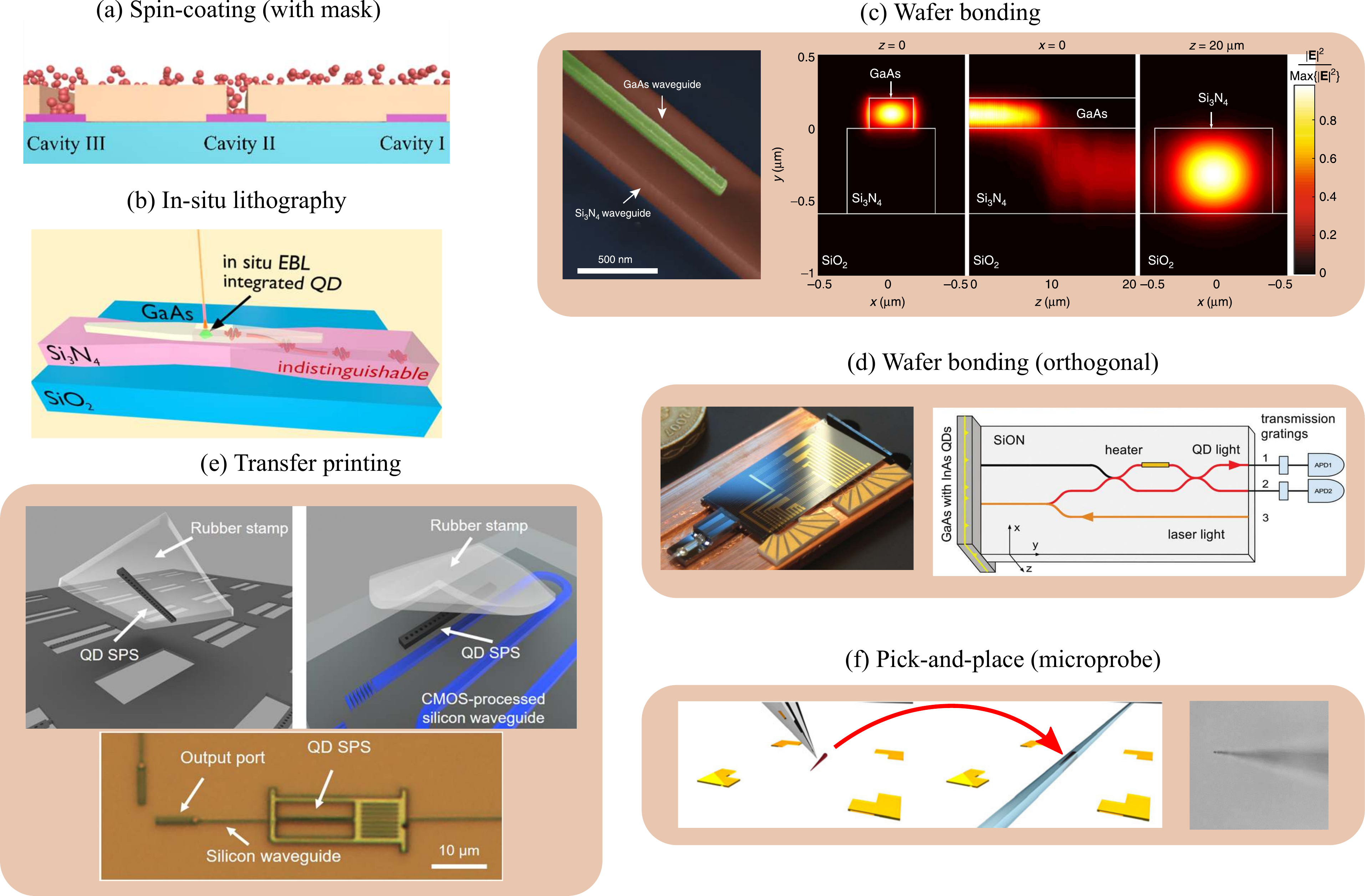}
	\caption{\label{fig:integ}Integration strategies for quantum dots (QDs) on photonic circuits.
		(a) Schematic of QDs deposited via spin-coating. A lithographic mask allows for deterministic positioning on top of photonic cavity devices. \ACSReprint{chen2018deterministic}{2018}
		(b) Schematic of a photonic waveguide patterned via \textit{in situ} electron beam lithography (EBL) around a pre-selected QD, ensuring optimal alignment between the QD and waveguide. \ACSReprint{schnauber2019indistinguishable}{2019} 
		(c) [Left] A GaAs nanobeam on a Si$_3$N$_4$ waveguide fabricated via EBL from a wafer-bonded GaAs/Si$_3$N$_4$ heterostructure. [Right] Simulation results showing the efficient coupling between the GaAs and Si$_3$N$_4$ waveguides. \CCBYFourReprint{davanco2017heterogeneous}
		(d) Optical image [left] and schematic [right] of InAs QDs integrated with a photonic chip via orthogonal wafer bonding. \AIPReprint{murray2015quantum}
		(e) [Top] Schematic of the transfer printing technique using a transparent rubber stamp. [Bottom] Optical image of a transfer-printed QD device on top of a silicon waveguide. \CCBYFourReprint{katsumi2019quantum}
		(f) [Left] Schematic of pick-and-place positioning of QD devices using a sharp microprobe. [Right] Electron microscope image of a nanowire QD attached to a microprobe. \CCBYFourReprint{elshaari2017chip}
	}

\end{figure*}

Aberrations caused by refractive index mismatch between diamond and air (or immersion oil) may be corrected by using adaptive optics such as membrane deformable mirror (DM) and/or spatial light modulators (SLM) that modify the light's wavefront to compensate for the refractive index mismatch. Indeed, NV centers have been successfully created using focused fs laser pulses with a SLM that had a transverse spatial resolution of $\approx$ 200 nm that was limited by the diamond's nitrogen concentration\cite{Chen2017Laser}. More recently, it was demonstrated that the same fs laser system may be used to both create a vacancy \textit{and} anneal the diamond (locally) by careful control of the laser pulse energy\cite{Chen2019High}. Coupled together with real-time monitoring of the fluorescence, NV centers at a single site could be generated with near-unity yield and statistically selective generation of NV centers with a particular orientation is even possible by monitoring the polarization pattern of the fluorescence (which is correlated with the NV center's orientation) and keeping the annealing pulse on until a desired polarization pattern is generated (NV centers with the ``wrong orientation'' can be destroyed after creation by keeping the annealing pulses on)\cite{Chen2019High} (see Fig. \ref{Fig:NVLaserWrite}). 

\subsubsection{In-situ Lithography}
\label{sec:in_situ_litho}
Photonic structures can be fabricated on adjacent layers to a QD sheet.
However, in the absence of site-controlled growth\cite{rigal2018single}, the self-assembled QDs would be randomly located\cite{nakamura2004ultra}
and thus would not be optimally placed with respect to the photonic structures for efficient coupling.
To circumvent this problem, the QDs can first be located and pre-selected, e.g. via cathode luminescence, and the waveguide structures can then be patterned and etched via an \textit{in situ} lithography technique\cite{schnauber2019indistinguishable} (see Fig. \ref{fig:integ}b). 

With this technique, 
very small systematic misalignments ($<$\,10\,nm) have been achieved, as well as minimal fabrication-induced spectral shifts ($\lesssim$\,1nm) which can be compensated via tuning of the QD\cite{pregnolato2019deterministic}. 

This approach has also been used for nanodiamonds, where DLSPP waveguides were fabricated with deterministic positioning to include pre-characterized nanodiamonds with NV-centers \cite{siampour+2:17, siampour+2:17:2}.The authors were able to control the in-plane position of the nanodiamond with the desired NV-center within about 30 nm.


\subsection{Wafer Bonding}

Quantum emitters embedded in a high-quality bulk crystalline material are able to produce stable single-photon emission with high purity and indistinguishability. Ideally, hybrid heterostructures of QDs and photonic components can be grown directly on a single wafer. However, growing such heterostructures directly often results in poor crystal quality due to the formation of antiphase boundaries and large mismatches in material lattice constants, thermal coefficients, and charge polarity\cite{kim2020hybrid}.

Wafer-to-wafer bonding is a method for integrating dissimilar material platforms\cite{tanabe2012iii}. Consider the transfer of a III-V material onto a silicon nitride photonics circuit: Each material is grown separately with optimized substrates and conditions, thus maintaining high crystal quality for both compounds. The III-V wafer is flipped and bonded onto the top surface of the photonic wafer; subsequent removal of the substrate of the transferred wafer leaves a thin membrane structure on top of the photonic circuit. Photonic structures are then patterned using lithographic techniques; for example, the coupling of the emission from InAs QDs in a GaAs waveguide cavity structure into an underlying silicon nitride waveguide has been observed, with an overall coupling efficiency of $\sim$0.2 \cite{davanco2017heterogeneous} (Fig. \ref{fig:integ}c). Similar techniques can be employed to integrate diamond with other material platforms. For example, numerous GaP-on-diamond platforms\cite{Fu2008Coupling, Thomas2014Waveguide, Gould2016Large} have been realized by bonding a thin epitaxial film of GaP to a diamond substrate via Van der Waals bonding.

However, random positioning of SPEs with respect to the photonic structures in such wafers will result in non-optimal coupling, leading to a low yield of efficiently coupled devices across the wafer. 
This can be improved via \textit{in situ} lithography (see section \ref{sec:in_situ_litho}) around pre-identified SPEs after the wafer bonding step\cite{schnauber2019indistinguishable}.

The wafer bonding can also be performed orthogonally (Fig. \ref{fig:integ}d) for optimized SPE out-of-plane emission (e.g. with distributed Bragg reflector cavities\cite{murray2015quantum,ellis2018independent}) to be efficiently coupled into photonic waveguides. 
However, since only devices at the wafer edge can be integrated (as opposed to across the entire wafer for non-orthogonal bonding), this approach appears to be less scalable.

\subsection{Pick-and-place\label{Sec:QDPickNPlace}}

Another approach that allows for precise positioning of emitters on the photonic circuit is to pick-and-place individual emitters (or the nanostructures they are embedded in) instead of having a single wafer-scale integration step. 
Emitters can be pre-characterized and pre-selected, and then selectively integrated at desired positions on the photonic circuit. For example, they can be either placed on top of existing waveguides\cite{elshaari2017chip,mnaymneh2020chip, Mouradian2015Scalable}, or at specific points relative to a marker for subsequent waveguide encapsulation, i.e. waveguide material is deposited and patterned over the emitter\cite{zadeh2016deterministic}. 
Besides ensuring optimal coupling of the emitters to the photonic circuit, this method also allows for a greater flexibility in the choice of emitter host material and device geometry. 

There are two common techniques for performing the pick-and-place transfer: transfer printing via an adhesive stamp, and using a sharp microprobe.

\subsubsection{Transfer Printing}
The transfer printing method typically uses a stamp made of an adhesive and transparent material, such as polydimethylsiloxane (PDMS) or Gelfilm from Gelpak,
which allows for precise alignment of the structures under an optical microscope during the transfer.
This has been successfully demonstrated for exfoliable layered crystals\cite{Castellanos_Gomez_2014} and QDs\cite{katsumi2018transfer,katsumi2019quantum,katsumi2020situ} (Fig. \ref{fig:integ}e).
In the case of 2D flakes, using a dry viscoelastic stamp is advantageous compared to wet processes\cite{doi:10.1021/nl1008037, nnano_2010_172, doi:10.1063/1.3665405}, since there are no capillary forces involved which could potentially collapse suspended material.

First, the emitter (e.g. QD nanowire or exfoliated 2D flake) is attached to the stamp.
Next, the emitter is brought to the sample surface using XYZ micromanipulators. 
To release the emitter, the stamp is pressed against the surface and then peeled off slowly. 
Due to the stamp's viscoelasticity, it behaves as an elastic solid at short timescales while slowly flowing at longer timescales. 
Consequently, the viscoelastic material can detach from the emitter by slowly peeling the stamp off the surface.
Different strategies to control the adhesion of the stamp are detailed in a separate review paper \cite{linghu2018transfer}. 

There are several challenges in using the transfer printing technique. 
The stamping process induces a force over a large sample area and may damage parts of the fragile photonic circuit, although this may be mitigated by using a sufficiently small stamp not much larger than the transferred material\cite{lee2017printed}. 
Moreover, it is difficult to re-position the emitters as the adhesion between the integrated structures is typically much stronger than their adhesion to the stamp.

Nevertheless, this method provides close to 100\% success rate for transfer onto atomically flat materials, though for rougher surfaces the yield is lower due to reduced adhesion forces.
With the aid of additional alignment markers, positioning accuracies better than 100\,nm have been achieved\cite{katsumi2018transfer}. 

\subsubsection{Microprobe\label{Sec:pick_and_place_microprobe}}
An alternative is to perform the pick-up and transfer using a sharp microprobe (Fig. \ref{fig:integ}f). A small amount of adhesive (e.g. PDMS) can be added to the probe tip to aid the transfer, analagous to a micro-stamp\cite{Li2015Nanofabrication}. Although this technique requires precise control of the microprobe, it is able to transfer small, fragile structures such as single nanowires with high accuracy and controllability when aided by an optical microscope\cite{zadeh2016deterministic,elshaari2017chip,elshaari2018strain}, and especially so when using an electron microscope\cite{kim2017hybrid}. Currently, the manual transfer of individual devices one by one can be very time-consuming, but there is great potential in automating the process and allowing for scalable fabrication of many integrated devices.

\begin{figure}
\includegraphics[width=0.47\textwidth]{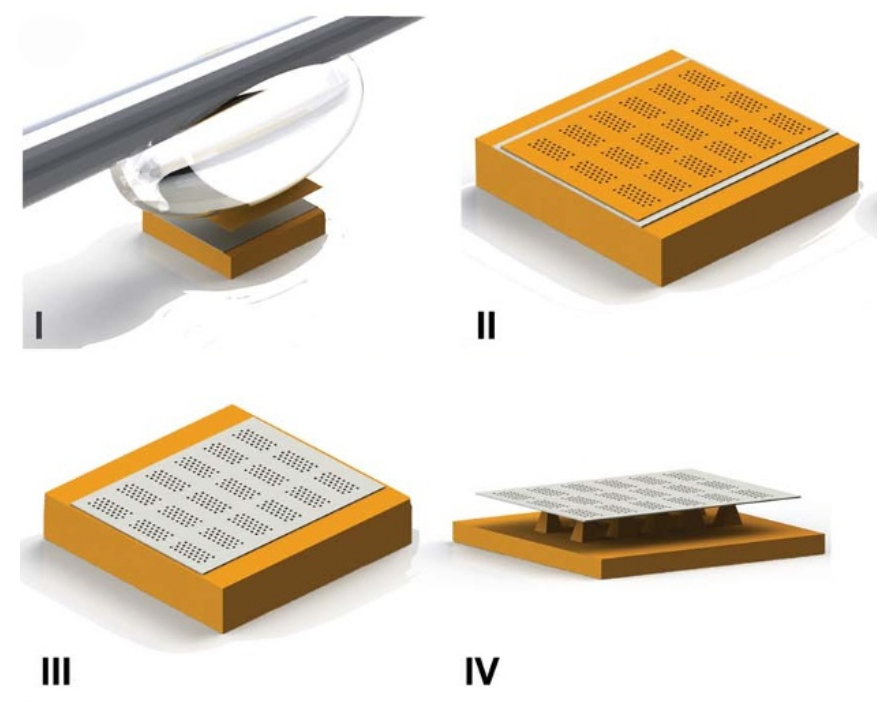}
\caption{Schematic of mask transfer technique. (I) A micro-PDMS adhesive on a tungsten probe tip is used to transfer a Si mask (orange) onto a diamond membrane (grey) on top of a Si substrate. (II) The Si mask serves as a etch mask for O$_2$ plasma RIE. (III) Pattern on Si mask is transferred to the diamond membrane after etching and removal of the Si mask. (IV) An isotropic SF$_6$ dry etch can then be used to undercut the Si substrate to create a suspended diamond photonic structure if desired. \CCBYNCNDFourReprint{Li2015Nanofabrication} }
\label{Fig:PDMSTransferSchematic}
\end{figure}

Besides picking up single nanowires, a microprobe enabled pick-and-place technique allows for the mechanical transfer and removal of Si masks onto diamond substrates (see Fig. \ref{Fig:PDMSTransferSchematic}). This allows for the fabrication of high quality Si masks due to existing mature Si processing technology, which when combined with a negligible Si etch rate during O$_2$ RIE etching of the diamond membrane, allows for tight fabrication tolerances of the diamond membrane. Moreover, such a mask can be reused multiple times due to the negligible Si etch rate. One significant advantage of this mask transfer technique is that the diamond membrane is never exposed to damaging irradiation that can adversely affect the NV$^-$ centers' properties. Indeed, using such a fabrication procedure, spin coherence lifetimes of $\sim$ 200 $\mu$s were measured for NV$^-$ centers coupled to suspended 1-D photonic crystals defects, which are similar to lifetimes measured in their parent CVD crystals\cite{Li2015Coherent}. Masks positioned in this way can be placed with sub-micron or even nm scale accuracy if integrated with an electron microscope\cite{Li2015Nanofabrication, Aoki2008Coupling}.

Another commonly used microprobe is the tip of an atomic force microscope (AFM). Combined with a confocal microscope, nanoparticles such as nanodiamonds containing desired color centers can be picked up and integrated with the optical devices such as photonic crystal cavity \cite{schell:11}. Scanning the AFM tip in intermittent contact mode over the focus of the confocal microscope allows identification of the pre-characterized nanodiamonds. The tip is then pressed on the center of the nanodiamond. A force is applied and surface adhesion allows the nanodiamond to be attached to the tip. After successful picking up, which could take over 50 trials, the tip is pressed over the new desired place to allow the nanodiamond to be integrated with the photonic structure. However, there is only limited success rate for the placing stage \cite{schell:11}.

\section{Conclusion and outlook\label{Sec:Conclusion}}

In this paper we have reviewed a variety of SPEs that are conveniently embedded within a solid. These emitters have promising properties and can be used to form the building blocks of future quantum networks. As discussed in section \ref{Sec:TheoryNV}, there are various important metrics to evaluate a SPE in terms of its suitability for various applications. 
For the sake of easy comparison, Table \ref{table:benchmarking} lists a selected range of integrated SPE, along with some of the metrics introduced in section \ref{Sec:TheoryNV}. 
Similarly, Table \ref{Table:resonator_comparison} tabulates various resonators that have been integrated with SPEs and characterizes them based on a few relevant properties.



\begin{table*}[!htb]
    \centering
    \begin{tabular*}{0.98\textwidth}{@{\extracolsep{\fill}} cccccccccc}
    \hline\hline
    SPE & Platform & $\lambda$\,(nm) & $g^{(2)}(0)$ & $R_\textrm{count}$ & $T_1$ (ns) & $\Gamma^\dagger$/2$\pi$ (GHz) & $\xi^\dagger$ & $V$ & Ref\\
    \hline
    
    InAs QD & GaAs WG on SiN WG & 1130 & 0.0(1) & - & 1.014(4) & - & - & - & \onlinecite{davanco2017heterogeneous}\\
    
    InAs QD & GaAs RR on SiN WG & 1110 & 0.52(8) & - & 0.263(7) & - & - & - & \onlinecite{davanco2017heterogeneous}\\ 
    
    InAsP QD & InP NW on SiN WG & 988 & 0.03 & 22.4\,MHz & 1.3 & 14.5 & 0.008 & - & \onlinecite{mnaymneh2020chip}\\ 
    
    InAs QD & GaAs WG on SiN WG & 916 & 0.11(4) & - & 1.39(4) & 2.20(19) & 0.052(6) & $0.89^{+0.11}_{-0.29}$ & \onlinecite{schnauber2019indistinguishable}\\ 
        
    InAs QD & InP NB on Si WG/GC & 1300 & 0.33 & 2.1\,MHz & 1.25 & - & - & - & \onlinecite{kim2017hybrid}\\
    
    InAs QD & GaAs NB/GC & 921 & - & - & 0.182(1) & 0.96(7) & 0.91(7) & - & \onlinecite{thyrrestrup2018quantum}\\ 
    
    InAs QD & GaAs NB/GC & 927 & 0.05 & 2.9\,MHz & 1.4 & - & - & - & \onlinecite{hummel2019efficient}\\
    
    InGaAs QD & GaAs NB & 921 & 0.006 & - & 0.185 & 1.12(3) & 0.77(2) & 0.94(1) & \onlinecite{kirvsanske2017indistinguishable}\\
        
    InAs QD & GaAs PCW/GC & 944-950 & - & - & 0.346(2) & 0.54 & 0.852(4) & - & \onlinecite{pedersen2020near}\\
    
    InAs QD & GaAs PCW/GC & 895 & 0.02(2) & - & 0.70(3) & 0.22(3) & 1.0(2) & 0.80(3) & \onlinecite{kalliakos2016enhanced}\\
        
    InAs QD & GaAs PCW to fiber & 931 & 0.20(8) & 8.2(1.7)\,MHz & 0.885 & - & - & - & \onlinecite{daveau2017efficient}\\
    
    InGaAs QD & GaAs WG & 941 & 0.009(2) & - & 0.50(1) & - & - & 0.975(5) & \onlinecite{dusanowski2019near}  \\
    
    \hline
    
    NV$^-$ & Diamond RR & 637 & - & - & 8.3 & 40 & 0.0005 & - & \onlinecite{Faraon2011Resonant} \\
    
    NV$^-$ & Diamond RR/WG/GC & 638 & - & 325 Hz & 8 & 51 & 0.0004 & - & \onlinecite{Faraon2013Quantum} \\
    
    NV$^-$ & Diamond RR/WG/GC & 637 & 0.24 & 15 kHz & - & - & - & - & \onlinecite{Hausmann2012Integrated} \\
    
    NV$^-$ & Diamond L3 PCC & 637 & 0.38 & 13.2 kHz & 4 & 10 & 0.004 & - & \onlinecite{Faraon2012Coupling} \\
    
    NV$^-$ & Diamond NB & 638 & 0.2 & - & - & 491 & - & - & \onlinecite{Hausmann2013Coupling} \\
    
    NV$^-$ & Diamond NB & 637 & 0.28 & 60 kHz & - & 15 & - & - & \onlinecite{Li2015Coherent} \\
    
    NV$^-$ & Diamond WG & 637 & 0.07 & 1.45 MHz & - & 0.323 & - & - & \onlinecite{Mouradian2015Scalable} \\
    \hline 
    NV$^-$ in ND & GaP PCC & 643 & $<0.5$  & 1 MHz & 12.7 & - & - & - & \onlinecite{englund+6:10} \\
    
    NV$^-$ in ND & HSQ WG/GC & 650-700 & $<0.5$ & - & 6 & - & - & - & \onlinecite{siampour+2:17} \\
    
     NV$^-$ in ND & HSQ Cavity/WG/GC & 700-750 & $<0.5$ & - & 3 & - & - & - & \onlinecite{siampour+2:17:2} \\
     
     NV$^-$ in ND & Ag Nanopatch antennas & 650 & 0.41 & 56.3\,MHz & 0.36 & - & - & - & \onlinecite{simeon:18} \\
     
     SiV in ND & Optical Microcavity& 737-759 & $<0.5$ & 0.106 & 0.46-1.97 & 21 & - & - & \onlinecite{benedikter+9:17} \\
     &to fiber &&&--\,1.78 MHz & \\
       
     GeV in ND & HSQ WG/GC & 602 & $<0.5$ & - & 3.8 & - & - & - & \onlinecite{siampour+3:18} \\
    \hline
    
    2D nanoflake & WSe$_2$ on silver NW & 736.74 & - & 30 kHz & 2.4 & 44.2 & 0.0015 & - & \onlinecite{journal69} \\
    2D nanoflake & WSe$_2$ on MIM WG & 737.1 & - & 300 Hz & 3.2(1.1) & 55.2 & 0.0009 & - & \onlinecite{journal70} \\
    2D nanoflake & WSe$_2$ on LiNbO$_3$ BS & 720-760 & - & 4 kHz & - & - & - & - & \onlinecite{journal71} \\
    2D nanoflake & WSe$_2$ on Si$_3$N$_4$ WG & 730-750 & 0.47 & 100 kHz & 7.99 & 725 & 27.5$\times10^{-6}$ & - & \onlinecite{journal72} \\

    \hline\hline
    \end{tabular*}
\caption{Comparison of selected integrated SPEs. $\lambda$ is the SPE's emission wavelength, $R_\textrm{count}$ gives the count rate and $V$ is the HOM visibility. $g^{(2)}(0)$ and $T_1$ are as defined in equations \eqref{eq:g2_reduced} and \eqref{Eq:P_def} respectively. $\Gamma^\dagger$ is the experimentally measured FWHM of the optical transition that may or may not be homogeneously broadened. $\xi^\dagger$ is computed using Eq. \eqref{Eq:P_def} with $\Gamma\to\Gamma^\dagger$. We note that although $\xi^\dagger\in[0,1]$ no longer predicts the size of a HOM dip (since $\Gamma^\dagger\neq\Gamma$ for an inhomogeneously broadened line), it still serves as a measure of indistinguishability, with $\xi^\dagger=1$ indicating perfect indistinguishability. We have made the following abbreviations for the sake of brevity: WG - waveguide, RR - ring resonator, NW - nanowire, NB - nanobeam, PCW - photonic crystal waveguide, PCC - photonic crystal cavity, GC - grating coupler, and BS - beamsplitter. Measurement uncertainties are given in brackets.}

\label{table:benchmarking}
\end{table*}

\begin{table*}[!htb]
	\centering
	\begin{tabular*}{0.95\textwidth}{@{\extracolsep{\fill}} cccccccc}
		\hline\hline
		Type & Material & $Q_{the}$ & $Q_{exp}$ & $V_{mode}$ $(\lambda/n)^3$ & $\lambda$ (nm) & $F_{exp}$ & Ref \\
		\hline
		Ring resonator & Diamond & - & 5500 & 15 & 638 & 12 & \onlinecite{Faraon2013Quantum} \\
		
		L3 PCC & Diamond & 6000 & 3000 & 0.88 & 637 & 70 & \onlinecite{Faraon2012Coupling} \\
		
		Ring resonator & Diamond & $>10^6$ & $\sim$ 4000 & 17 - 32 & 637 & 12 & \onlinecite{Faraon2011Resonant} \\
				
		1D nanobeam & Diamond & $10^4$ & 1635 & 3.7 & 638 & 7 & \onlinecite{Hausmann2013Coupling} \\
		
		Ring resonator + WG & Diamond & - & $(3.2\pm0.4)\times10^3$ & - & 665.9 & - & \onlinecite{Hausmann2012Integrated} \\
		
		1D nanobeam & Diamond & - & 7000 & 0.47 & 637 & 22 & \onlinecite{Lee2014Deterministic} \\
		
		
		1D nanobeam & Diamond & - & 3300 & - & 637 & 62 & \onlinecite{Li2015Coherent} \\
		
		Disk resonator & GaP & - & $2500-10,000$ & - & 637 & - &  \onlinecite{Gould2016Large} \\
		
		L3 PCC & ND+GaP & 1000 & 603 & 0.75 & 639.5 & 12.1 &  \onlinecite{wolters:10} \\
		
		L3 PCC & ND+GaP & 6000 & 610 & 0.74 & 643 & 7.0 &  \onlinecite{englund+6:10} \\
		
		Microdisk cavity & ND+Si$\text{O}_2$ & 340000 & 170000 & 82 & 637 & - &  \onlinecite{barclay+5:09} \\

        Optical microcavity & ND+Planar mirror & - & 19000 & 3.4 & $737-759$ & 9.2 &  \onlinecite{benedikter+9:17} \\
        		&+ SM optical fiber & \\

		Microdisk resonantor & ND+3C-SiC & 150000 & 2700 & 5.5 & 737 & $2-5$ &  \onlinecite{marina:19} \\
		
		\hline\hline
	\end{tabular*}
\caption{Comparison of various integrated resonators. $Q_{the}$ and $Q_{exp}$ are the theoretical and experimental quality factors respectively. $V_{mode}$, $\lambda$ and $F_{exp}$ are the mode volume, resonance wavelength and experimental Purcell factor respectively. }
\label{Table:resonator_comparison}
\end{table*}

As tables \ref{table:benchmarking} and \ref{Table:resonator_comparison} show, there are a few ways of integrating SPEs with on-chip optical structures and thus realizing functional quantum devices. Indeed, these interfaces enhance the light-matter interaction and allow efficient interaction and entanglement between the distant emitters. However, translation from proof-of-concept laboratory demonstrations of individual components to full-scale quantum devices is still quite immature. Considerable efforts are required to overcome issues associated with the material incompatibility of quantum emitters, photonic circuits and other required components on the same chip. Furthermore there are still challenges in developing  high-throughput and reliable integration techniques.  We propose the following four critical steps to address these challenges: 

First, quantum photonic devices should be thoroughly designed, fabricated and tested since quantum information processing imposes stringent demands on loss and fabrication accuracy. These demands, which are at the limits of conventional silicon photonic technology, might require fabrication for quantum applications to be achieved at the expense of scalable fabrication by, for example, using time-consuming electron beam lithography instead of photolithography. 

Second, the coupling between quantum systems and resonant photonic cavities should be optimized through the accurate positioning of the emitter in the cavity. This calls for further improvements in the reliability and throughput of methods such as AFM maniupulation, nano-patterning, and various transfer techniques that have been employed for accurate positioning of these quantum emitters.

Third, full scalability implies integration with on-chip single-photon sensitive detectors and lasers on-chip. The development of on-chip active devices would eliminate the need for using bulk optics and allow a significantly smaller footprint for the photonic platform. Design, fabrication, and characterization of on-chip photodetectors and lasers operating at desired performance levels is a challenging task since it involves multiple fabrication steps involving various materials that require state-of-the-art clean room facilities and a comprehensive strategy for heat management and integrating associated optoelectronics.

Fourth, a crucial building block for quantum networks is the realization of quantum-mechanical interaction and entanglement between two separate quantum nodes on the same optical chip. Photons emitted by two independent nodes should be able to coherently interfere on a beamsplitter and produce an interference signal. This task requires demanding control of quantum emitters and photonic elements but demonstration of such an interaction would lead to more complicated schemes of quantum networking, including the interaction of a large number of quantum emitters on the same chip.

Clearly, a consolidation of new technologies is required to address these challenges and to demonstrate a platform for quantum networking that is scalable and amenable to mass manufacturing. This program should leverage on a broad collaboration between experts in quantum physics, integrated photonics, material science, and electronics. 



%
%

%

\section*{Data availability statement}
Data sharing is not applicable to this article as no new data were created or analyzed in this study.

\begin{acknowledgments}
This work was supported by NRF-CRP14-2014-04, ``Engineering of a Scalable Photonics Platform for Quantum Enabled Technologies'' and the Quantum Technologies for Engineering program of A*STAR. The authors also acknowledge  support from the Quantum Technologies for Engineering (QTE) program of A*STAR project \# A1685b0005. 
\end{acknowledgments}

\newpage
\section*{References}

\bibliography{References}

\end{document}